\documentclass[%
prx,
aps,
amsmath,amssymb,
longbibliography,
superscriptaddress,
reprint,
nofootinbib,
citeautoscript
]{revtex4-2}
\usepackage{mathtools,amsmath,amssymb,amsthm,bm,mathrsfs}
\usepackage{float}
\usepackage{graphicx}
\usepackage[dvipsnames]{xcolor}
\usepackage{hyperref,url}
\usepackage[capitalize]{cleveref}
\definecolor{citeblue}{HTML}{0055cc}
\hypersetup{
breaklinks,
colorlinks=true,
linkcolor=citeblue,
citecolor=citeblue,
urlcolor=citeblue,
filecolor=citeblue
}
\usepackage[T1]{fontenc}
\usepackage[utf8]{inputenc}
\usepackage{newunicodechar}
\newunicodechar{−}{-}
\newunicodechar{‖}{\ensuremath{\Vert}}
\usepackage{lmodern}
\usepackage{microtype}
\usepackage[linesnumbered,ruled,vlined]{algorithm2e}
\usepackage{mhchem}
\makeatletter 
    
\renewcommand\onecolumngrid{% <<<<<<
\do@columngrid{one}{\@ne}%
\def\set@footnotewidth{\onecolumngrid}% <<<<<<<<<<<<<<<<
\def\footnoterule{\kern-6pt\hrule width 1.5in\kern6pt}%
}

\renewcommand\twocolumngrid{% <<<<<<
        \def\footnoterule{% restore rule
        \dimen@\skip\footins\divide\dimen@\thr@@
        \kern-\dimen@\hrule width.5in\kern\dimen@}
        \do@columngrid{mlt}{\tw@}
}%
\newcommand{\FeS}{[Fe$_2$S$_2$(SCH$_3$)$_4$]$^{2-}$}

\makeatother

\newtheorem{theorem}{Theorem}[section]
\newtheorem{lemma}[theorem]{Lemma}
\newtheorem{corollary}[theorem]{Corollary}
\newtheorem{proposition}[theorem]{Proposition}

\theoremstyle{definition}
\newtheorem{remark}[theorem]{Remark}
\newtheorem{definition}[theorem]{Definition}

\SetKwInput{KwPromise}{Promise}

\newcommand{\mHa}{\textup{m}E_\textup{h}}
\newcommand{\R}{\mathbb{R}}
\newcommand{\C}{\mathbb{C}}
\newcommand{\I}{\mathbb{I}}
\newcommand{\zero}{\mathbb{O}}
\newcommand{\T}{\mathrm{T}}
\newcommand{\calO}{\mathcal{O}}
\newcommand{\scrL}{\mathscr{L}}
\newcommand{\scrM}{\mathscr{M}}
\newcommand{\scrR}{\mathscr{R}}
\renewcommand{\Re}{\operatorname{Re}}
\renewcommand{\Im}{\operatorname{Im}}

\newcommand{\Orth}{\mathrm{O}}
\newcommand{\SO}{\mathrm{SO}}
\newcommand{\U}{\mathrm{U}}
\newcommand{\Sym}{\mathrm{S}}
\newcommand{\B}{\mathrm{B}}
\newcommand{\Gr}{\mathrm{Gr}}
\DeclareMathOperator{\tr}{tr}
\DeclareMathOperator{\pf}{pf}
\DeclareMathOperator{\rank}{rank}

\DeclareMathOperator{\inv}{inv}
\DeclareMathOperator{\diag}{diag}
\DeclareMathOperator*{\E}{\mathbb{E}}
\DeclareMathOperator{\V}{Var}
\DeclareMathOperator{\spn}{span}

\newcommand{\ket}[1]{| #1 \rangle}
\newcommand{\bra}[1]{\langle #1 |}
\newcommand{\braket}[1]{\langle #1 \rangle}
\newcommand{\ketbra}[2]{\ket{#1}\!\bra{#2}}
\newcommand{\vac}{\mathbf{0}}
\newcommand{\Nsamp}{N_{\mathrm{s}}}
\newcommand{\Nwalk}{N_{\mathrm{w}}}
\newcommand{\Ntau}{N_{\tau}}

\newcommand{\Ntrain}{N_{\mathrm{train}}}

\newcommand{\Mchan}{\mathcal{M}}
\newcommand{\Pdeg}{\mathcal{P}}
\newcommand{\Qtilde}{\widetilde{Q}}
\newcommand{\Sbar}{\overline{S}}
\newcommand{\even}{\mathrm{even}}

\SetKwInput{KwInput}{Input}
\SetKwInput{KwOutput}{Output}
\SetKwComment{Comment}{$\triangleright$\ }{}
\newcommand{\inter}{\mathrm{inter}}
\newcommand{\intra}{\mathrm{intra}}

\newcommand{\CZ}{\mathrm{CZ}}
\newcommand{\LUCJ}{\mathrm{LUCJ}}
\newcommand{\CompileMatchgatePermutation}{\mathsf{CompileMatchgatePermutation}}
\newcommand{\ReducePairs}{\mathsf{ReducePairs}}
\newcommand{\SortPermutation}{\mathsf{SortPermutation}}
\newcommand{\CompileSwaps}{\mathsf{CompileSwaps}}
\newcommand{\SortFour}{\mathsf{SortFour}}
\newcommand{\changed}{\textup{\textsc{Changed}}}
\newcommand{\true}{\textup{\textsc{True}}}
\newcommand{\false}{\textup{\textsc{False}}}

\renewcommand{\l}[1]{\mathopen{}\left#1}
\renewcommand{\r}[1]{\right#1\mathclose{}}

\input{splitrefs}

\NewDocumentCommand{\az}{s m}{%
  \IfBooleanTF{#1}
    {{\color{magenta}#2}}
    {{\color{magenta}AZ: #2}}
}

\setcitestyle{super,open={},close={}}

\makeatletter
\def\super@cite@check{%
  \ifcat\noexpand\@let@token.%
    \expandafter\super@cite@swap
  \else
    \expandafter\super@cite@end
  \fi}
\newcommand\hidemaintoc{%
  \let\SIsaved@l@section\l@section       \let\l@section\@gobbletwo
  \let\SIsaved@l@subsection\l@subsection \let\l@subsection\@gobbletwo}
\newcommand\showmaintoc{%
  \let\l@section\SIsaved@l@section
  \let\l@subsection\SIsaved@l@subsection}
\makeatother

\begin{document}

\title{Machine learning quantum data to accelerate quantum Monte Carlo}

\newcommand{\affHarvard}{\affiliation{Department of Chemistry and Chemical Biology, Harvard University, Cambridge, MA 02138, USA}}
\newcommand{\affGoogle}{\affiliation{Google Quantum AI, Santa Barbara, CA 93111, USA}}
\newcommand{\affGooglesecond}{\affiliation{Google Research, Mountain View, CA 94043, USA}}
\newcommand{\affSandiaCA}{\affiliation{Quantum Algorithms and Applications Collaboratory, Sandia National Laboratories, Livermore, CA 94550, USA}}
\newcommand{\affSandiaNM}{\affiliation{Quantum Algorithms and Applications Collaboratory, Sandia National Laboratories, Albuquerque, NM 87185, USA}}
\newcommand{\affMIT}{\affiliation{Research Laboratory of Electronics, Massachusetts Institute of Technology, Cambridge, 02139 MA, USA}}
\newcommand{\affcquic}{\affiliation{Center for Quantum Information and Control (CQuIC), University of New Mexico, Albuquerque, NM 87106}}

\author{Tong Jiang}
\thanks{These authors contributed equally.}
\affHarvard
\author{Andrew Zhao}
\thanks{These authors contributed equally.}
\affSandiaCA
\author{Shashwat Kumar}
\affGoogle
\author{Jigyasa Nigam}
\affHarvard
\author{Eliott Rosenberg}
\affGoogle
\author{Ryan Babbush}
\affGoogle
\author{Andrew D. Baczewski}
\affSandiaNM
\author{William J. Huggins}
\affGoogle
\author{Joonho Lee}
\email{joonholee@g.harvard.edu}
\affHarvard
\affGoogle

\date{\today}

\begin{abstract}
Simulating strongly correlated electrons in molecules and materials remains a central goal for quantum computing. A promising strategy uses quantum processors to prepare trial wavefunctions that guide classical quantum Monte Carlo (QMC). However, the enormous computational cost of repeatedly processing quantum measurements has severely limited the practical scalability of this approach. Here, we overcome this bottleneck by compressing quantum measurement data into machine-learning models that efficiently provide the wavefunction overlaps needed by QMC. This accelerates quantum-classical QMC by orders of magnitude, enabling chemically accurate simulations on up to 40 qubits, more than doubling the scale of previous hardware demonstrations with comparable accuracy. We demonstrate our approach on hydrogen chains and strongly correlated iron--sulfur clusters using measurements from Google's Willow processor. By enabling efficient reuse of quantum-state information on the classical coprocessor, our framework addresses a fundamental bottleneck in hybrid quantum algorithms that will persist even into the fault-tolerant era.
\end{abstract}

\maketitle
\addtocontents{toc}{\protect\hidemaintoc}
{\it Introduction.} Accurately simulating correlated electrons remains a central challenge in computational science, with applications ranging from molecular discovery to high-temperature superconductivity.~\cite{heather2021,RevModPhys.90.011001}
Quantum computers offer a fundamentally different way to represent electronic correlation, motivating extensive efforts to develop quantum algorithms for chemistry and materials science.~\cite{bauer2020algorithms,chien2026simulating,bidart2025quantum,valahu2023direct}
Among these approaches, quantum--classical hybrid quantum Monte Carlo (QC-QMC) has emerged as a promising strategy for achieving high accuracy.~\cite{huggins2022unbiasing}
In QC-QMC, a quantum processor prepares a correlated trial wavefunction whose overlaps with sampled many-body configurations (walkers) guide the classical Monte Carlo projection and help control the fermionic sign problem.
The key promise of QC-QMC is that expressive quantum circuits can provide trial wavefunctions beyond the reach of classical approximations, while the subsequent Monte Carlo projection relaxes the accuracy requirements on quantum state preparation.
Realizing this promise, however, requires an efficient interface between quantum overlap evaluations and classical QMC calculations.

\begin{figure}[!t]
    \centering
    \includegraphics[width=0.42\textwidth]{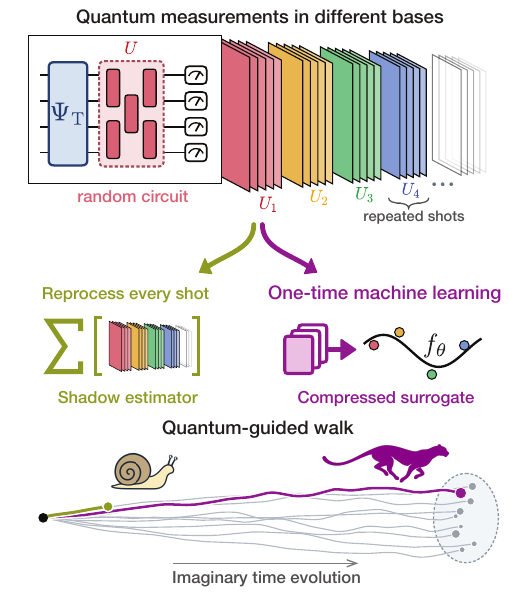}
    \caption{\textbf{Learning from quantum data to accelerate QC-QMC.}
    A quantum processor prepares a trial state $|\Psi_\mathrm{T}\rangle$ and collects randomized measurements (``classical shadows''), each repeated for multiple shots. Conventional QC-QMC repeatedly processes these measurements to estimate trial-state overlaps with walkers throughout imaginary-time propagation. In our approach, we compress the measurement data into a machine-learning surrogate $f_\theta$, which efficiently provides the overlaps QMC requires without repeated shadow postprocessing.
    In the quantum-guided walk, each path is a QMC walker propagating in imaginary time from the initial state (black circle); the final walkers (dots, sized by weight) form an ensemble (shaded oval) that samples the ground state.
    }
    \label{fig:qcqmc-overview}
\end{figure}

To avoid the latency of repeated quantum--classical feedback, QC-QMC collects a reusable ensemble of randomized measurements of the trial state, from which a classical coprocessor estimates the overlaps required by QMC.~\cite{huang2020predicting,huggins2022unbiasing,zhao2021fermionic,wan2023matchgate}
Yet, repeatedly reconstructing these overlaps for the many walkers encountered throughout QMC propagation incurs enormous classical computational cost~\cite{kiser2024cost,huang2024evaluating}.
This postprocessing bottleneck has severely limited hardware demonstrations, which have so far reached only 16 qubits~\cite{huggins2022unbiasing,zhao2026trapped,nappi2026practicability} and consumed over 13,000 GPU hours for a single calculation~\cite{zhao2026trapped}. Meanwhile, a 40-qubit calculation was estimated to require millions of CPU hours~\cite{nappi2026practicability}.

Here, we overcome this bottleneck by developing a machine learning (ML) framework (Fig.~\ref{fig:qcqmc-overview}) that compresses quantum measurement data into a classical surrogate.~\cite{carleo2017solving,choo2020fermionic,hermann2020deep,pfau2020abinitio,wei2024neural,huang2025certifying}
Rather than reconstructing the full quantum state, we learn its overlaps with the many-body configurations relevant to QMC.
Once trained, the surrogate efficiently supplies these overlaps and their derivatives, bypassing repeated classical postprocessing of quantum measurements.
This reduces the computational cost by orders of magnitude and makes end-to-end QC-QMC on 40 qubits of Google's Willow processor affordable, more than doubling the scale of previous hardware demonstrations.
At this scale, QC-QMC gives energies within chemical accuracy, for hydrogen chains with a mean-field trial and for an iron--sulfur cluster with a correlated trial.

{\it Theory and algorithms.}
In projector QMC, the exact ground state, $|\Psi_0\rangle$, is obtained by imaginary-time propagation, which is approximated by a stochastically sampled ensemble of wavefunctions,
\begin{equation}
    |\Psi_0\rangle \propto
    \lim_{\tau\rightarrow\infty}e^{-\tau\hat{H}}|\Phi_0\rangle,
    \qquad
    |\Psi(\tau)\rangle \approx
    \sum_{i=1}^{\Nwalk} w_i|\phi_i\rangle.
\end{equation}
The initial state $|\Phi_0\rangle$ is assumed to have nonzero overlap with the ground state, and $\{|\phi_i\rangle\}$ are statistical sample wavefunctions, known as walkers, with associated weights $\{w_i\}$.
In unconstrained QMC, cancellations among walker weights cause the signal-to-noise ratio to decay exponentially with increasing imaginary time and system size, a manifestation of the notorious fermionic sign problem~\cite{PhysRevB.55.7464}. 
Practical QMC methods control this instability by guiding the calculation with constraints imposed by a trial wavefunction $|\Psi_\T\rangle$. In this work, we use auxiliary-field QMC,~\cite{zhang2003phaseless,motta2018ab} but our technique is applicable to other QMC methods~\cite{jiang2025walking,blunt2025quantum}. 
The trial wavefunction enters QMC through its overlaps with the walkers, $\langle\Psi_\T|\phi_i\rangle$, and their derivatives, which determine the walker evolution and energy estimates (Supplementary Sec.~\ref{app:afqmc}).

To obtain these overlaps from a quantum-prepared trial state~\cite{huggins2022unbiasing}, we perform randomized measurements via Clifford matchgate circuits, yielding classical snapshots that we store and reuse to estimate overlaps with the walkers~\cite{wan2023matchgate}. Using an improved algorithm for processing matchgate shadows, we can process each snapshot per overlap with $\calO(N^3)$ cost, where $N$ is the number of qubits (Supplementary Sec.~\ref{app:overlap-algorithm}).
This operation must be repeated for $\Nsamp$ measurements, $\Nwalk$ walkers, and $\Ntau$ imaginary-time steps, resulting in a cost of $\calO(N^3\Nsamp\Nwalk\Ntau)$ for the overlaps alone, before accounting for the derivatives needed for the importance sampling and energy estimates.

Our central idea is to decouple quantum-measurement processing from repeated overlap evaluations during QMC.
We first estimate complex-valued overlaps for a much smaller set of $\Ntrain \ll \Nwalk\Ntau$ training determinants and use them to train a differentiable surrogate overlap model $f_\theta(\phi)\propto\langle\Psi_{\mathrm{T}}|\phi\rangle$.
This introduces a one-time labeling cost of $\calO(N^3\Nsamp\Ntrain)$. Subsequent QMC propagation queries only the trained surrogate, whose evaluation cost is independent of $\Nsamp$ and whose derivatives can be evaluated directly without reprocessing the quantum measurements.
Further details are provided in Supplementary Sec.~\ref{app:method}.

In principle, the surrogate can be any differentiable model of the overlap trained on the measurement data.
We train $f_\theta$ by supervised learning on overlap labels representative of those used by QMC.
Because the training labels are obtained from noisy quantum measurements, care must be taken to prevent the model from fitting overlap uncertainty due to the measurement noise and a finite number of measurements~\cite{torlai2018neural,carrasquilla2019reconstructing,PhysRevResearch.5.013026}.
We quantify the relative statistical error of the experimental labels by a noise floor $F$, estimated from fluctuations across subsets of the measurement circuits (Supplementary Sec.~\ref{app:shadow_noise} and Fig.~\ref{fig:noise_floor_LUCJ}).
Since the labels have an uncertainty level on the order of $F$ itself, a learned model whose relative error on the validation labels (Eq.~\eqref{eq:v}) falls below $F$ may be fitting uncertainty.
We therefore use this as a stopping criterion for the training.

In general, the quantum trial states may be prepared through variational quantum algorithms~\cite{cerezo2021variational}, high-fidelity quantum state preparation~\cite{berry2025rapid}, or quantum implementations of sophisticated classical wavefunctions. We demonstrate our ML-surrogate framework in two settings. First, we use mean-field trial wavefunctions for hydrogen chains containing up to 20 atoms (\ce{H20}, 40 qubits) to benchmark our proposed approach. To enable ML inference for arbitrary QMC walker configurations, we extend neural quantum states (NQS)~\cite{carleo2017solving} to accept arbitrary Slater determinants as inputs. Training NQS with randomized quantum measurements has shown promise~\cite{huang2025certifying,wei2024neural}, but previous models were restricted to bit-string amplitudes, rather than the overlaps with rotated Slater determinants required by auxiliary-field QMC. 

Second, for strongly correlated iron--sulfur complexes, preparing an accurate trial wavefunction directly on near-term quantum hardware remains prohibitively expensive. For our hardware demonstration, we instead assume access to an accurate wavefunction and approximate it with shallow quantum circuits, whose parameters we optimize classically to reproduce overlaps with representative training walkers. We then execute these circuits on quantum hardware, collect randomized measurements, and train our ML surrogate based on nonorthogonal configuration interaction (NOCI)~\cite{jimenez2013multi,giuliani2026precise}. This enables chemically accurate QC-QMC calculations for \FeS{} in 28- and 40-qubit active spaces, despite hardware noise.

\begin{figure}[t]
    \centering
    \includegraphics[width=0.48\textwidth]{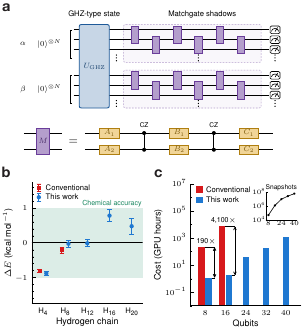}
    \caption{\textbf{Accurate and accelerated QC-QMC using experimental randomized measurements.}
    \textbf{a}.~Quantum circuit: a GHZ-type state is prepared on the $\alpha$ (spin-up) and $\beta$ (spin-down) registers. Independent random matchgate circuits then act on each register before computational-basis measurements. The lower diagram illustrates the decomposition of a matchgate block into single-qubit Clifford and CZ gates.
    \textbf{b}.~QC-QMC energy deviations from the noiseless, infinite-shot reference for hydrogen chains from \ce{H4} to \ce{H20} (8 to 40 qubits) at an interatomic separation of $3.0$ Bohr in the STO-6G basis. Results from conventional shadow postprocessing (red) are compared with those from the ML surrogate (blue). 
    The shaded region indicates chemical accuracy ($\pm1$~kcal\,mol$^{-1}$). ML results are averaged over independently trained models (Fig.~\ref{fig:seeds_dependence}), with error bars indicating standard errors of the mean, including reference uncertainty.
    \textbf{c}.~Total classical computational cost of conventional QC-QMC and our ML-accelerated approach. At 8 and 16 qubits, both methods use the same quantum measurement data, yielding speedups of 190- and 4,100-fold, respectively. The ML costs include training-label generation, surrogate training, and QMC propagation. The inset shows the total number of measurement snapshots (circuits $\times$ shots) used at each system size (Supplementary Table~\ref{tab:cz_counts}).
    }
    \label{fig:hchain}
\end{figure}

{\it Results and discussion.}
The experiments in this work were performed on a Google Willow quantum processor~\cite{google2025willow}, which has over a hundred qubits, two-qubit gate error rates on the order of $10^{-3}$ (Supplementary Fig.~\ref{fig:hardware_error}), and reported repetition rates of approximately $6\times10^4$ shots per second. 
This enables us to collect millions of classical shadow measurements every few minutes.
To suppress the effects of hardware noise, we employ randomized compiling, dynamical decoupling, parity postselection, and qubit-assignment averaging (Supplementary Sec.~\ref{app:error-mitigation}).

Hydrogen chains are a standard benchmark for strongly correlated electronic structure methods~\cite{motta2017towards,motta2020ground}, as the degree of electronic correlation can be tuned by varying the interatomic separation. 
Increasing the chain length also provides a systematic way to assess how computational methods scale with system size.
As a controlled benchmark of our ML framework, we use a spin-unrestricted Hartree--Fock (HF) trial state for hydrogen chains up to \ce{H20}, performing the experiments and QMC calculations in the HF basis.
We show the circuit in Fig.~\ref{fig:hchain}a: a GHZ-type state is prepared, and each spin register is then measured through its own random Clifford matchgate circuit.
Measuring the two registers separately suffices because each QMC walker is a product of an $\alpha$ and a $\beta$ determinant~\cite{zhao2024group} (Supplementary Sec.~\ref{app:spin-block}).
Together with an improved compilation (Supplementary Sec.~\ref{app:matchgate_compilation}), this keeps the circuit sizes compact (Supplementary Fig.~\ref{fig:Hchain_matchgate_counts}).
At every size, our matchgate circuits need under a third of the $\CZ$ gates and under half the depth of the full matchgate circuits used in previous experiments~\cite{zhao2026trapped}. Shallower circuits accumulate less hardware noise per snapshot, which is crucial for extracting useful overlap information (Supplementary Fig.~\ref{fig:SI_hchain_conv}).

We train an expressive NQS surrogate directly on noisy measurement data, starting from a random initialization with no classical initial guess.
Fig.~\ref{fig:hchain}b shows the QC-QMC energies obtained with the trained surrogate as the QMC trial, relative to the ideal limit (infinite number of shots and noiseless devices) of QMC with HF trial.
At every size, the deviation averaged over networks trained with different random seeds (Fig.~\ref{fig:seeds_dependence}) lies within chemical accuracy (1~kcal\,mol$^{-1}$).
Fig.~\ref{fig:hchain}c compares the cost of our approach with that of conventional QC-QMC, which evaluates every overlap, force bias, and local energy directly from the snapshots, using our optimized GPU implementation (Supplementary Fig.~\ref{fig:gpu_timimg}).
At 8 and 16 qubits, conventional QC-QMC needs 100 and 2,500 circuits, respectively, at 500 shots each, to reach chemical accuracy.
With the same data, our ML model reduces QC-QMC cost by 190-fold (8 qubits) and 4,100-fold (16 qubits). This comparison includes the cost of labeling the training data, training the NQS model, and running QMC end-to-end (Table~\ref{table:hchain-cost}).
At 24 qubits and beyond, the estimated cost of processing all measurement data using conventional QC-QMC exceeds $10^{5}$ GPU hours, making direct benchmarking prohibitively expensive.
At these sizes, the one-time cost of producing the training labels dominates the total cost of our ML approach (>90\%), which is still orders of magnitude smaller than a conventional QC-QMC run.

\begin{figure*}[!t]
    \centering
    \includegraphics[width=0.9\textwidth]{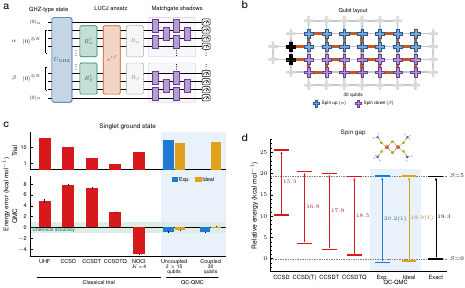}
    \caption{\textbf{Chemically accurate QC-QMC for \FeS{} in a 28-qubit active space.}
    The active space contains 18 electrons in 14 orbitals (18e, 14o).
    \textbf{a}.~Quantum circuit consisting of GHZ-type state preparation, a shallow local unitary cluster Jastrow (LUCJ) ansatz, and randomized matchgate measurements. We consider spin-coupled and spin-uncoupled variants. The former includes $\alpha$--$\beta$ Jastrow interactions, whereas the latter operates on two independent 15-qubit spin registers, each containing one ancilla to ensure even fermion parity.
    Full circuits in Supplementary Figs.~\ref{fig:LUCJ_circuit_14o-coupled} and~\ref{fig:LUCJ_circuit_20o-uncoupled}.
    \textbf{b}.~Qubit layout on the Willow processor.
    Each spin register ($\alpha$, blue; $\beta$, purple) runs along a two-row snake of dark orange couplers, which both circuits use; black crosses mark the ancilla qubit of each register. The light orange couplers between the registers are used only by the spin-coupled circuit, for its $\alpha$--$\beta$ Jastrow pairs. Grey qubits and couplers are unused. Error rates in Supplementary Fig.~\ref{fig:hardware_error}.
    \textbf{c}.~Singlet ground-state energy errors relative to the exact result, comparing classical trials (red) with quantum-circuit trials (blue and yellow). The upper panel shows trial-state errors, while the lower panel shows the corresponding QMC errors. Experimental QC-QMC results (blue) use measurements from Willow; ideal results (yellow) use noiseless circuits in the infinite-shot limit. The shaded region indicates chemical accuracy ($\pm1$~kcal\,mol$^{-1}$).
    \textbf{d}.~Energies of the singlet ($S=0$) and high-spin ($S=5$) states relative to the exact singlet energy, comparing unrestricted coupled-cluster methods, experimental and ideal QC-QMC, and exact results. Vertical lines indicate the corresponding spin gaps.
    }
\label{fig:fes}
\end{figure*}

We next turn to a widely studied strongly correlated system: the iron--sulfur cluster. This class of transition-metal complexes facilitates electron transfer and catalysis across a wide range of biological processes~\cite{beinert1997iron}. 
Their electronic structure is challenging owing to competing spin couplings among iron $d$ electrons, which produce many nearly degenerate spin states~\cite{sharma2014fes,li2019electronic,zhai2026femoco,reiher2017elucidating}, and charge fluctuations between iron and sulfur atoms.
Achieving high accuracy at low computational cost for this compound and related systems has been challenging for both classical QMC~\cite{kjonstad2026ironafqmc} and near-term quantum methods such as sample-based quantum diagonalization (SQD)~\cite{robledo2025chemistry,reinholdt2025critical,danilov2025enhancing,kanno2026quantum}.
This provides a more demanding test of our framework than the hydrogen chain benchmark.

For this demonstration, we use a shallow, hardware-efficient ansatz based on the local unitary cluster Jastrow (LUCJ)~\cite{huggins2022unbiasing,motta2023lucj}, as in the SQD study~\cite{robledo2025chemistry} but with a different parameterization (Supplementary Sec.~\ref{app:lucj}).
We optimize the LUCJ circuit parameters to reproduce overlaps between the reference wavefunction and training walkers.
Interestingly, direct variational energy minimization of the LUCJ wavefunction does not yield accurate QC-QMC results (Table~\ref{tab:lucj_exact}).
Here, we use a compact NOCI expansion~\cite{jimenez2013multi,giuliani2026precise} as the surrogate, initialized from an inexpensive classical calculation and subsequently trained on quantum measurements. This provides a direct comparison with a classically optimized NOCI trial containing the same number of determinants, allowing us to assess the benefit of learning from quantum data.
We study two active spaces, (18-electron, 14-orbital) and (30-electron, 20-orbital), hereafter denoted as (18e, 14o) and (30e, 20o), respectively, which map onto 28 and 40 qubits. 

In Fig.~\ref{fig:fes}a, we show the quantum circuit for (18e, 14o): a GHZ-type preparation followed by a shallow LUCJ ansatz, measured with spin-blocked randomized measurements~\cite{zhao2024group} (matchgate shadows). 
Each spin register runs along a two-row snake on Willow (Fig.~\ref{fig:fes}b).
The spin-coupled circuit places its $\alpha$--$\beta$ Jastrow pairs on the couplers (light orange) between the two snakes, whereas the spin-uncoupled circuit omits them and runs as two independent 15-qubit circuits including ancillas.

Fig.~\ref{fig:fes}c compares classical QMC and QC-QMC for the singlet ground state. To illustrate the difficulty of this problem, we show the coupled-cluster (CC) hierarchy up to singles (S), doubles (D), triples (T), and quadruples (Q), where the trial energy improves as one adds more excitations, but the QMC error does not quickly converge to chemical accuracy.~\cite{kjonstad2026ironafqmc} Even CCSDTQ, whose own energy is within chemical accuracy, leaves an AFQMC error of 2.8~kcal\,mol$^{-1}$ when its linearized wavefunction is used as the trial.
In contrast, our quantum trial achieves chemical accuracy on Willow both with and without the $\alpha$--$\beta$ Jastrow couplings, with errors of $-0.86(6)$ and $-0.85(11)$~kcal\,mol$^{-1}$, respectively (Supplementary Fig.~\ref{fig:si_afqmc_lucj_K}).
This experiment collected $1.3\times10^{8}$ and $2.5\times10^{7}$ postselected snapshots for the spin-coupled and -uncoupled circuits, respectively (Supplementary Table~\ref{tab:cz_counts}).
By contrast, a classical NOCI trial with the same number of determinants (four) as our spin-coupled surrogate, optimized for its variational energy, has a lower variational energy than our quantum trials yet yields an AFQMC energy about 5~kcal\,mol$^{-1}$ below the exact result, and classical NOCI trials with 2--8 determinants all miss chemical accuracy (Supplementary Fig.~\ref{fig:noci_fes}).
This highlights the role of quantum data in generating an accurate QMC trial wavefunction.

In Fig.~\ref{fig:fes}d, we calculate the energy gap between the singlet and the $S = 5$ high-spin state. CC methods systematically improve and approach the exact value at CCSDTQ. QC-QMC, with both spin states guided by the measurement-trained trial surrogate, gives 20.2(1)~kcal\,mol$^{-1}$ against the exact 19.3~kcal\,mol$^{-1}$, within chemical accuracy and on par with CCSDTQ.
Our experimental QC-QMC closely reproduces the ideal (noiseless, infinite-shot) QC-QMC results, highlighting the noise robustness of QC-QMC combined with the ML surrogate.
Notably, the spin-uncoupled trial is as accurate as the spin-coupled one, so the $\alpha$--$\beta$ Jastrow pairs are not needed here. 
Without explicit opposite-spin Jastrow couplings in the trial, the circuit has no gates between the spin registers and splits into two independent halves, enabling experiments with reduced circuit complexity.
For $S=5$, the $\alpha$-spin sector is fully occupied, so a nontrivial $\alpha$-spin circuit is unnecessary (Supplementary Sec.~\ref{app:fes-highspin}).

\begin{figure}
    \centering
    \includegraphics[width=0.45\textwidth]{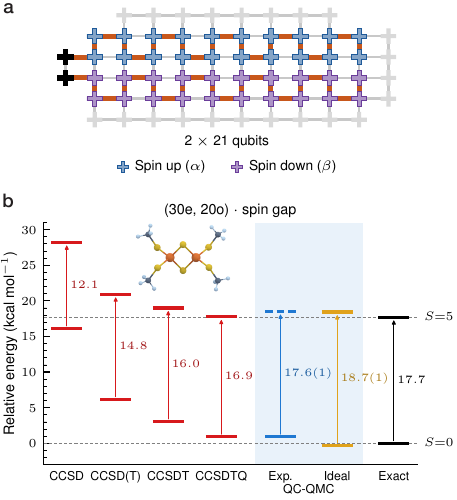}
    \caption{\textbf{QC-QMC spin gap of \FeS{} in a 40-qubit active space.}
    \textbf{a}.~Qubit layout for the singlet ($S=0$) calculation on Willow. The 40 spin orbitals map to 40 qubits, with one additional ancilla per spin register to ensure even fermion parity. We implement the spin-uncoupled ansatz as two independent 21-qubit circuits operating in parallel on 42 physical qubits.
    Error rates in Supplementary Fig.~\ref{fig:hardware_error}.
    \textbf{b}.~Energies of the singlet ($S=0$) and high-spin ($S=5$) states relative to the exact singlet energy. We compare unrestricted coupled-cluster results with QC-QMC and exact results. Vertical lines indicate the spin gaps. The QC-QMC (Exp.) result combines experimental classical shadows for the singlet state with classically simulated shadows for the high-spin state (dashed line). The QC-QMC (ideal) result uses noiseless circuits in the infinite-shot limit for both states.
    }
    \label{fig:fes_gap_40q}
\end{figure}

Finally, we use the spin-uncoupled trial wavefunction for the (30e, 20o) active space, corresponding to a 40-qubit total system run as two 21-qubit circuits (Fig.~\ref{fig:fes_gap_40q}). 
Using $3.0\times10^{8}$ postselected snapshots and a NOCI surrogate, QC-QMC achieves chemical accuracy for the singlet ground state ($0.91(8)$~kcal\,mol$^{-1}$) relative to the exact energy.
Unlike the singlet, where nearest-neighbor Jastrow pairs suffice, the $S=5$ state requires pairs beyond nearest neighbors. Routing these pairs along the qubit line makes the circuit too deep to run at sufficiently high fidelity on the current device, so we classically emulate the quantum data using fast quantum circuit simulators~\cite{isakov2021simulations}. 
We train an ML model on these emulated data as we do for the experimental data (Supplementary Table~\ref{tab:cz_counts} and Fig.~\ref{fig:noise_floor_LUCJ}c).
Combining the experimental singlet results with the simulated high-spin data yields the spin gap within chemical accuracy.

{\it Conclusions and outlook.}
Our work overcomes a major computational bottleneck in hybrid quantum--classical computing by compressing quantum measurement data into reusable machine-learning models. This accelerates QC-QMC by orders of magnitude, enabling chemically accurate simulations on up to 40 qubits, more than doubling the scale of previous hardware demonstrations. Because the classical postprocessing bottleneck persists even on fault-tolerant quantum computers, our approach addresses a fundamental challenge for both near-term and future quantum computing. 

A remaining challenge is the statistical measurement cost of resolving overlaps that decay exponentially with system size~\cite{mazzola2022challenges,lee2022response}. Although this limitation has not prevented our present calculations, even for strongly correlated systems such as iron--sulfur clusters, understanding its impact at larger scales will be crucial for identifying problems where QC-QMC can achieve practical quantum advantage. More broadly, our results establish a pathway toward scalable hybrid algorithms in which quantum information can be efficiently exploited by classical many-body methods.

{\it Acknowledgments.}
This work was primarily supported by the U.S.\ Department of Energy, Office of Fusion Energy Sciences, ``Foundations for quantum simulation of warm dense matter'' project (TJ, AZ, ADB, JL).
TJ and JL were partially supported by the startup funds from Harvard University and Wellcome Leap through the Quantum for Bio (Q4Bio) program.
TJ also acknowledges support from the Gordon and Betty Moore Foundation Fellowship.
We thank Pedram Roushan for valuable feedback on the presentation of this work, Jinghong Zhang and Nishad Maskara for their engagement at the beginning of the project, and Ryan Byrne, Mohammed Alghadeer, and Ricky Oliver for their support in running the quantum experiments.

This research used computational resources of the Oak Ridge Leadership Computing Facility at the Oak Ridge National Laboratory through the INCITE Project No.\ MAT281, which is supported by the Office of Science
of the U.S.\ Department of Energy under Contract No. DE-AC05-00OR22725, and the FAS Research Computing cluster supported by the Faculty of Arts and Sciences (FAS) Division of Science Research Computing Group at Harvard, and National Nuclear Security Administration’s Advanced Simulation and Computing Program.
JN, TJ and JL acknowledge the support of NSF ACCESS through allocation CHE250121 from the Advanced Cyberinfrastructure Coordination Ecosystem: Services \& Support (ACCESS) program, which is supported by U.S. National Science Foundation grants 
\#2138259, \#2138286, \#2138307, \#2137603, and \#2138296.

We collected the hardware data presented in this work remotely on a 105-qubit Willow processor, with access provided via Google’s Quantum Engine through the Willow Early Access Program. Calibration and support were provided by the Quantum Hardware Residency Program. We thank the Google Quantum AI team for providing the quantum systems and support that enabled these results. The views expressed in this work are solely those of the authors and do not reflect the policy of Google or the Google Quantum AI team.

This article has been authored by an employee of National Technology \& Engineering Solutions of Sandia, LLC under Contract No.\ DE-NA0003525 with the U.S.\ Department of Energy (DOE). The employee owns all right, title, and interest in and to the article and is solely responsible for its contents. The United States Government retains, and the publisher, by accepting the article for publication, acknowledges that the United States Government retains a non-exclusive, paid-up, irrevocable, worldwide license to publish or reproduce the published form of this article or allow others to do so, for United States Government purposes. The DOE will provide public access to these results of federally sponsored research in accordance with the DOE Public Access Plan \url{https://www.energy.gov/downloads/doe-public-access-plan}.

{\it AI statement.} Gemini 4, Claude (Fable 5.1 and Opus 5.5) and Codex (Astra 6.0 and Sol 6.1) assisted the authors with algorithmic implementations, programming, critical review, and manuscript preparation. The authors supervised all use and validated the outputs.

{\it Code and data availability.} The data supporting this study are provided in the main text and Supplementary Information. The code and raw measurement records are available from the corresponding author upon request and will be made publicly available in the future.

{\it Author contributions.}
TJ and JL conceived the idea and designed the study.
TJ performed classical numerics and QC-QMC calculations with contributions from AZ, JN, and ADB. AZ developed improved algorithms for classical shadows and circuit compilation, and performed quantum experiments with contributions from TJ, SK, and ER. 
TJ and AZ developed and optimized the GPU implementation of the QC-QMC postprocessing.
TJ, AZ, and JL drafted the manuscript.
JL supervised the overall project.
All authors participated in discussions, manuscript writing, and data analysis.

\MainReferences{main}

\clearpage
\onecolumngrid
\begin{center}
\textbf{\large Supplementary Material}
\end{center}

\setcounter{secnumdepth}{2}
\setcounter{tocdepth}{2}
\setcounter{section}{0}
\setcounter{equation}{0}
\setcounter{figure}{0}
\setcounter{table}{0}

\makeatletter
\@addtoreset{equation}{section}
\@addtoreset{figure}{section}
\@addtoreset{table}{section}
\makeatother

\renewcommand{\thesection}{S\arabic{section}}
\renewcommand{\theequation}{S\arabic{section}.\arabic{equation}}
\renewcommand{\thefigure}{S\arabic{section}.\arabic{figure}}
\renewcommand{\thetable}{S\arabic{section}.\arabic{table}}
\renewcommand{\theHsection}{S\arabic{section}.\arabic{section}}
\renewcommand{\theHequation}{S\arabic{section}.\arabic{equation}}
\renewcommand{\theHfigure}{S\arabic{section}.\arabic{figure}}
\renewcommand{\theHtable}{S\arabic{section}.\arabic{table}}

\addtocontents{toc}{\protect\showmaintoc}
\tableofcontents
\clearpage

\section{Quantum-state preparation and measurements}
\label{app:experiment}

\subsection{Hardware implementation}
\paragraph{Qubit layouts and error rates.}

\begin{figure}
    \centering
    \includegraphics[width=\textwidth]{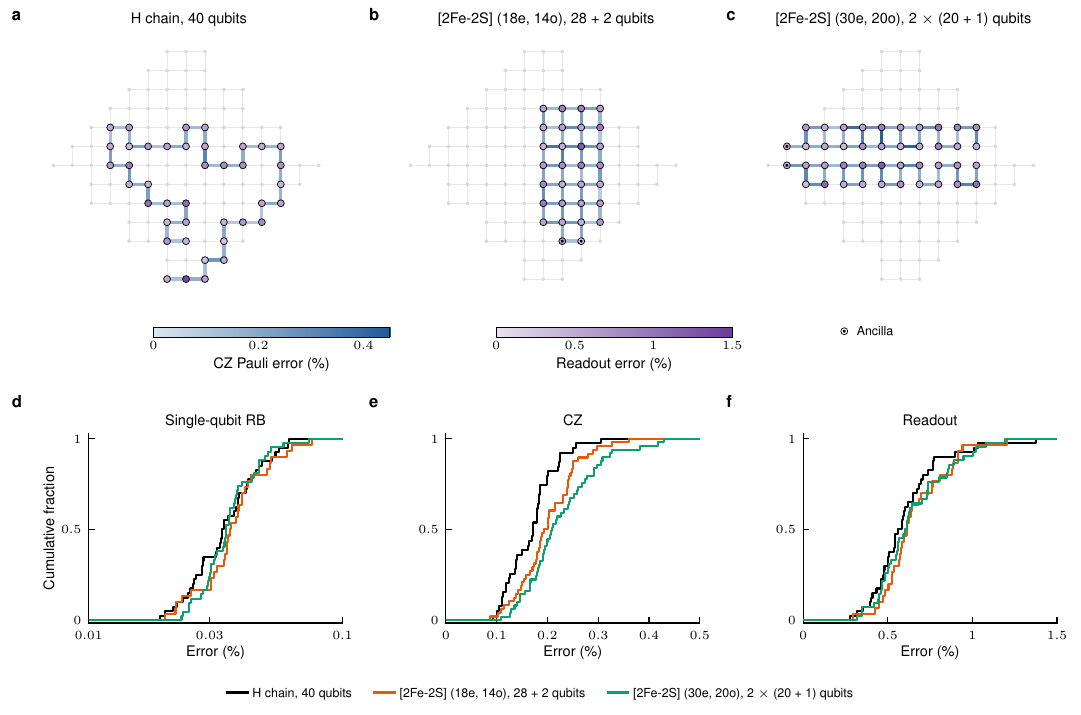}
    \caption{\textbf{Representative Willow error rates.} \textbf{a}, Example of a 40-qubit experiment for a hydrogen chain mapped to a one-dimensional geometry. The choice of qubits depends on the chain length and device drift, determined by a greedy search to maximize expected fidelity. \textbf{b}, The $(28+2)$-qubit two-dimensional grid, also determined by a similar type of greedy search. \textbf{c}, The (40+2)-qubit geometry was fixed by a hardware constraint. \textbf{d--f}, Empirical cumulative distribution functions estimated from single-qubit randomized benchmarking (RB), inferred CZ Pauli error from cross-entropy benchmarking (XEB), and readout error (averaged evenly over $0\to1$ and $1\to0$ errors).}
    \label{fig:hardware_error}
\end{figure}

\Cref{fig:hardware_error} shows the qubit layouts used on the Willow processor for the 40-qubit hydrogen chain and for \FeS{} in (18e, 14o) and (30e, 20o), together with their single-qubit, two-qubit (CZ) and readout error rates.
Single-qubit and readout errors are similar for the three layouts.
CZ errors are lower for the hydrogen chain than for the two \FeS{} layouts, since a one-dimensional path can avoid poor couplers more easily than a compact two-dimensional grid.

\paragraph{Circuits, gate counts, and shots.}

Each hardware circuit consists of a GHZ-type preparation, the trial circuit, and a random matchgate circuit for the shadow measurement. 
Each spin sector of $N$ orbitals is mapped by the Jordan--Wigner transformation onto a line of $N$ qubits. When a sector has an odd number of electrons in the trial, we prepend one ancilla mode, so both physical registers have even fermion parity. 
This is essential for the matchgate shadows overlap formula~\cite{wan2023matchgate}. 
Initialized in a GHZ-type state, each circuit prepares $(\ket{\vac}+\ket{\Psi_\T})/\sqrt{2}$ due to particle-number symmetry. 
For the hydrogen chain experiments, $\ket{\Psi_\T} = \ket{1^{N/2},0^{N/2}}_\alpha \otimes \ket{1^{N/2},0^{N/2}}_\beta$. We map this to a one-dimensional line of $2N$ qubits by ordering the spin sectors head-to-head, i.e., orbital indices ascending from the middle of the line outwards.

\begin{figure}
    \centering
    \includegraphics[width=0.5\linewidth]{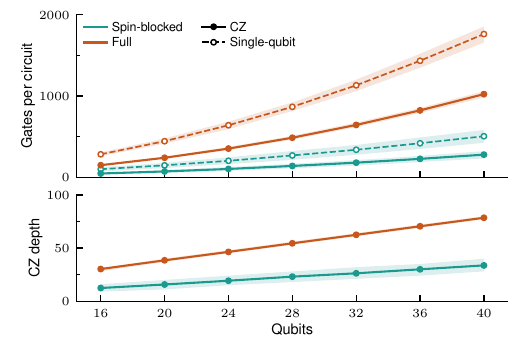}
    \caption{\textbf{Comparison of our Clifford matchgate circuits to prior art.}
    Gates per circuit (top; CZ solid, single-qubit dashed) and CZ depth (bottom) against the total number of qubits, for spin-blocked circuits, in which each spin register is measured through its own matchgate circuit (teal), and for full matchgate circuits over both registers, as in previous experiments.
    Shading spans the 5th--95th percentiles across circuits.
    Counts exclude state preparation and measurement, and depth counts two-qubit layers.}
    \label{fig:Hchain_matchgate_counts}
\end{figure}

\begin{figure}
    \centering
    \includegraphics[width=0.7\textwidth]{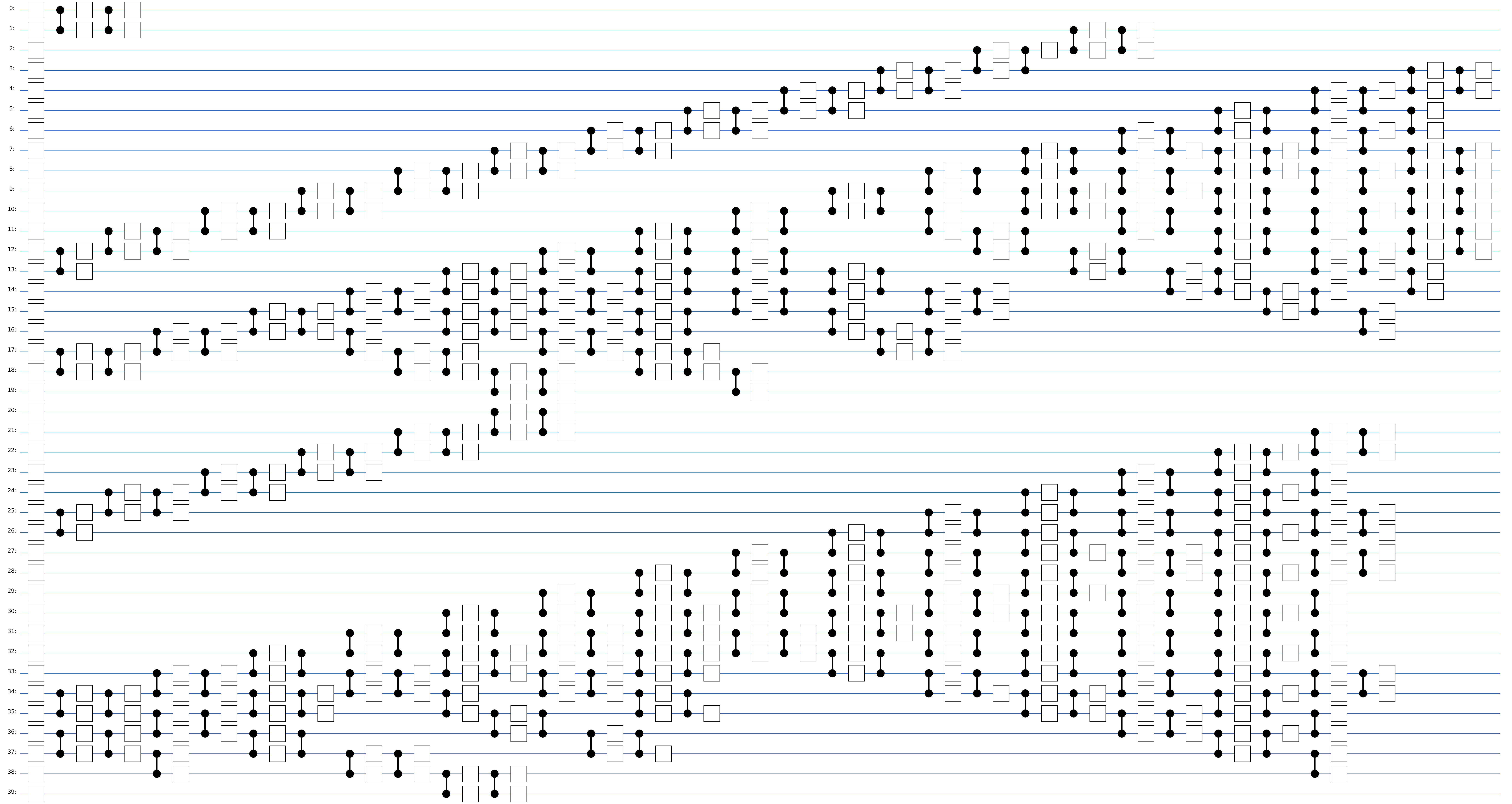}
    \caption{\textbf{A random spin-blocked Clifford matchgate circuit on 40 qubits.} Qubits 0--19 form the $\alpha$ register and 20--39 the $\beta$ register;~no gate couples the two. Each spin register receives a uniform, independent random permutation compiled according to Algorithm~\ref{alg:compile_shadow_local}. Boxes are single-qubit (Clifford) gates and two-qubit gates are $\CZ$s.}
\label{fig:matchgate_circuit_40q}
\end{figure}

For the hydrogen chains, only the GHZ-type preparation and measurement circuit run on the device; for \FeS{}, we insert a single-layer LUCJ ansatz between the GHZ preparation and measurement.
Here we record the gate and shot counts for each circuit. In \cref{tab:cz_counts} we give a breakdown of the two-qubit (CZ) gates in each circuit, according to the GHZ-type preparation, the LUCJ ansatz (if applicable), and the Clifford matchgate measurements (on average over 100 random circuits). 
In \cref{fig:Hchain_matchgate_counts} we display the matchgate circuit sizes in terms of both gates and depth. In particular, we compare them with full (non-spin-blocked) matchgates considered in previous experiments.
We display an example of a spin-blocked matchgate circuit diagram in \cref{fig:matchgate_circuit_40q}, and of the LUCJ circuits in \cref{fig:LUCJ_circuit_14o-coupled,fig:LUCJ_circuit_20o-uncoupled}.

\begin{table*}[t]
\centering
\caption{
\textbf{Quantum resource counts.}
Left: final two-qubit CZ gate counts of each quantum trial, broken down into the GHZ-type preparation, the post-Hartree--Fock circuit
(LUCJ) and the matchgate measurement circuits; ``Total'' is the count after concatenating all components and merging adjacent gates
whenever possible. Matchgate and total counts are the mean $\pm$ sample standard deviation over 100 random circuits.
Right: measurements behind the labels of Fig.~\ref{fig:noise_floor_LUCJ}: random-basis circuits $C$, summed over all hardware runs or
simulation seeds; shots (after parity post-selection for hardware); and the noise floor $F$ at $C$.
The spin $S=(N_\alpha-N_\beta)/2$ distinguishes the LUCJ ansatze of the singlet $S=0$ and high-spin $S=5$ sectors of the iron--sulfur
cluster. UHF and the coupled 14-orbital trial use one joint GHZ preparation across both spin sectors; the remaining trials use two
independent GHZ circuits. Qubit counts include one ancilla per spin register whenever the electron number in that register is odd.
}
\label{tab:cz_counts}
\setlength{\tabcolsep}{4pt}
\begin{tabular}{lcrrrrrrrr}
\hline
 & & & \multicolumn{4}{c}{CZ gates} & \multicolumn{3}{c}{Measurement} \\
\cline{4-7}\cline{8-10}
Trial & $S$ & Qubits & GHZ & LUCJ & Matchgates & Total (merged) & Circuits & Total shots & $F$ \\
\hline
UHF (H$_4$)    & --- & 8  & 3  & --- & $10.40 \pm 3.31$   & $9.33 \pm 1.96$    & 100    & $5.00\times10^{4}$ & 0.227 \\
UHF (H$_8$)    & --- & 16 & 7  & --- & $43.78 \pm 8.50$   & $36.29 \pm 5.38$   & 2,500  & $1.25\times10^{6}$ & 0.0714 \\
UHF (H$_{12}$) & --- & 24 & 11 & --- & $100.53 \pm 13.41$ & $79.34 \pm 9.15$   & 10,000 & $1.00\times10^{7}$ & 0.0533 \\
UHF (H$_{16}$) & --- & 32 & 15 & --- & $178.62 \pm 18.27$ & $138.88 \pm 13.02$ & 20,000 & $2.76\times10^{7}$ & 0.0968 \\
UHF (H$_{20}$) & --- & 40 & 19 & --- & $279.29 \pm 28.92$ & $217.20 \pm 20.34$ & 50,000 & $6.42\times10^{7}$ & 0.213 \\
\hline
LUCJ (14o, uncoupled) & $0$ & 30 & 18 & 230 & $158.71 \pm 17.24$ & $356.08 \pm 12.77$ & 50,000 & $2.50\times10^{7}$ & 0.0439 \\
LUCJ (14o, coupled)   & $0$ & 30 & 19 & 244 & $158.71 \pm 17.24$ & $372.25 \pm 12.69$ & 100,000 & $1.34\times10^{8}$ & 0.369 \\
LUCJ (20o, uncoupled) & $0$ & 42 & 30 & 374 & $308.96 \pm 29.27$ & $618.90 \pm 24.20$ & 229,632 & $3.04\times10^{8}$ & 0.0353 \\
\hline
LUCJ (14o, $\beta$ only) & $5$ & 14 & 3 & 293 & $68.38 \pm 13.58$  & $343.24 \pm 10.03$ & 60,000 & $1.49\times10^{8}$ & 0.00803 \\
LUCJ (20o, $\beta$ only) & $5$ & 20 & 9 & 666 & $138.06 \pm 18.81$ & $772.64 \pm 14.68$ & 100,000$^{a}$ & $1.00\times10^{8}$$^{a}$ & 0.0115$^{a}$ \\
\hline
\end{tabular}
\par\vspace{2pt}{\raggedright\footnotesize $^{a}$Noiseless classical simulation of the same circuits.\par}
\end{table*}

\subsection{Error mitigation}
\label{app:error-mitigation}

Our experiments use relatively few algorithmic quantum error mitigation strategies. Here we describe the circuit-level error mitigation techniques used in this work.

\begin{figure}[t]
    \centering
    \includegraphics[width=0.8\textwidth]{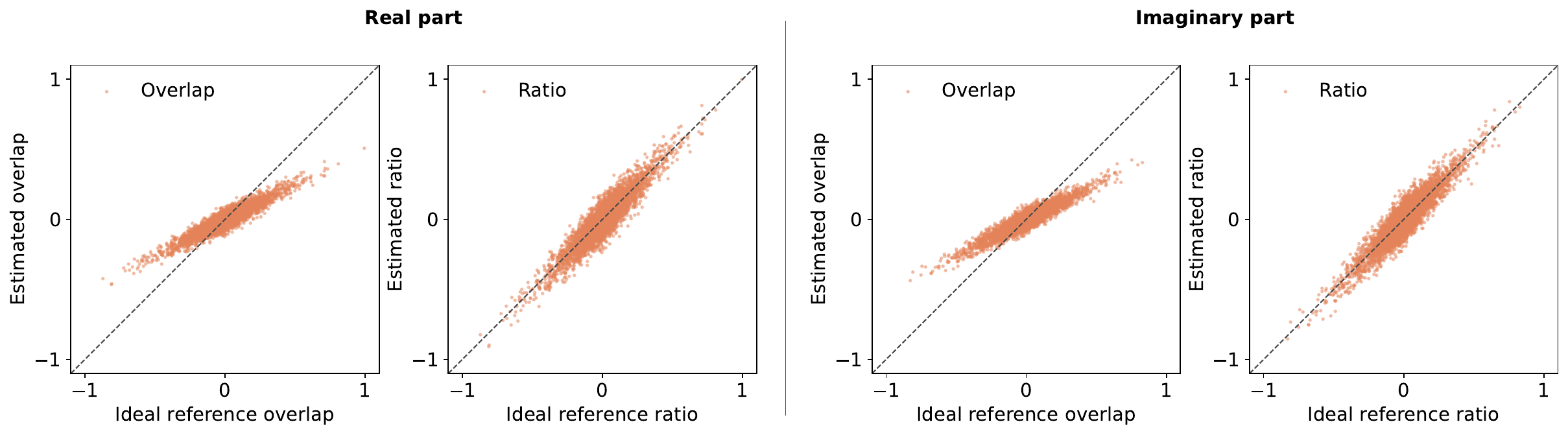}
    \caption{\textbf{Demonstration of overlap ratio error cancellation in (spin-blocked) matchgate shadows.} We plot estimated overlaps and overlap ratio for the 40-qubit UHF experiment. The walkers are a sample of 5,000 training determinants, and we estimate the overlaps over 40,000 circuits, each with 500 shots. For demonstration, we take ratios by dividing by a fixed overlap, the UHF determinant itself. In practice, we take ratios between walkers at different imaginary times. Values are uniformly normalized to arbitrary units such that they lie within $[-1, 1]$.
    }
    \label{fig:overlap_ratio_error-mitigation}
\end{figure}

\paragraph{Overlap ratios with shadows.} 
By and large, the most significant contribution to mitigating errors in QC-QMC is the inherent twirling of noise channels by classical shadows~\cite{chen2021robust}. For global Clifford shadows, this manifests as an overall scale factor that is uniform over all traceless observables. Since AFQMC only requires overlap \textit{ratios}, this scale factor ideally cancels;~this phenomenon was precisely observed in the original QC-QMC demonstration~\cite{huggins2022unbiasing}. Note that exact cancellation requires certain assumptions on the device noise (namely, Markovian and gate-independent) which are not obeyed in practice. The deviation in the presence of gate-dependent noise has been studied previously~\cite{zhao2024group,brieger2025stability}.
The previous discussion concerns global Clifford shadows;~for matchgate shadows, the noise twirl does not yield just two irreps, but rather $n+1$ irreps~\cite{zhao2024group}. The cancellation of errors due to taking overlap ratios using matchgate shadows was studied by Huang \textit{et al}~\cite{huang2024evaluating}. They observed that despite the more delicate nature of the irreps of the matchgate group, a cancellation still persists under the same Markovian, gate-independent assumptions. We observe the same phenomenon in our experiments; Fig.~\ref{fig:overlap_ratio_error-mitigation} plots a sample of the training overlaps for the 40-qubit UHF data. The noisy overlaps show clear attenuation, which we can approximately correct by taking ratios. Note that these are overlaps computed directly from the shadows, before any ML training.

\paragraph{Qubit error suppression.}
We utilize randomized compiling (RC)~\cite{wallman2016noise} and dynamical decoupling (DD)~\cite{viola1999dynamical} for each random circuit that we run. RC involves inserting random single-qubit Clifford gates around each two-qubit gate such that the modified circuit is equivalent to the original circuit in the noiseless limit. In the presence of device noise, RC primarily converts coherent errors into incoherent errors, whose effects are easier to manage. This is especially amenable to the inherent noise twirling of classical shadows~\cite{chen2021robust}. DD involves a sequence of $X$ and $Y$ gates whenever a qubit idles during circuit layers, with the effect being identity without noise. In the presence of errors, DD suppresses $T_2$ dephasing by toggling the qubit's frame while it idles.
Note that due to the dramatic difference in rates of repeating a single circuit $N$ times vs.\ changing $N$ different circuits, it is infeasible to compile a new RC and DD sequence for each shot per circuit. Thus the sequences are fixed over all shots of the same circuit.

\paragraph{Parity postselection.}
Because each trial has an even electron number (enforced in some cases by appending dummy qubits), and because we specifically choose every matchgate to preserve parity, the parity of every circuit output
\begin{equation}
    U_Q \l( \frac{\ket{\vac} + \ket{\Psi_\T}}{\sqrt{2}} \r), \qquad \det(Q) = 1,
\end{equation}
is even in the absence of noise. Note that particle number itself is not generally preserved by the matchgates, so we can only postselect on parity. However, with spin-blocked matchgate shadows, we have a further symmetry that the parity in each spin sector is even; this is what we postselect on. Some representative postselection rates are displayed in \cref{table:postselection}.
\cref{app:convergence} discusses its effect on the overlap estimates, shown in \cref{fig:SI_hchain_conv}.

\begin{table}[t]
\centering
\caption{
\label{table:postselection}\textbf{Representative spin-blocked postselection rates.} A random sample of 1, 100, and 10,000 circuits (trial with random spin-blocked matchgates) are chosen, each sampled on the Willow processor at 4,000 shots per circuit. Columns 4--7 indicate the percentage of total shots where the bit strings in spin registers $\alpha/\beta$ have even/even parity, even/odd, etc. The final two columns indicate the marginal percentage of each spin register's even-parity rate. Boldface percentages denote the pass criterion for each trial;~the (40+2)-qubit LUCJ ansatz factorizes across the spin sectors so its criterion differs from the other two.}
\setlength{\tabcolsep}{5pt}
\begin{tabular}{lllrrrrrr}
\hline
Trial state & Circuits & Total shots & Even/even & Even/odd & Odd/even & Odd/odd
    & $\alpha$ even & $\beta$ even \\
\hline
UHF, 40 qubits & 1 & 4,000
    & \textbf{54.3\%} & 22.6\% & 16.4\% & 6.7\% & 76.9\% & 70.7\% \\
------ & 100 & 400,000
    & \textbf{52.4\%} & 20.4\% & 19.4\% & 7.7\% & 72.9\% & 71.8\% \\
------ & 10,000 & 40,000,000
    & \textbf{51.5\%} & 20.6\% & 19.9\% & 8.0\% & 72.1\% & 71.4\% \\
\hline
LUCJ, 28+2 qubits & 1 & 4,000
    & \textbf{36.7\%} & 20.2\% & 27.9\% & 15.2\% & 57.0\% & 64.6\% \\
------ & 100 & 400,000
    & \textbf{35.7\%} & 22.7\% & 25.2\% & 16.4\% & 58.4\% & 60.9\% \\
------ & 10,000 & 40,000,000
    & \textbf{35.5\%} & 22.8\% & 25.0\% & 16.6\% & 58.4\% & 60.5\% \\
\hline
LUCJ, 40+2 qubits & 1 & 4,000
    & 30.1\% & 20.9\% & 29.5\% & 19.5\% & \textbf{51.0\%} & \textbf{59.6\%} \\
------ & 100 & 400,000
    & 29.0\% & 24.9\% & 25.2\% & 20.9\% & \textbf{53.9\%} & \textbf{54.2\%} \\
------ & 10,000 & 40,000,000
    & 29.0\% & 24.7\% & 25.3\% & 21.0\% & \textbf{53.6\%} & \textbf{54.3\%} \\
\hline
\end{tabular}
\end{table}

\paragraph{Qubit assignment averaging.}
Finally, we apply qubit assignment averaging (QAA), as in previous experiments~\cite{google2020hubbard}. This involves rotating/flipping the assignment of simulated spin orbitals to physical qubits on the chip across the many measurements collected during the experiment. This reduces anisotropic error-rate effects across the chip that bias certain observables. This improves the inherent noise resilience of classical shadows, due to the noise-twirling property, in classical simulations of noise models~\cite{zhao2024group}.
For the hydrogen chains, where the physical qubit layout is also a one-dimensional chain, there are only two assignments:~if $1\alpha, 2\alpha, \ldots$ denote the canonical ordering of $\alpha$ spin orbitals, and similarly $1\beta, 2\beta, \ldots$, then the two variants are (let $N$ be the number of molecular orbitals)
\begin{align}
    &N\alpha, \ldots, 1\alpha, 1\beta, \ldots, N\beta,\\
    &N\beta, \ldots, 1\beta, 1\alpha, \ldots, N\alpha.
\end{align}
For LUCJ circuits, we can flip lengthwise and switch which block is $\alpha$ or $\beta$. That is, we have the four assignments ($0\alpha, 0\beta$ denote the ancilla orbitals):
\begin{align}
\begin{array}{c|c}
    \begin{aligned}
        &0\alpha, 1\alpha, \ldots, N\alpha\\
        &0\beta, 1\beta, \ldots, N\beta
    \end{aligned} & \begin{aligned}
        &N\alpha, \ldots, 1\alpha, 0\alpha\\
        &N\beta, \ldots, 1\beta, 0\beta
    \end{aligned}\\
     \hline
    \begin{aligned}
        &0\beta, 1\beta, \ldots, N\beta\\
        &0\alpha, 1\alpha, \ldots, N\alpha
    \end{aligned} & \begin{aligned}
        &N\beta, \ldots, 1\beta, 0\beta\\
        &N\alpha, \ldots, 1\alpha, 0\alpha
    \end{aligned}
\end{array}
\end{align}
All four are used for the $(28+2)$-qubit LUCJ circuit;~for the $(40+2)$-qubit circuit, because of a minor hardware constraint, we could not flip the qubit assignment left to right, so we only used the two assignments that swap between $\alpha$ and $\beta$ registers.

\subsection{Local unitary cluster Jastrow ansatz}
\label{app:lucj}

We follow the circuit structure of the local unitary cluster Jastrow (LUCJ) ansatz~\cite{huggins2022unbiasing,motta2023lucj}. We use the geometry of two adjacent snaked qubit chains (one for each spin type) embedded into the two-dimensional square lattice geometry of the Willow chip;~see Figs.~3 and 4 in the main text (see also \cref{fig:hardware_error}) for the exact arrangement. 

We use a single LUCJ layer with a final orbital rotation, acting on the broken-symmetry unrestricted Hartree--Fock (BS-UHF) determinant $\ket{\Phi_0}$,
\begin{equation}
    \ket{\Psi_\LUCJ}=e^{K_{\mathrm f}}\,e^{K}e^{iJ}e^{-K}\ket{\Phi_0},
    \label{eq:lucj}
\end{equation}
with
\begin{equation}
    K=\sum_{\sigma}\sum_{pq}\kappa^{\sigma}_{pq}\,a_{p\sigma}^\dagger a_{q\sigma},
    \qquad
    K_{\mathrm f}=\sum_{\sigma}\sum_{pq}\kappa^{\mathrm f,\sigma}_{pq}\,a_{p\sigma}^\dagger a_{q\sigma},
    \qquad
    J=\sum_{\sigma\tau}\sum_{(p,q)\in\mathcal M_{\sigma\tau}}J^{\sigma\tau}_{pq}\,n_{p\sigma}n_{q\tau},
\end{equation}
where $\kappa^{\sigma}$ and $\kappa^{\mathrm f,\sigma}$ are general anti-Hermitian matrices and $J^{\sigma\tau}_{pq}$ are real. The orbital rotations are independent for the two spins ($\kappa^{\alpha}\neq\kappa^{\beta}$): the reference is a broken-symmetry determinant whose $\alpha$ and $\beta$ orbitals differ, and the rotations of each spin are written in that spin's BS-UHF molecular-orbital basis, so that setting all parameters to zero returns $\ket{\Phi_0}$. The interaction mask $\mathcal M$ follows the qubit connectivity. The same-spin terms couple nearest neighbors along each chain, $(p,p+1)$. The spin-coupled circuit adds opposite-spin on-site terms $(p,p)$ at the sites where the two chains are adjacent; the spin-uncoupled circuit has no opposite-spin terms, so its state is a product of an $\alpha$ and a $\beta$ part, and each part is prepared on its own register.
The two outer rotations in Eq.~\eqref{eq:lucj} act on a walker as a one-body rotation and keep it a Slater determinant. We therefore move them from the circuit to the walkers,
\begin{equation}
    \braket{\Psi_\LUCJ|\phi}=\braket{\Psi_{\mathrm{circ}}|\tilde\phi},
    \qquad
    \ket{\Psi_{\mathrm{circ}}}=e^{iJ}e^{-K}\ket{\Phi_0},
    \qquad
    \ket{\tilde\phi}=e^{-K}e^{-K_{\mathrm f}}\ket{\phi},
\end{equation}
and prepare only $\ket{\Psi_{\mathrm{circ}}}$ on the processor: a determinant preparation by nearest-neighbor fermionic Givens rotations followed by the Jastrow layer, with no gates after it. 

\begin{figure}
    \centering
    \includegraphics[width=\textwidth]{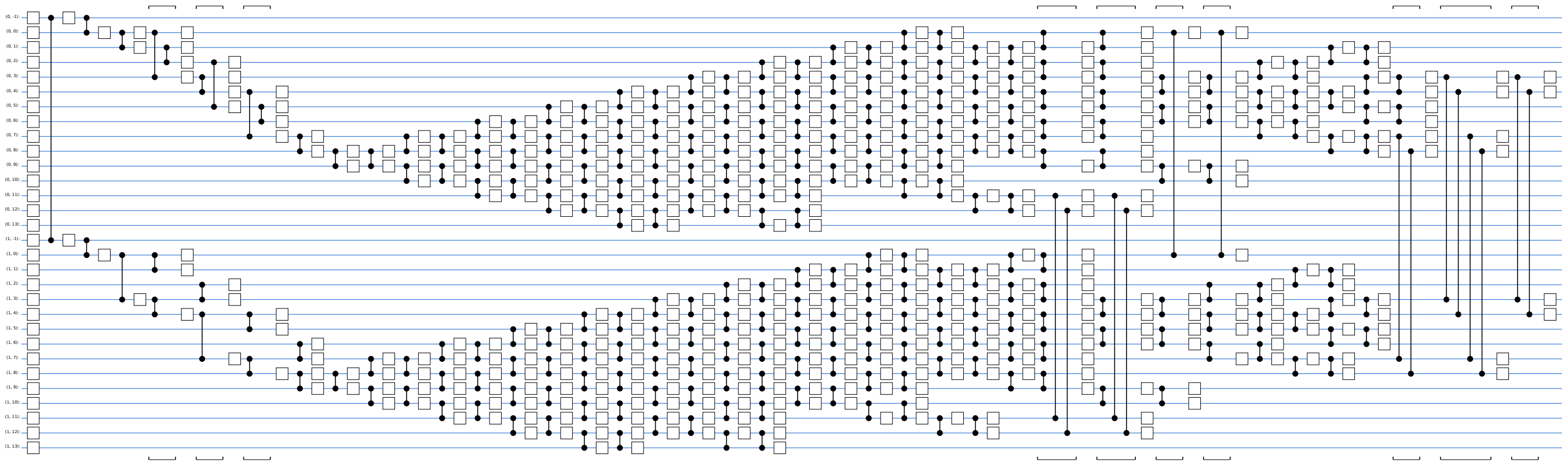}
    \caption{\textbf{The 14-orbital spin-coupled LUCJ ansatz.} This ansatz features Jastrow-paired CZ gates coupling spin sectors. Includes the two ancilla qubits for a total of 30 qubits, as well as the GHZ preparation circuit on 20 total qubits ($9+9$ physical electrons, 2 dummy electrons). The final one-body rotation after the Jastrow terms is omitted, instead to be classically applied to the walkers.}
    \label{fig:LUCJ_circuit_14o-coupled}
\end{figure}

\begin{figure}
    \centering
    \includegraphics[width=\textwidth]{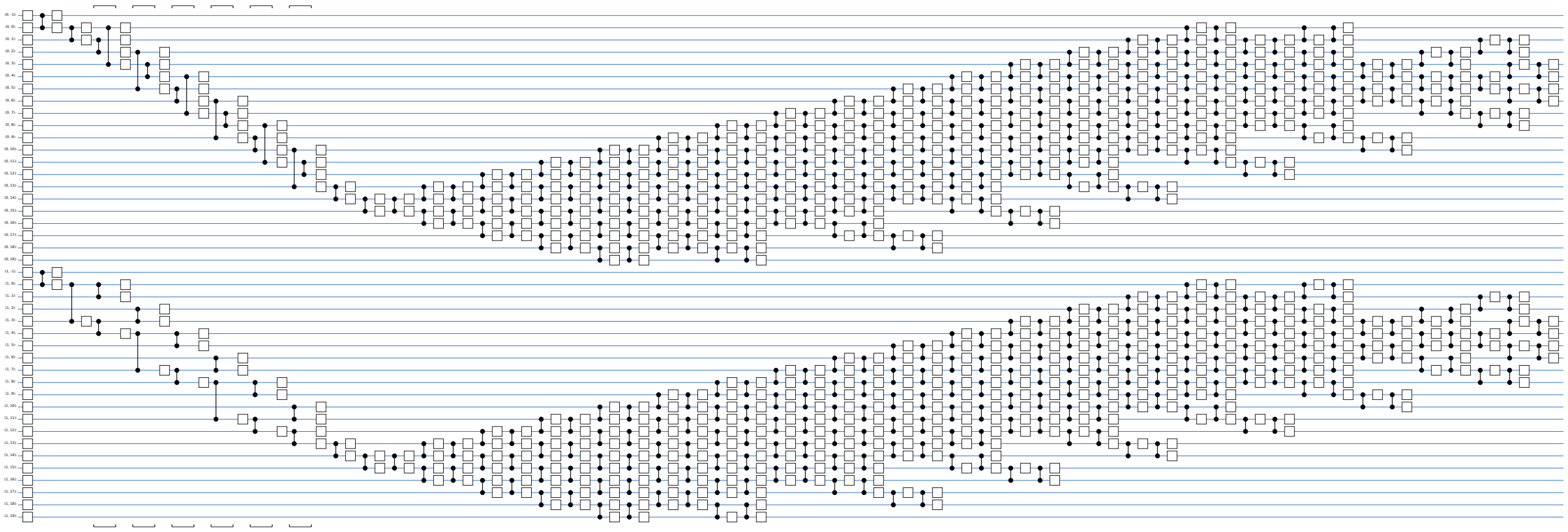}
    \caption{\textbf{The 20-orbital spin-independent LUCJ ansatz.} Each spin sector is decoupled from the other, including the GHZ preparation. Includes the two ancilla qubits for a total of 42 qubits, and each GHZ preparation circuit prepares two independent cat states ($15+15$ physical electrons, 2 dummy electrons). The final one-body rotation after the Jastrow terms is omitted, instead to be classically applied to the walkers.}
    \label{fig:LUCJ_circuit_20o-uncoupled}
\end{figure}

The rotated walkers $\tilde\phi$ are computed classically, and all overlaps reported in this work refer to the full state of Eq.~\eqref{eq:lucj}.
In \cref{fig:LUCJ_circuit_14o-coupled,fig:LUCJ_circuit_20o-uncoupled} we show the full ansatz, including the initial GHZ-type state preparation but excluding the random matchgates. Two-qubit gates are exclusively $\CZ$ gates, and arbitrary single-qubit rotations (represented as phased XZ gates in Willow's native gate set) are denoted by empty squares. For the 14-orbital system, we prepared two circuits: a spin-coupled one (14o-C), in which both the GHZ preparation and the Jastrow $\CZ$ pairs couple the two spin registers, and a spin-uncoupled one (14o-U), with two independent spin registers. The 20-orbital system used the spin-uncoupled form.

\paragraph{Optimization of the circuit parameters.}
We do not set the parameters following a coupled-cluster calculation, as originally proposed~\cite{motta2023lucj}. Instead, we optimize them classically in two stages. 
First, we minimize the variational energy~\cite{belagali2026efficient}. Second, starting from the energy-optimized state, we \emph{teach} the circuit a target wavefunction $\ket{\Psi_{\mathrm{tgt}}}$ with FCI by minimizing the overlap loss function (Eq.~\eqref{eq:train-loss}) with walkers sampled by BS-UHF-guided phaseless AFQMC. 
The optimization uses Adam followed by geometric direct minimization with a line search~\cite{van2002geometric}. 
We keep the parameters with the lowest validation loss $S$ in a separate set of validation walkers; we never use reference energy, reference fidelity, or reference AFQMC results to choose them.
\Cref{tab:lucj_exact} compares the two stages: teaching raises the variational energy by 14 and 26$~\mHa{}$ but brings the AFQMC error from about $-11~\mHa{}$ to within $1~\mHa{}$ for both active spaces.

\begin{table}[t]
\centering
\caption{
\label{table:lucj_energy_vs_overlap}\textbf{Energy-optimized versus taught spin-uncoupled LUCJ circuits.} Error of the variational energy of each circuit and of phaseless AFQMC with the exact (noiseless) circuit state as the trial, in mEh, each relative to FCI for its own active-space Hamiltonian. Teaching starts from the energy-optimized circuit; the taught circuits are the ones run on Willow. Teaching raises the variational energy but brings the AFQMC energy close to FCI.}
\setlength{\tabcolsep}{6pt}
\begin{tabular}{llrr}
\hline
Active space & Circuit & Variational & AFQMC \\
\hline
(18e, 14o) & Energy-optimized & $+16.73$ & $-11.23(11)$ \\
           & Taught           & $+30.66$ & $-0.80(11)$ \\
\hline
(30e, 20o) & Energy-optimized & $+26.89$ & $-10.88(10)$ \\
           & Taught           & $+52.84$ & $-0.43(11)$ \\
\hline
\end{tabular}\label{tab:lucj_exact}
\end{table}

\section{Quantum-classical hybrid QMC (QC-QMC) with machine-learned surrogate}
\label{app:method}
\subsection{Overview of QC-QMC}
\label{app:afqmc}
We use phaseless auxiliary-field quantum Monte Carlo (AFQMC)~\cite{zhang2003phaseless,motta2018ab,huggins2022unbiasing} to sample the imaginary-time projection
\begin{equation}
    \ket{\Psi_0}
    \propto
    \lim_{\tau\to\infty} e^{-\tau H}\ket{\Phi_0},
\end{equation}
where the initial state $\ket{\Phi_0}$ has nonzero overlap with the ground state in the targeted symmetry sector. 
Introducing a trial wavefunction $\ket{\Psi_{\mathrm{T}}}$, the importance-sampled walker representation is
\begin{equation}
    \ket{\Psi(\tau)}
    \simeq
    \sum_{i=1}^{N_{\mathrm{w}}}
    w_i(\tau)
    \frac{\ket{\phi_i(\tau)}}{
        f_{\mathrm{T}}(\phi_i(\tau))
    },
\end{equation}
\begin{equation}
f_{\mathrm{T}}(\phi) = \braket{\Psi_{\mathrm{T}}|\phi}    
\end{equation}
where each walker $\ket{\phi_i}$ is a Slater determinant
and $w_i$ is its importance-sampling weight.
The hybrid quantum-classical quantum Monte Carlo (QC-QMC) uses a trial state $|\Psi_\mathrm{T}\rangle$ that is prepared on a quantum processor to guide the QMC walk~\cite{huggins2022unbiasing}.
A short-time factorization and Hubbard--Stratonovich
transformation express the interacting propagator as an
integral over one-body propagators. At each time step
$\Delta\tau$, Gaussian auxiliary fields $\mathbf{x}_i$
are sampled and shifted by a trial-dependent force bias
$\bar{\mathbf{x}}_i$, giving
\begin{equation}
    \ket{\phi_i'}
    =
    B_{\Delta\tau}
    (\mathbf{x}_i-\bar{\mathbf{x}}_i)
    \ket{\phi_i}.
\end{equation}
Because $B_{\Delta\tau}$ is a one-body propagator, the updated walker remains a Slater determinant. The force bias reduces fluctuations in the importance weights. To control the fermionic phase problem, the phaseless approximation maintains nonnegative walker weights and applies the projection factor
\begin{equation}
    \max\{0,\cos\Delta\theta_i\},
    \qquad
    \Delta\theta_i
    =
    \arg\!\left[
        \frac{f_{\mathrm{T}}(\phi_i')}
             {f_{\mathrm{T}}(\phi_i)}
    \right].
\end{equation}
This factor multiplies the nonnegative branching weight obtained from the importance-sampled propagation. The constraint introduces a systematic bias that depends on the trial wavefunction. The total energy is evaluated using the mixed estimator (local energy),
\begin{equation}
\begin{aligned}
    E_{\mathrm{L}}(\phi)
    &=
    \frac{\bra{\Psi_{\mathrm{T}}}H\ket{\phi}}
         {f_{\mathrm{T}}(\phi)},\\
    E(\tau)
    &=
    \frac{
        \sum_i w_i(\tau)
        \operatorname{Re}E_{\mathrm{L}}(\phi_i(\tau))
    }{
        \sum_i w_i(\tau)
    }.
\end{aligned}
\end{equation}
The reported energy is estimated from the equilibrated
imaginary-time sequence through reblocking analysis.

The trial-dependent quantities required by QC-QMC can be obtained from derivatives of the overlap with respect to the walker orbitals~\cite{jiang2025unbiasing}. This provides a common formulation for classical and quantum trials, and differentiable surrogate models.
Let $\Phi$ denote the occupied-orbital matrix of the walker
$\ket{\phi}$, and write
$f_{\mathrm{T}}(\Phi)=\braket{\Psi_{\mathrm{T}}|\phi}$.
For a one-body operator
$\hat A=\sum_{pq}A_{pq}a_p^\dagger a_q$, define
\begin{equation}
    F_A(t)=f_{\mathrm{T}}(e^{tA}\Phi).
\end{equation}
The action of $e^{t\hat A}$ on a Slater determinant is
equivalent to transforming its orbital matrix by $e^{tA}$.
Consequently,
\begin{equation}
\begin{aligned}
    F_A'(0)
    &= \bra{\Psi_{\mathrm{T}}}\hat A\ket{\phi},\\
    F_A''(0)
    &= \bra{\Psi_{\mathrm{T}}}\hat A^2\ket{\phi}.
\end{aligned}
\label{eq:app-directional}
\end{equation}
The transformed determinant is generally unnormalized, so we must retain its normalization factors during differentiation. We write the factorized Hamiltonian as
\begin{equation}
    H=E_{\mathrm{core}}+\hat v_0
      -\frac{1}{2}\sum_\gamma\hat v_\gamma^2,
    \qquad
    \hat v_\gamma
    =\sum_{pq}(v_\gamma)_{pq}a_p^\dagger a_q,
\end{equation}
where $\hat v_0$ includes the one-body correction associated
with the two-body factorization. For an auxiliary-field
propagator containing
$\sqrt{\Delta\tau}(x_\gamma-\bar x_\gamma)\hat v_\gamma$,
the force bias and local energy are
\begin{align}
    \bar x_\gamma
    &=
    -\sqrt{\Delta\tau}\,
    \frac{F_{v_\gamma}'(0)}{f_{\mathrm{T}}(\Phi)},
    \\
    E_{\mathrm{L}}(\Phi)
    &=
    E_{\mathrm{core}}
    +\frac{F_{v_0}'(0)}{f_{\mathrm{T}}(\Phi)}
    -\frac{1}{2}\sum_\gamma
    \frac{F_{v_\gamma}''(0)}{f_{\mathrm{T}}(\Phi)}.
    \label{eq:app-local-energy}
\end{align}
Thus, first derivatives determine the force bias and one-body energy, while second derivatives determine the remaining contribution to the local energy. In QC-QMC, these derivatives can be evaluated by classically computing the shadow-overlap estimator while holding the measurement record fixed. The same construction applies when the overlap estimator is replaced by a trained surrogate $f_\theta$, yielding approximate force biases and local energies from its derivatives. Here the model parameters $\theta$ remain fixed, and differentiation includes the full dependence of the model on the walker orbitals.

\subsection{Matchgate shadow estimates and noise floor}
\label{app:shadow_noise}

\paragraph{Overlap estimation from shadows.}
Let $\ket{\chi}=(\ket{\vac}+\ket{\Psi_\T})/\sqrt{2}$ and write $\rho=\ketbra{\chi}{\chi}$.
For any $\eta$-electron determinant $\ket{\phi}$, its overlap with $\ket{\Psi_\T}$ is determined via~\cite{huggins2022unbiasing}
\begin{equation}
\braket{\Psi_\T | \phi} = 2 \tr(\rho \ketbra{\phi}{\vac}).
\end{equation}
It was shown that $M$ such overlaps can be estimated to additive $\epsilon$ precision from $\calO(\sqrt{N} \log(N) \log(M)/\epsilon^2)$ matchgate shadow samples~\cite{wan2023matchgate}. 
That work also gave an algorithm with runtime $\calO(N^4)$ per sample per determinant to evaluate the overlap estimator.
This is based on polynomial interpolation, which exhibits numerical instability~\cite{kiser2024cost}. 
We introduce two new algorithms with improved $\calO(N^3)$ runtime, based on a Paige--Van Loan reduction~\cite{paige1981schur} and a generalized eigenvalue solver~\cite{moler1973algorithm}, respectively.
Note that independent and concurrent work also derived a cubic-scaling algorithm that, similar to our first contribution (Sec.~\ref{app:overlap-algorithm}), applies to almost all overlap instances~\cite{nappi2026practicability}. 
Our second contribution, however, applies to all instances, albeit with a slightly larger runtime prefactor.
We describe all algorithms and prove their properties in Supplementary Sec.~\ref{app:overlap-algorithm}.
For the numerical implementation of these routines, we make bespoke refinements based on system size and GPU architecture to maximize performance in practice;~the details are recorded in Supplementary Sec.~\ref{app:overlap-implementation}.

\begin{figure}[htbp]
    \centering
    \includegraphics[width=0.8\linewidth]{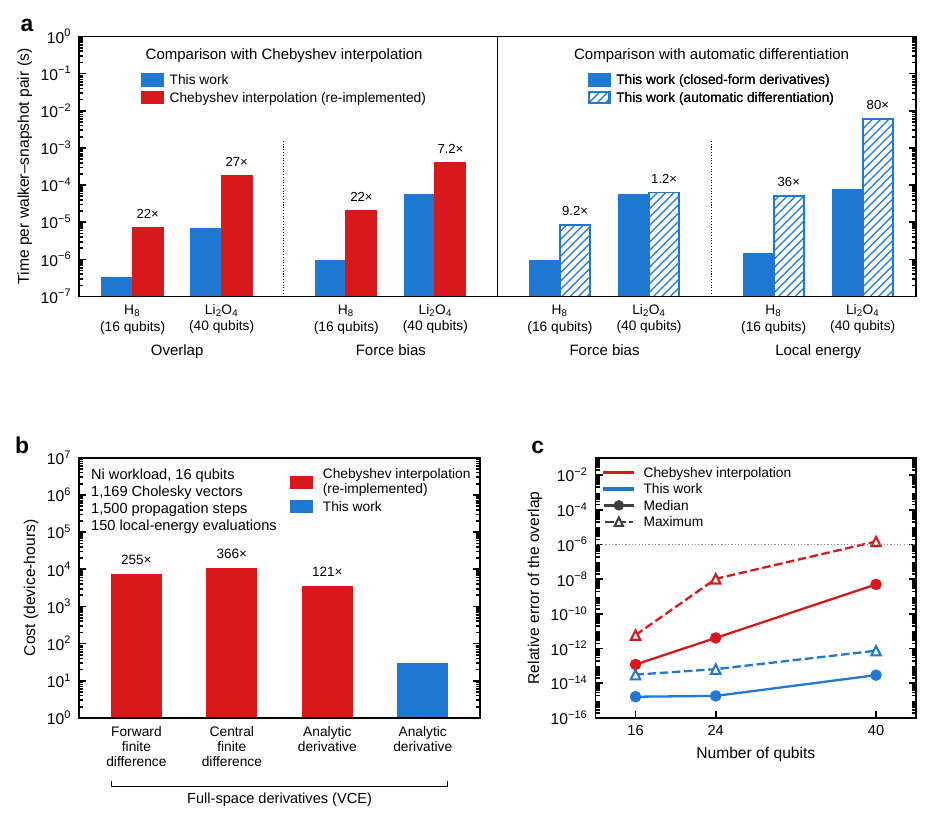}
    \caption{\textbf{Cost and accuracy of brute-force matchgate shadows.}
    All timings we measured are medians of three runs on one AMD MI250X GCD; solid blue bars show our improved algorithm with its production fused GPU kernels.
    \textbf{a},~Time per walker--snapshot pair for H$_8$ (16 qubits, 8 electrons, 15 Cholesky vectors) and Li$_2$O$_4$ (40 qubits, 26 electrons, 126 Cholesky vectors).
    Left: overlap and force bias of our algorithm (blue) compared with the Chebyshev-interpolation scheme~\cite{zhao2026trapped} (red; re-implemented in JAX in our code base).
    Right: force bias and local energy from our closed-form derivatives (Sec.~\ref{app:shadow-derivatives}) compared with automatic differentiation of the same overlap estimator (hatched), with Haar-random shadows.
    \textbf{b},~Cost of the full Ni complex run (16 qubits; 1,280 walkers, 60,000 snapshots, 1,169 Cholesky vectors; 1,500 propagation steps and 150 local-energy evaluations).
    Solid red bars show the interpolation scheme on one GCD, with the full-space derivatives of the overlap in the virtual-correlation-energy (VCE) factorization computed by finite differences, or with the analytic derivatives; the blue bar shows our improved shadow algorithm with analytic derivatives.
    \textbf{c},~Relative error of single-snapshot overlap estimates against a 40-digit reference, median and maximum over walker--snapshot pairs (16 walkers $\times$ 64 snapshots at 16 and 24 qubits, 16 $\times$ 32 at 40 qubits).
    }
    \label{fig:gpu_timimg}
\end{figure}

Fig.~\ref{fig:gpu_timimg} compares our implementation with the previous interpolation scheme~\cite{wan2023matchgate}.
For benchmarking purposes, we re-implemented the interpolation scheme with Chebyshev polynomials~\cite{zhao2026trapped} and compared it against our kernels.
We compared the overlap and force bias time cost (Fig.~\ref{fig:gpu_timimg}a) and the end-to-end time cost for a 16-qubit Ni complex in a full orbital space by making use of the virtual correlation energy (VCE) algorithm~\cite{jiang2025unbiasing} (Fig.~\ref{fig:gpu_timimg}b).
Apart from the acceleration, we also note that the traditional interpolation scheme loses accuracy as the system grows: its median per-pair relative error rises from $1\times10^{-13}$ at 16 qubits to $5\times10^{-9}$ at 40 qubits, and its maximum from $6\times10^{-12}$ to $1.5\times10^{-6}$, while ours stays below $10^{-12}$ (Fig.~\ref{fig:gpu_timimg}c).
This reflects the poor conditioning of interpolation at real nodes followed by conversion to the monomial basis~\cite{kiser2024cost}; interpolation in the Fourier basis, which we use, does not suffer from it (Supplementary Sec.~\ref{app:overlap-implementation}).
Within our own algorithm, we also compared the derivative calculation time cost between automatic differentiation and closed-form derivatives for the force bias and the local energy.
We also note that for the single-walker benchmark of ref.~\citenum{nappi2026practicability} (40 qubits, $10^5$ snapshots, one propagation step), with a reported 737~s on ten AMD EPYC 7763 cores, our implementation takes 37~s on one MI250X GCD, reflecting the benefit of implementing fast matchgate shadow algorithms on GPUs.
Beyond cost, our overlap algorithm applies to all instances, so no near-singular snapshots need to be discarded (Supplementary Sec.~\ref{app:overlap-algorithm}).
The algorithmic choices behind these gains are described in Supplementary Secs.~\ref{app:overlap-implementation} and \ref{app:shadow-derivatives}.

\begin{figure}
    \centering
    \includegraphics[width=1.0\linewidth]{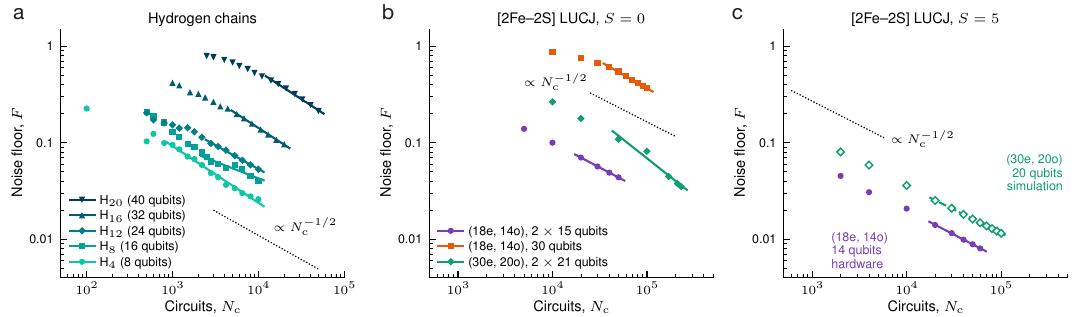}
    \caption{\textbf{Convergence of the noise floor with the number of measurement circuits.}
    Noise floor $F$ (Eq.~\eqref{eq:noise_floor}) against the number of circuits $N_\mathrm{c}$. Markers are measured values; lines are power-law fits $F \propto N_\mathrm{c}^{-p}$; the dotted line shows $N_\mathrm{c}^{-1/2}$.
    \textbf{a}, Hydrogen chains H$_4$--H$_{20}$.
    \textbf{b}, LUCJ circuits of \FeS{} at $S = 0$ on hardware: spin-uncoupled and spin-coupled circuits for (18e, 14o), both on 30 qubits, and the spin-uncoupled circuit for (30e, 20o) on 42 qubits. 
    \textbf{c}, LUCJ circuits at $S=5$, where only the $\beta$ register is run (14 and 20 qubits): (18e, 14o) on hardware; (30e, 20o) from classically simulated shadows.}
    \label{fig:noise_floor_LUCJ}
\end{figure}

\paragraph{Noisy classical-shadow estimates.}
The classical-shadow estimates of the overlaps between the prepared state $|\Psi_\mathrm{T}\rangle$ and a set of walker single determinants $\{|\phi_w\rangle\}$ are noisy data,
\begin{equation}
    y_w = \frac{1}{N_c} \sum_{c=1}^{N_c} \hat{y}_{w,c} ,
    \qquad
    \mathbb{E}\bigl[ \hat{y}_{w,c} \bigr]
      = \lambda_0 \langle \Psi_\mathrm{T} | \phi_w \rangle ,
\label{eq:label}
\end{equation}
where $\hat{y}_{w,c}$ is the estimate from circuit $c$, with its shots averaged. The common scalar $\lambda_0$ is the effect of the hardware noise after twirling by matchgate shadows, assuming Markovian, gate-independent errors~\cite{huang2024evaluating}. 
In practice, the device noise does not obey those assumptions exactly, but we can nonetheless implicitly fit for it. 
We absorb it into our fitting parameters below.
Each label $y_w$ in Eq.~\eqref{eq:label} is an unbiased estimate, with a statistical error that decreases as circuits are added.
Measurement noise therefore comes from two sources: a finite number of shots for a given circuit and hardware noise. 
This is in addition to the intrinsic shadow error due to a finite number of circuits.
We split the walkers into training and validation sets and compute every label (overlap) from the same circuits, so both sets share the same realization of measurement noise.
To quantify these sources of error, we divide the $N_c$ circuits into $n$ groups (the result is insensitive to the group size once each group contains at
least 100 circuits and beyond 500 shots per circuit in our practice), and denote the label obtained from group $g$ by $y_w^{(g)}$. The standard error of $y_w$ is
\begin{equation}
  \sigma_w^2 = \frac{1}{n(n-1)} \sum_{g=1}^{n} \bigl| y_w^{(g)} - y_w \bigr|^2
  \label{eq:sigma_mean}
\end{equation}
which measures the statistical uncertainty of the labels, from both the random measurement circuits and the finite number of shots; as such, it does not exhibit systematic errors common to all groups.
For spin-uncoupled circuits, $y_w = y_w^{\alpha} y_w^{\beta}$, so we apply Eq.~\eqref{eq:sigma_mean} to each spin-sector and combine the two to first order,
\begin{equation}
  \sigma_w^2 = |y_w^{\beta}|^2 \sigma_{w,\alpha}^2
             + |y_w^{\alpha}|^2 \sigma_{w,\beta}^2
             + 2\,\mathrm{Re}\bigl[ (y_w^{\alpha})^{*} y_w^{\beta}\, c_w \bigr] ,
  \label{eq:sigma_prod}
\end{equation}
where
$c_w = \frac{1}{n(n-1)} \sum_{g} (y_w^{\alpha,(g)} - y_w^{\alpha})(y_w^{\beta,(g)} - y_w^{\beta})^{*}$ is the covariance of the two half labels, kept because both halves are measured in the same circuits. In practice, we evaluate $\sigma_w$ by a jackknife over the same groups, which agrees with Eq.~\eqref{eq:sigma_prod} to leading order in $1/n$. 
We define the noise floor on the validation walkers,
\begin{equation}\label{eq:noise_floor}
  F = \frac{\lVert \sigma \rVert_{\mathrm{val}}}{\lVert y \rVert_{\mathrm{val}}}
    = \left( \frac{\sum_{w \in \mathrm{val}} \sigma_w^2}
                  {\sum_{w \in \mathrm{val}} |y_w|^2} \right)^{1/2} .
\end{equation}
Fig.~\ref{fig:noise_floor_LUCJ} shows the convergence of the noise floor of hydrogen chains and different LUCJ circuits for \FeS{}  (including state preparation and the matchgate shadow measurements).

For a given surrogate $\Psi_\theta$ with parameters $\theta$, we measure its fit by
\begin{equation}
  v = \left( \frac{\sum_{w \in \mathrm{val}}
        \bigl| \lambda \langle \Psi_\theta | \phi_w \rangle - y_w \bigr|^2}
       {\sum_{w \in \mathrm{val}} |y_w|^2} \right)^{1/2} ,
  \qquad
  \lambda = \frac{\sum_{w \in \mathrm{train}}
        \langle \Psi_\theta | \phi_w \rangle^{*} y_w}
       {\sum_{w \in \mathrm{train}} |\langle \Psi_\theta | \phi_w \rangle|^2} ,
  \label{eq:v}
\end{equation}
where $\lambda$ is the least-squares estimate of $\lambda_0$ on the training walkers, so that no parameter is adjusted to the validation labels.
Writing $y_w = \lambda_0 \langle \Psi_\mathrm{T} | \phi_w \rangle + \epsilon_w$ with $\mathbb{E}\bigl[ |\epsilon_w|^2 \bigr] = \sigma_w^2$, the circuit state itself gives $\mathbb{E}[v^2] \simeq F^2$: from the labels alone, no surrogate can be shown to agree with the circuit state more closely than $F$. A surrogate at relative distance $d$ from the noise-free labels gives $\mathbb{E}[v^2] \simeq d^2 + F^2$ when the noise is independent of the surrogate. Therefore, we define the signal-to-noise ratio (SNR) of a starting state with validation error $v_0$ as
\begin{equation}\label{eq:snr}
  \mathrm{SNR} = \frac{\sqrt{v_0^2 - F^2}}{F} \simeq \frac{d}{F} ,
\end{equation}
which measures how much of its error the labels can resolve. Because the training and validation walkers share the same hardware noise source, training can lower $v$ below $F$ by fitting to the noise; we therefore use $v \le F$ as a stopping criterion.

\subsection{Surrogate models and training strategy}\label{app:surrogates}
\begin{figure}[htbp]
    \centering
    \includegraphics[width=0.7\linewidth]{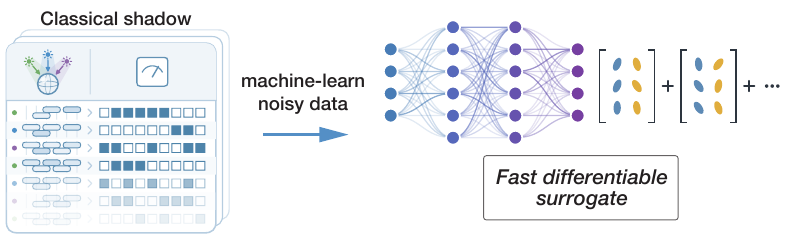}
    \caption{\textbf{Learning a surrogate from classical shadows.}
    Random matchgate measurements on the device give classical shadows, from which the overlaps of the trial state with a fixed set of training walkers are estimated.
    A differentiable classical model is fitted to these noisy estimates and then supplies the overlap and its derivatives at every step of QMC.}
\label{fig:si_shadow_ml}
\end{figure}

During QC-QMC, we replace the shadow estimator with a differentiable model trained on the shadow estimates (Fig.~\ref{fig:si_shadow_ml}), which returns the overlap and its first and second derivatives for any walker at a cost independent of the number of snapshots.
In this work we used two such models separately: a backflow neural quantum surrogate for the hydrogen chains and a nonorthogonal configuration interaction (NOCI) surrogate for \FeS{}.
We deliberately start the NQS from a random initialization, with no classical initial guess, so that the hydrogen-chain trial is learned from the noisy measurements alone.
The NOCI surrogate instead starts from a cheap classical NOCI calculation.
For harder problems, they can be combined, for example by starting from a NOCI expansion and letting a backflow network learn walker-dependent corrections to its orbitals, so that the chemical prior sets the starting point and the network adds flexibility and expressivity where the data can resolve it.

\paragraph{Backflow neural quantum surrogate.}
We propose a neural quantum surrogate, which returns a single determinant whose orbitals depend on the input walker $\phi=(\phi_\alpha,\phi_\beta)$ through its spin projectors,
\begin{equation}
\begin{split}
f_\theta(\phi)&=\prod_{\sigma\in\{\alpha,\beta\}}
\det\!\left[G^{\sigma}_\theta(P^{\alpha}_\phi,P^{\beta}_\phi)^{\dagger}\phi_\sigma\right],\\
P^{\sigma}_\phi&=\phi_\sigma\left(\phi_\sigma^\dagger\phi_\sigma\right)^{-1}\phi_\sigma^\dagger ,
\end{split}\label{eq:backflow_density}
\end{equation}
where the function
\begin{align}
G^{\sigma}_\theta &: \Gr(\eta_\alpha,N) \times \Gr(\eta_\beta,N) \to \C^{N\times \eta_\sigma},\\
\Gr(\eta, N) &= \{P \in \C^{N\times N} : P^2=P=P^\dagger, \space \tr(P) = \eta\}, \notag
\end{align}
outputs a set of $\phi$-dependent orbitals.
The network sees the walker only through its projector $P^{\alpha}_\phi\oplus P^{\beta}_\phi$, a point on the
complex Grassmann manifold $\Gr(\eta_\alpha,N) \times \Gr(\eta_\beta,N)$. 
$P^{\sigma}_\phi$ is invariant under $\phi_\sigma\to\phi_\sigma U_\sigma$, so the surrogate transforms as an
overlap with a fixed state, $f_\theta(\phi U)=\det(U_\alpha) \det(U_\beta) \, f_\theta(\phi)$.
For singlet calculation, we use spin projection where the walkers are spin-restricted ($\phi_\alpha=\phi_\beta$, hence $P^{\alpha}_\phi=P^{\beta}_\phi$),
the network can just take a single projector as input while still outputting separate $\alpha$ and $\beta$ orbitals, which allows for spin-broken determinants.

\paragraph{Derivatives of the neural surrogate.}
\label{app:nqs-derivatives}
We take every derivative along a unitary orbital rotation, $\Phi_B(s)=e^{isB}\Phi$ with $B$ Hermitian and $s$ real, and define
\begin{equation}
  \langle\hat B\rangle_\theta=\frac{-i\,\partial_s f_\theta(\Phi_B(s))|_{s=0}}{f_\theta(\Phi)},\qquad
  \langle\hat B^2\rangle_\theta=\frac{-\partial_s^2 f_\theta(\Phi_B(s))|_{s=0}}{f_\theta(\Phi)},
\end{equation}
for the one-body operator $v_0$ and the Cholesky vectors $L_\gamma$ (in Eq.~\eqref{eq:app-local-energy}, $v_\gamma=iL_\gamma$).
Only derivatives up to second order are needed, so we use $\Phi_B(s)=(I+isB-\tfrac12 s^2B^2)\Phi$.
All derivatives are obtained by automatic differentiation of the scalar overlap. One reverse-mode pass gives
\begin{equation}
  G_{pq}=\frac{1}{f_\theta(\Phi)}\sum_k\left(\frac{\partial f_\theta}{\partial\Phi_{pk}}\Phi_{qk}
  -\frac{\partial f_\theta}{\partial\Phi^{*}_{pk}}\Phi^{*}_{qk}\right),
\end{equation}
with $\langle\hat B\rangle_\theta=\sum_{pq}B_{pq}G_{pq}$ for real symmetric $B$, so the force bias along every Cholesky
vector follows from a single pass. For a fixed state, the second term vanishes, and $G$ is the mixed one-body Green's
function. The one-body energy uses one forward-mode derivative along $h$. We evaluate $\langle\hat L_\gamma^2\rangle_\theta$ for each Cholesky vector by nested forward-mode derivatives along $\Phi_{L_\gamma}(s)$.

\paragraph{Nonorthogonal configuration interaction surrogate.}
The surrogate can also be built from a quantum-chemistry wavefunction ansatz, which brings chemical prior knowledge into the model and serves as a good initial guess for training. Here we use NOCI~\cite{jimenez2013multi,giuliani2026precise}, a linear combination of $K$ Slater determinants whose orbitals are optimized independently and need not be orthogonal between determinants,
\begin{equation}
    \ket{\Psi_\theta}=\sum_{k=1}^{K}c_k\ket{D_k},
    \qquad
    f_\theta(\phi)=\braket{\Psi_\theta|\phi}
    =\sum_{k=1}^{K}c_k^{*}\prod_{\sigma\in\{\alpha,\beta\}}
    \det\!\left(W_{k\sigma}^{\dagger}\phi_\sigma\right),
    \label{eq:noci}
\end{equation}
where $W_{k\sigma}\in\C^{N\times\eta_\sigma}$ holds the occupied orbitals of spin $\sigma$ in $\ket{D_k}$ and $\theta=\{c_k,W_{k\sigma}\}$.
Unlike the backflow surrogate, $\ket{\Psi_\theta}$ is a fixed state that does not depend on the walker. 
The $\alpha$ and $\beta$ orbitals of each determinant are independent, so every $\ket{D_k}$ may break spin symmetry, as a broken-symmetry unrestricted Hartree--Fock (BS-UHF) determinant does.
Each determinant is parameterized by a Thouless chart~\cite{thouless1960stability} around a fixed origin $O_{k\sigma}\in\C^{N\times\eta_\sigma}$,
\begin{equation}
    W_{k\sigma}=O_{k\sigma}+V_{k\sigma}Z_{k\sigma},
    \qquad
    Z_{k\sigma}\in\C^{(N-\eta_\sigma)\times\eta_\sigma},
    \label{eq:noci-chart}
\end{equation}
where the columns of $V_{k\sigma}$ form an orthonormal basis of the orthogonal complement of the columns of $O_{k\sigma}$ ($V_{k\sigma}^{\dagger}O_{k\sigma}=0$). Every determinant with nonzero overlap with the origin is reached by some $Z_{k\sigma}$, up to a norm and phase that $c_k$ absorbs. Occupied--occupied rotations, which change only the norm and phase of $\ket{D_k}$, are excluded, and no orthonormality constraint is imposed.
For a walker with orbital matrices $\Phi_\sigma$ define, for each determinant,
\begin{equation}
    S_{k\sigma}=W_{k\sigma}^{\dagger}\Phi_\sigma,
    \qquad
    \Theta_{k\sigma}=\Phi_\sigma S_{k\sigma}^{-1}W_{k\sigma}^{\dagger},
    \qquad
    o_k=c_k^{*}\prod_\sigma\det S_{k\sigma},
\end{equation}
so that $f_\theta(\Phi)=\sum_k o_k$. Here $(\Theta_{k\sigma})_{qp}=\bra{D_k}a_{p\sigma}^{\dagger}a_{q\sigma}\ket{\phi}/\braket{D_k|\phi}$ is the mixed one-body Green's function of determinant $k$ (in transposed form), and $\Theta_{k\sigma}^2=\Theta_{k\sigma}$. Applying Jacobi's formula $\partial_t\det M=\det M\,\tr(M^{-1}\partial_tM)$ to the directional derivatives of Eq.~\eqref{eq:app-directional} for a spin-free one-body operator $\hat A$ gives
\begin{equation}
\begin{aligned}
    F_A'(0)&=\sum_k o_k\sum_\sigma\tr\!\left(A\Theta_{k\sigma}\right),\\
    F_A''(0)&=\sum_k o_k\Bigl\{\Bigl[\sum_\sigma\tr\!\left(A\Theta_{k\sigma}\right)\Bigr]^2
    +\sum_\sigma\tr\!\left[A\Theta_{k\sigma}A\left(1-\Theta_{k\sigma}\right)\right]\Bigr\},
\end{aligned}
\label{eq:noci-wick}
\end{equation}
which is the generalized Wick theorem applied to each determinant. Inserted into Eq.~\eqref{eq:app-local-energy}, Eq.~\eqref{eq:noci-wick} makes the force bias and local energy overlap-weighted averages over determinants, with weights $o_k/f_\theta(\Phi)$. All quantities are closed-form, and no automatic differentiation is needed during propagation. The half-rotated operators $W_{k\sigma}^{\dagger}v_\gamma$ are computed once per trial, so the cost per walker is $O(KN_\gamma N\eta^2)$, set by the exchange term, where $N_\gamma$ is the number of auxiliary fields.

For $K=1$ the origin is the BS-UHF determinant. For $K\ge2$ the origins and coefficients come from a variation-after-projection (VAP) calculation~\cite{jimenez2013multi}, which uses the Hamiltonian only and no shadow data. The orbitals minimize the singlet-projected energy
\begin{equation}
    E_0[\Psi]=\frac{\bra{\Psi}\hat H\hat P_0\ket{\Psi}}{\bra{\Psi}\hat P_0\ket{\Psi}},
    \qquad
    \hat P_0=\frac{1}{2}\int_0^{\pi}\!d\beta\,\sin\beta\,e^{-i\beta\hat S_y},
    \label{eq:vap}
\end{equation}
where $\hat P_0$ projects $S_z=0$ states onto total spin $S=0$, and the integral is evaluated by quadrature in $\beta$. For fixed orbitals, the coefficients are the lowest solution of the projected generalized eigenproblem $\sum_l\bra{D_k}\hat H\hat P_0\ket{D_l}c_l=E\sum_l\bra{D_k}\hat P_0\ket{D_l}c_l$. 
We add new determinants by starting from a perturbed copy of existing ones, and we reoptimize the whole expansion at every stage. The QMC trial and the starting point of the learning procedure are the unprojected state (Eq.~\eqref{eq:noci}) with the VAP orbitals and coefficients.

\paragraph{Machine learning quantum measurements from a NOCI prior.}
The labels $y_w$, their standard errors $\sigma_w$, the validation error $v$, the noise floor $F$ and the signal-to-noise ratio (SNR) are defined in Sec.~\ref{app:shadow_noise}. The SNR of the starting state, Eq.~\eqref{eq:snr}, compares the part of its error that the labels can resolve with the label noise, and we use it to decide how many parameters to fit. If $\mathrm{SNR}>1$, the labels resolve the error of the starting state and we teach all parameters (i). If $\mathrm{SNR}\le1$, the starting state already agrees with the labels to within their noise; fitting all orbital parameters would mostly fit the noise, so we keep the VAP orbitals and refit only the $K$ coefficients (ii).
The walkers come in pairs $p=(\phi_{p0},\phi_{p1})$ taken from the same random walk at two imaginary times, and both walkers of every pair are labeled.

(i) \emph{Teaching.} Starting from the VAP state ($Z_{k\sigma}=0$), all orbital parameters $Z_{k\sigma}$ and coefficients $c_k$ are trained by minimizing the overlap error and overlap ratios. We optimize with Adam and keep the first checkpoint with $v\le F$, or the one with the lowest $v$ if $F$ is never reached.

(ii) \emph{Coefficient refit.} With the orbitals frozen at the VAP origins, Eq.~\eqref{eq:noci} is linear in $c_k^{*}$, $f_\theta(\phi_w)=\sum_k\mathcal O_{wk}c_k^{*}$ with $\mathcal O_{wk}=\prod_\sigma\det(O_{k\sigma}^{\dagger}\phi_{w\sigma})$. Absorbing the global label factor into the coefficients turns the fit into a linear least-squares problem over both walkers of every training pair,
\begin{equation}
    c^{*}=\operatorname*{arg\,min}_{x\in\C^K}\sum_{w\in\mathrm{train}}\Bigl|\sum_k\mathcal O_{wk}x_k-y_w\Bigr|^2,
    \label{eq:noci-refit}
\end{equation}
whose solution is unique when $\mathcal O$ has full column rank and needs no optimizer or checkpoint selection.

The training set consists of $N_{\mathrm{train}}$ walker pairs drawn from UHF-guided phaseless AFQMC, one time step apart,
\begin{equation}
\mathcal{D}=\{(\phi_i,\phi_i')\}_{i=1}^{N_{\mathrm{train}}},\qquad \phi_i=\phi(\tau_i),\quad \phi_i'=\phi(\tau_i+\Delta\tau).
\end{equation}
We find no significant difference between training on walkers sampled under UHF guidance and on walkers sampled under
correlated-trial guidance, so the training set can be generated with an inexpensive classical trial before the correlated
trial is available.
Each label is the averaged shadow overlap of the trial state with the walker,
\begin{equation}
y(\phi)=\frac{1}{S}\sum_{s=1}^{S}\hat o_s(\phi),\qquad \mathbb{E}\big[\hat o_s(\phi)\big]=\langle\Psi_T|\phi\rangle,
\end{equation}
where $\hat o_s(\phi)$ is the overlap estimator from the $s$-th of $S$ shadow snapshots.
Labels are divided by the modulus of the largest training label, and the model output is divided by its modulus on the same walker,
\begin{equation}
\tilde y(\phi)=\frac{y(\phi)}{|y(\phi_{\mathrm{ref}})|},\qquad
\tilde f_\theta(\phi)=\frac{\Psi_\theta(\phi)}{|\Psi_\theta(\phi_{\mathrm{ref}})|},\qquad
\phi_{\mathrm{ref}}=\operatorname*{arg\,max}_{\phi\in\mathcal{T}}|y(\phi)|,
\end{equation}
where $\mathcal{T}$ is the set of $2N_{\mathrm{train}}$ training walkers (both members of every pair).
The loss is a log-cosh penalty on the real and imaginary parts of the complex residual
$r_\theta(\phi)=\tilde f_\theta(\phi)-\tilde y(\phi)$, each scaled by its spread over the training set,
\begin{equation}
\mathcal{L}_{\mathrm{ov}}(\theta)=\frac{1}{2N_{\mathrm{train}}}\sum_{\phi\in\mathcal{T}}
\left[\ln\cosh\frac{\operatorname{Re}r_\theta(\phi)}{s_{\mathrm{R}}}+\ln\cosh\frac{\operatorname{Im}r_\theta(\phi)}{s_{\mathrm{I}}}\right],
\end{equation}
with $s_{\mathrm{R}}$ and $s_{\mathrm{I}}$ the standard deviations of $\operatorname{Re}\tilde y$ and $\operatorname{Im}\tilde y$ over $\mathcal{T}$.
Each surrogate is trained by minimizing an overlap term and, optionally, a pair-ratio term,
\begin{equation}
\mathcal L(\theta)=\mathcal L_{\mathrm{ov}}(\theta)+\beta\,\mathcal L_{R}(\theta).
\label{eq:train-loss}
\end{equation}
The overlap term penalizes the residual between the surrogate and the labels over the training walkers,
after the unknown global factor of the labels is removed, either by normalizing labels and surrogate at a
reference walker or by a least-squares complex scale. The ratio term exploits the fact that the two walkers
of a pair are one time step apart and share the same shadow snapshots, so most of the noise cancels in their
ratio $R_i=y(\phi_i')/y(\phi_i)$. With $R_{\theta,i}=f_\theta(\phi_i')/f_\theta(\phi_i)$,
\begin{equation}
\mathcal L_R(\theta)=\frac{\sum_i\omega_i\Bigl[\Bigl(\frac{\ln|R_{\theta,i}|-\ln|R_i|}{s_{\ln}}\Bigr)^2
+\Bigl(\frac{\operatorname{wrap}(\arg R_{\theta,i}-\arg R_i)}{s_{\arg}}\Bigr)^2\Bigr]}{\sum_i\omega_i},
\label{eq:train-ratio}
\end{equation}
where the weights $\omega_i$ suppress pairs whose label ratio is dominated by noise.
These ratios set the importance weights and the phaseless angle of the walker.
One can also employ derivative information including force bias and local energies to perform Sobolev training.
We choose the checkpoint on walkers not used for training, with the noise floor $F$ as the reference.

\section{Benchmarks and error analysis}
\label{app:shadow-convergence}
\subsection{Noiseless simulations}\label{app:hchain-sim}
\begin{figure}[h]
    \centering
    \includegraphics[width=0.9\textwidth]{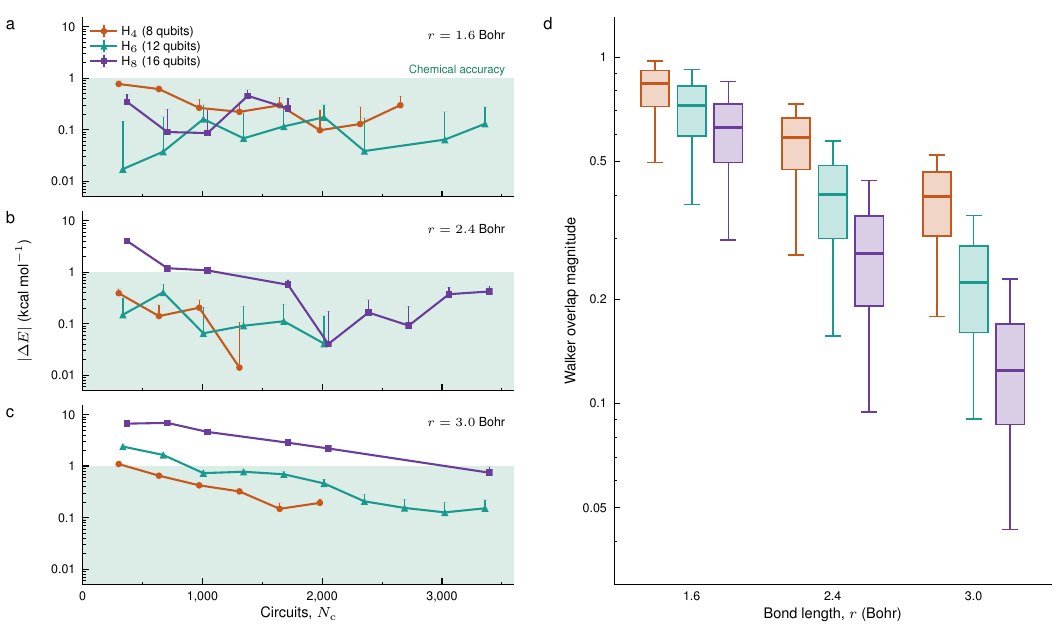}
    \caption{\textbf{QC-QMC with simulated classical shadows for hydrogen chains.} Noiseless simulated shadows at 50 shots per circuit for H$_4$, H$_6$ and H$_8$ (8, 12 and 16 qubits).
    \textbf{a--c}, Absolute deviation $|\Delta E|$ of the total AFQMC energy from UHF-AFQMC against the number of circuits $N_\mathrm{c}$, at $r=1.6$, $2.4$ and $3.0$ Bohr, respectively. 
    \textbf{d}, Walker-overlap magnitudes from the same runs, grouped by bond length. Each box pools 12800 walker values, weighted equally and without AFQMC population weights. Boxes span the interquartile range, lines mark the median, and whiskers extend to the 5th and 95th percentiles.}
    \label{fig:si-hchain-sim}
\end{figure}

We examine how shadow-reconstructed trial states and AFQMC energies depend on measurement budget and simulated readout noise for hydrogen chains with 8, 12, and 16 qubits. Simulated results and walker-overlap distributions are shown at interatomic spacings of $r=1.6$, $2.4$, and $3.0$ Bohr.
\Cref{fig:si-hchain-sim} shows the AFQMC energy deviation against the number of circuits for noiseless simulated shadows, and the walker-overlap magnitudes at each bond length.

\subsection{Simulated readout errors}\label{app:hchain-readout}
\begin{figure}
    \centering
    \includegraphics[width=0.85\textwidth]{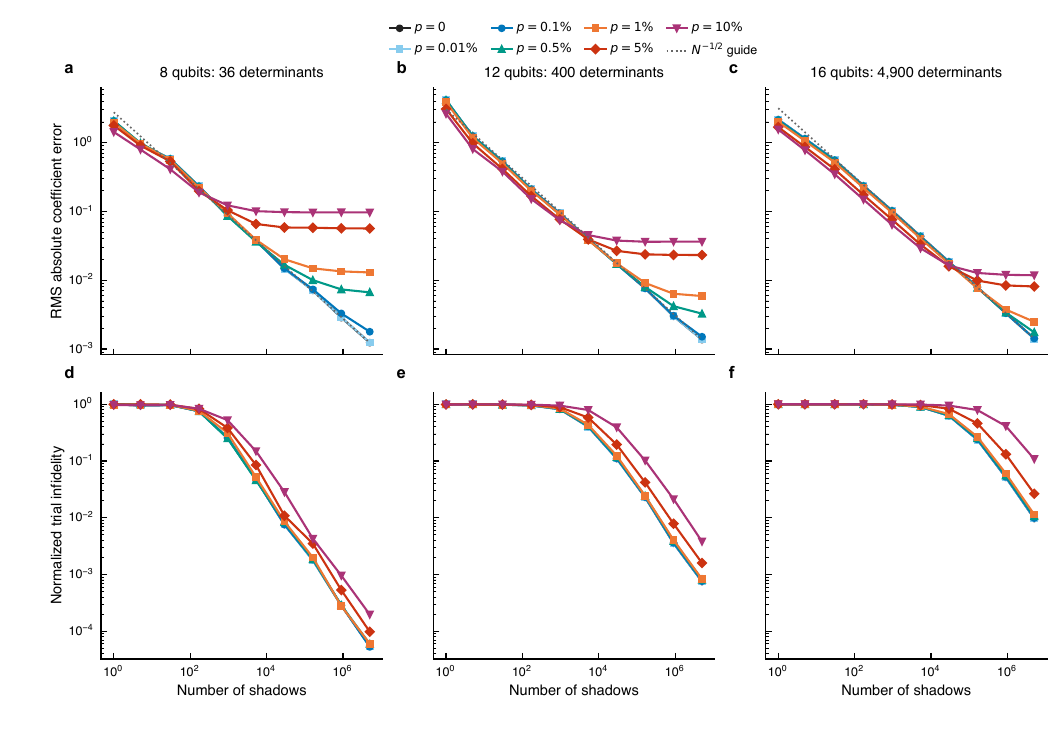}
    \caption{\textbf{Shadow coefficient convergence under simulated readout noise.}
    \textbf{a--c}, Root-mean-square absolute coefficient errors for 8, 12, and 16 qubits, respectively.
    \textbf{d--f}, Corresponding normalized trial-state infidelities, $1-|\langle\Psi_{\mathrm{UHF}}|\widehat{\Psi}\rangle|^2/(\langle\Psi_{\mathrm{UHF}}|\Psi_{\mathrm{UHF}}\rangle\langle\widehat{\Psi}|\widehat{\Psi}\rangle)$.
    Colors and markers denote readout flip probabilities $p$. Dotted lines show $N^{-1/2}$ scaling guides anchored to the final noiseless coefficient error. Absolute coefficient errors retain overall attenuation, whereas normalized infidelity is invariant to a global scale factor. 
    Each curve uses prefixes of a single sequence of shadows, so no error bars are shown.
    }
    \label{fig:si-hchain-coefficients}
\end{figure}

\begin{figure}
    \centering
    \includegraphics[width=0.85\textwidth]{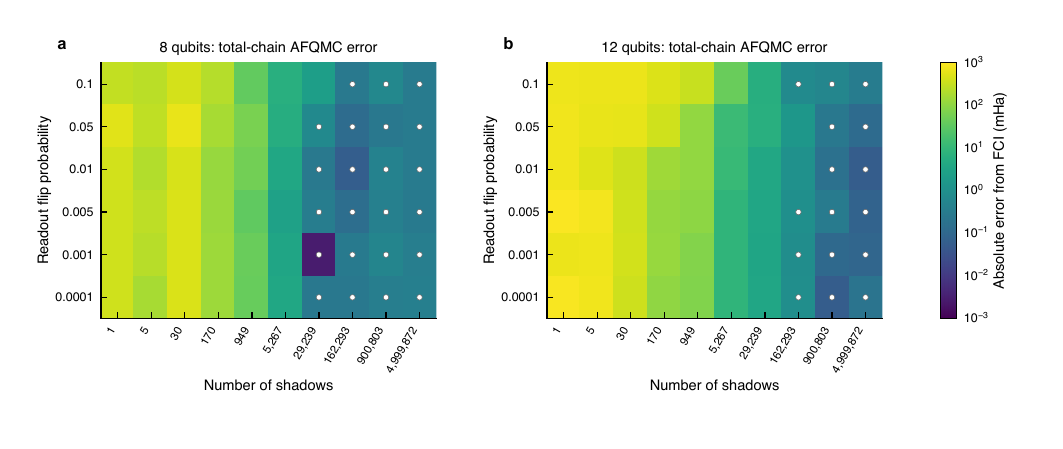}
    \caption{\textbf{AFQMC accuracy across simulated readout noise and shadow budget.}
    Absolute deviation of the total AFQMC energy from FCI for (\textbf{a}) 8 and (\textbf{b}) 12 qubits, using the shadow-estimated state as the trial, on a shared logarithmic color scale. White circles mark deviations below 1~$\mHa$.
    }
    \label{fig:si-hchain-noise}
\end{figure}

\Cref{fig:si-hchain-coefficients} illustrates this in simulation for UHF trials on 8, 12 and 16 qubits, where the full coefficient vector can be compared with shadow estimates under symmetric readout flips with probability $p$ up to 10\%, a range that includes the readout errors of the device (\cref{fig:hardware_error}).
Without noise, the coefficient error falls as $N^{-1/2}$ with the number of shadows $N$.
Readout noise leaves a floor in the absolute coefficient error that rises with $p$, but the normalized infidelity, which is insensitive to a global factor, keeps decreasing with $N$, more slowly only at the largest $p$.
Readout noise therefore acts mainly as a global attenuation, which Eq.~\eqref{eq:label} absorbs into $\lambda_0$.
\Cref{fig:si-hchain-noise} carries the same simulation through to the energy, using the shadow-estimated state as the AFQMC trial for 8 and 12 qubits.
Because AFQMC uses the trial only through overlap ratios, a global factor cancels (\cref{app:error-mitigation}), and the energy error is set mainly by the number of shadows rather than by the readout noise. With $9\times10^5$ or more shadows, the error is below 1~$\mHa$ at every flip probability shown, up to $p=10\%$, for both sizes; at 8 qubits, $1.6\times10^5$ shadows already suffice at every $p$, and $3\times10^4$ for $p\le 5\%$.

\subsection{Experimental convergence of matchgate-shadow overlaps}\label{app:convergence}
\begin{figure}
    \centering
    \includegraphics[width=0.85\textwidth]{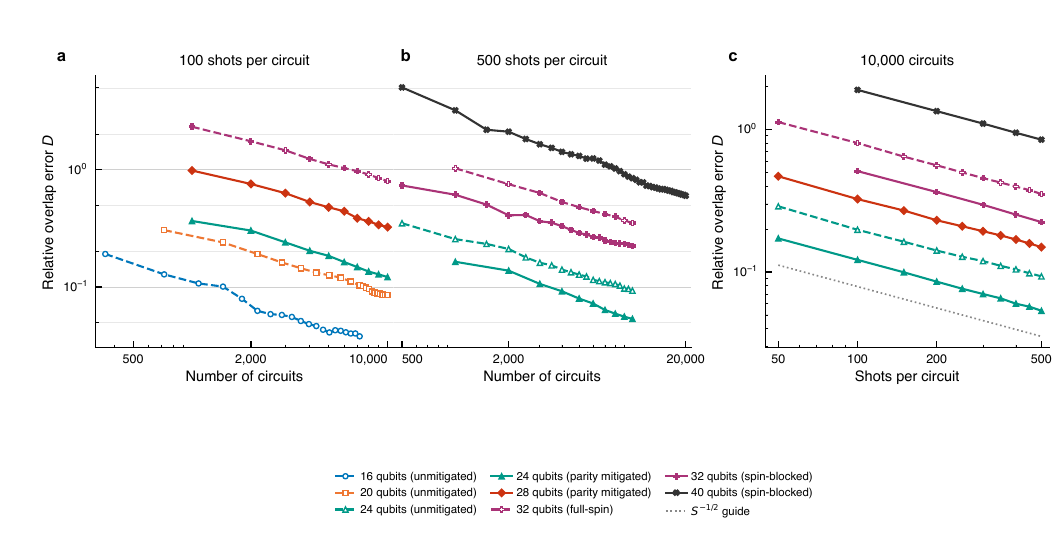}
\caption{\textbf{Experimental convergence of matchgate-shadow overlap estimates for QMC walkers.}
\textbf{a,b}, Relative overlap error versus the number of circuits at 100 and 500 shots per circuit, respectively. Panel \textbf{a} shows 16, 20, 24, 28 and 32 qubits; panel \textbf{b} compares 24, 32 and 40 qubits, including parity mitigation at 24 qubits and full-spin versus spin-blocked shadows at 32 qubits.
\textbf{c}, Convergence versus shots per circuit at a fixed budget of 10,000 circuits for full-spin shadows (24, 28 and 32 qubits) and spin-blocked shadows (32 and 40 qubits).
The dimensionless error is $D=\|\hat{\mathbf a}-\mathbf a\|_2/\|\mathbf a\|_2$, where $\mathbf a$ contains the analytic walker overlaps and $\hat{\mathbf a}$ contains shadow estimates normalized by the shadow-estimated UHF reference at the same measurement budget.
Colors identify qubit counts; dashed and solid curves distinguish the acquisition variants indicated in the legends.
All axes are logarithmic, and panels \textbf{a,b} share the same error scale. The gray dotted line in \textbf{c} is an $S^{-1/2}$ guide, where $S$ is the number of shots per circuit.}
    \label{fig:SI_hchain_conv}
\end{figure}

\begin{figure*}
    \centering
    \includegraphics[width=0.9\linewidth]{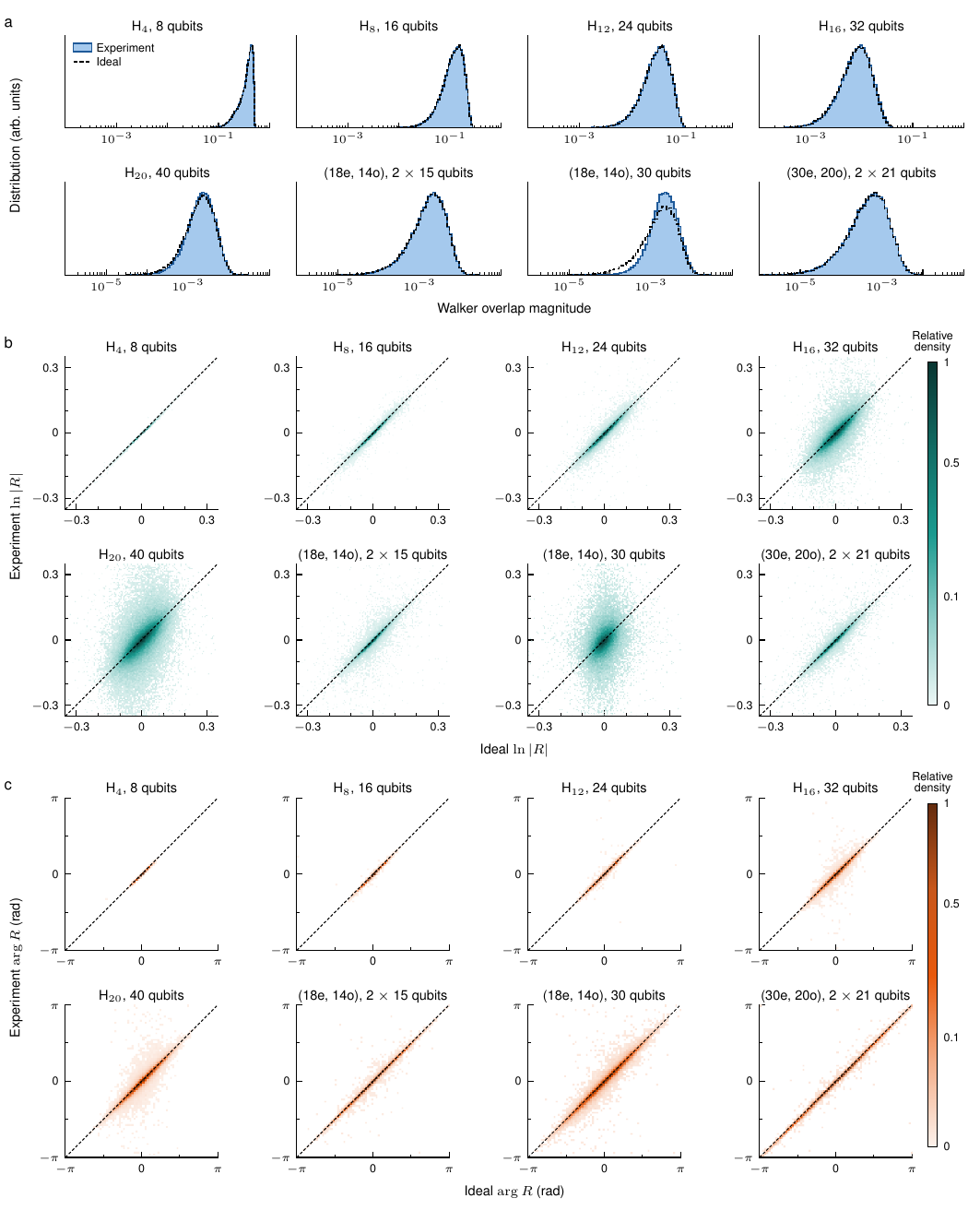}
    \caption{\textbf{Experimental shadows capture the walker-overlap signal.}
    Experiment (overlaps estimated from the hardware classical shadows) against ideal values for the hydrogen chains and the \FeS{} LUCJ circuits: \textbf{a}, distributions of the walker-overlap magnitude (experiment filled, ideal dashed); \textbf{b}, \textbf{c}, log-magnitude ($\ln|R|$) and phase ($\arg R$) of the overlap ratio $R$ for each walker pair, experiment against ideal.}
    \label{fig:si-exp-overlaps}
\end{figure*}

\begin{figure}[b]
    \centering
    \includegraphics[width=0.6\linewidth]{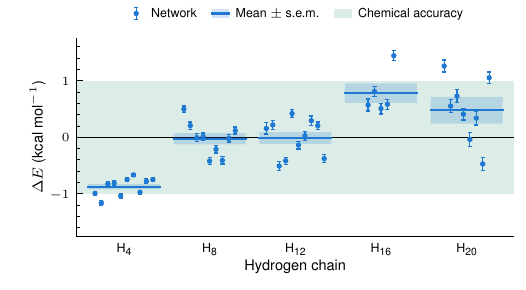}
    \caption{\textbf{Training-seed dependence of the hydrogen-chain energies.}
    Deviation of QC-QMC with the NQS surrogate for each network that passes the health checks, with the statistical error of its AFQMC run.
    }
    \label{fig:seeds_dependence}
\end{figure}

\Cref{fig:SI_hchain_conv} shows how hardware estimates of the walker overlaps for the hydrogen chains converge with the number of circuits and with the shots per circuit, and how two choices in the experiment affect this convergence.
Parity postselection (\cref{app:error-mitigation}) lowers the relative overlap error at 24 qubits by a factor of about 1.7 at every budget shown.
Spin-blocked shadows (\cref{app:spin-block}) lower it at 32 qubits by a factor of about 1.6 relative to full-spin shadows, consistent with their shallower circuits (\cref{fig:Hchain_matchgate_counts}).
At a fixed number of circuits, all curves fall as $S^{-1/2}$ with the number of shots $S$, so both choices reduce the prefactor rather than changing the rate. A factor of 1.6--1.7 in error corresponds to roughly 2.5--3 times fewer shots for the same accuracy. \Cref{fig:si-exp-overlaps} compares the overlaps estimated from the hardware shadows with their ideal value. The experiment reproduces both the distribution of walker-overlap magnitudes and the overlap ratios of individual walker pairs.

\subsection{NQS training, error, and cost analysis}

\begin{table*}[t]
\centering
\caption{
\label{table:hchain-cost}\textbf{Cost of conventional QC-QMC (optimized algorithm) and of QC-QMC with the neural surrogate for hydrogen chains.} GPU hours are hours on one graphics compute die (GCD) of an AMD Instinct MI250X GPU on the Frontier supercomputer at the Oak Ridge Leadership Computing Facility (OLCF). An AFQMC run has 640 walkers and 5,000 blocks, each block being 25 propagation steps (overlap and force bias) and one local-energy evaluation. Shadows: number of circuits, shots per circuit, and their product. The surrogate is trained on $N$ walker pairs plus 2,560 stopping-set pairs ($N$ = 3,200, 3,200, 19,200, 19,200 and 38,400 from 8 to 40 qubits), so $2(N + 2560)$ walkers are labeled, each with its overlap estimated from all the shadows.
The cost of this work covers the full protocol: Training is the sum of ten trainings (one per network seed), Run is ten AFQMC runs, one with each trained network as the trial, and Total is their sum with labeling. Speedup is the conventional cost over the total cost of this work (10 independent trained models and AFQMC). 
}
\small
\setlength{\tabcolsep}{4pt}
  \begin{tabular}{lrrrrrrrrrrr}
  \hline
   & & \multicolumn{3}{c}{Shadows} & Labeled & \multicolumn{4}{c}{This work (GPU hours)} & Conventional & \\
  \cline{3-5}\cline{7-10}
  System & Qubits & Circuits & Shots & Total & walkers & Labeling & Training & Run & Total & (GPU hours) & Speedup \\
  \hline
  H$_4$    & 8  & 100    & 500   & 50,000     & 11,520 & 0.004   & 0.40 & 0.76 & 1.17    & 218       & 187$\times$   \\
  H$_8$    & 16 & 2,500  & 500   & 1,250,000  & 11,520 & 0.33    & 0.35 & 1.13 & 1.80    & 7,367     & 4,082$\times$ \\
  H$_{12}$ & 24 & 10,000 & 1,000 & 10,000,000 & 43,520 & 37.9    & 0.45 & 1.76 & 40.1    & --   & -- \\
  H$_{16}$ & 32 & 20,000 & 1,380 & 27,600,000 & 43,520 & 189.9   & 0.46 & 2.55 & 192.9   & --   & -- \\
  H$_{20}$ & 40 & 50,000 & 1285 & 64,250,000 & 81,920 & 1,188.2 & 0.40 & 4.21 & 1,192.8 & -- & -- \\
  \hline
  \end{tabular}
\end{table*}

For each chain length (8, 16, 24, 32 and 40 qubits, i.e.\ H$_4$--H$_{20}$) we train ten neural-network trial wavefunctions on classical-shadow overlap data.
The network maps a walker $\phi=(\phi_\alpha,\phi_\beta)$ to walker-dependent orbitals $\psi_\sigma(\phi;\theta)\in\mathbb{C}^{n\times n_\sigma}$
and returns the single-determinant overlap
\begin{equation}
  \Psi_\theta(\phi)=\det\!\big[\psi_\alpha(\phi;\theta)^\dagger\phi_\alpha\big]\,\det\!\big[\psi_\beta(\phi;\theta)^\dagger\phi_\beta\big].
\end{equation}

One QMC run propagates 640 walkers for 5{,}000 blocks of 25 steps.
Each step requires the trial overlap and force bias of every walker, and each block requires one local-energy evaluation.
In conventional QC-QMC, every overlap, force bias, and local energy is averaged over all $N_\mathrm{s}$ snapshots with GPU matchgate-shadow kernels. Its cost is the measured cost of one block on the full shadow data, multiplied by 5000. 
At 8 and 16 qubits, the circuit counts are those at which brute-force shadow AFQMC reproduces the noiseless reference within chemical accuracy: 100 and 2{,}500 circuits at 500 shots, $N_\mathrm{s}=5\times10^4$ and $1.25\times10^6$.
For the surrogate, we count three parts: the shadow labels, network training, and the AFQMC run with the trained network. The labels are the overlaps of $2N+5120$ walkers: both walkers of the $N$ training pairs and of 2{,}560 stopping-set pairs, each averaged over all $N_\mathrm{s}$ snapshots with the same kernels.
For the surrogate training, each orbital $p$ is a token whose features are the real and imaginary parts of the $p$th row of $P$ (Eq.~\eqref{eq:backflow_density}). The tokens are linearly embedded to width $d$ with a learned orbital embedding, processed by a transformer encoder (SiLU activations) and a full linear output head, with Xavier-normal initialization scaled by 0.01.
We perform 10 independent trainings, and each trained network must pass a health check before its AFQMC run.
The check uses no benchmark energy and is evaluated on an external panel of validation walkers not used in training or early stopping. A network fails if its validation error exceeds the label noise floor, $v/F > 1$. Among the networks in a run that pass, we also remove a network if its mean local energy from the trained model exceeds the median over those networks by more than 1.6~mHa (chemical accuracy); only differences between networks enter, so we use no reference energy. 

In the AFQMC run, the trained network returns the overlap, and automatic differentiation computes its force bias and local energy.
Costs are given in GPU hours on OLCF Frontier, where one GPU is one graphics compute die (GCD) of an AMD Instinct MI250X accelerator (four accelerators and eight GCDs per node).
We count only these trial-wavefunction evaluations, since walker propagation costs the same in both methods.
The shadow data comprise $N_\mathrm{s}$ snapshots, the number of circuits times the shots per circuit (Supplementary Table~\ref{table:hchain-cost}).

\begin{figure}[h]
    \centering
    \includegraphics[width=\linewidth]{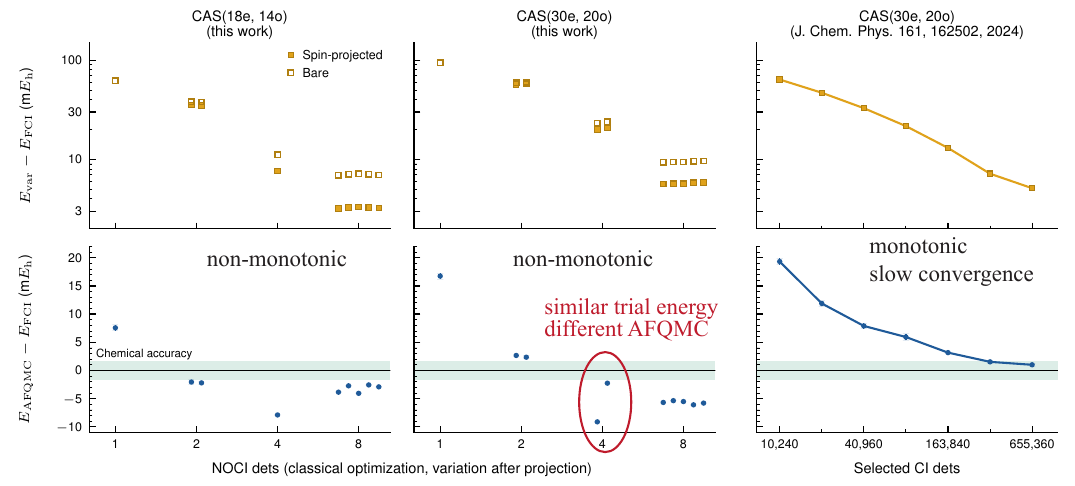}
    \caption{\textbf{AFQMC for \FeS{} with classical trial wavefunctions: energy-optimized NOCI trials and selected-CI trials.} 
    Top: variational energy of the trial. Bottom: AFQMC energy with that trial. Both are relative to FCI for the Hamiltonian of each column.
    Left and middle (this work): nonorthogonal CI trials with $K$ determinants, optimized classically by variation after spin projection (VAP), for (18e, 14o) and (30e, 20o). $K=1$ is the broken-symmetry UHF (BS-UHF) determinant. At each $K\ge2$, five optimizations were started from different random seeds; those converging to the same state (infidelity $\lesssim10^{-6}$) are shown as one point, and distinct states at the same $K$ are offset horizontally. Filled squares: spin-projected energy, the quantity VAP minimizes. Open squares: energy of the bare NOCI expansion.
    Right: selected-CI multi-Slater-determinant trials for (30e, 20o), 10,240 to 655,360 determinants, from previous work~\cite{jiang2024improved}.
    The green band marks chemical accuracy ($\pm1.6$~$\mHa$).
    }
    \label{fig:noci_fes}
\end{figure}

\subsection{Low-spin state of \FeS{}}
For [Fe$_2$S$_2$(SCH$_3$)$_4$]$^{2-}$ we use the geometry reported previously~\cite{sharma2014fes}. This 40-qubit Hamiltonian is reconstructed following the previously reported active-space procedure~\cite{li2017spin}, available at \url{https://github.com/jiangtong1000/fe2s2-hamiltonian-reconstruction-}. The 28-qubit Hamiltonian is the previously reported (18e, 14o) model~\cite{kjonstad2026ironafqmc}.
We optimized compact NOCI expansions by variation after spin projection (VAP), in which the orbitals and coefficients of $K$ determinants minimize the energy of the spin-projected state. For each $K\in\{2,4,8\}$ we ran five optimizations from different random seeds, grouped those that converged to the same state, and used each distinct state as the trial in phaseless AFQMC (\cref{fig:noci_fes}, left and middle).
Expansions are grown by doubling: each added determinant starts as a copy of an existing one rotated by small random angles (principal angles of 4--12$^\circ$) with zero coefficient, after which all orbitals and coefficients are re-optimized. For 20o, each seed is an independent lineage: starting from the spin-projected broken-symmetry UHF determinant ($K=1$), it is doubled to $K=2$, 4, and 8, each time from its own state at the previous $K$. For 14o, all five seeds at a given $K$ start from the same state: the broken-symmetry UHF determinant for $K=2$ and the seed-0 state of the previous $K$ for $K=4$ and 8, and differ only in the random rotations that initialize the added determinants. Each optimization used Adam followed by geometric direct minimization (GDM)~\cite{van2002geometric} with a line search to a gradient root mean square (RMS) below $10^{-6}$. We grouped seeds whose states coincide (infidelity below $10^{-6}$).
Although the variational energy decreases steadily with $K$, the AFQMC energy does not follow this trend. The error is neither monotonic in $K$ nor fixed by the variational energy; in particular, for (18e, 14o) at $K=8$ it ranges from $-4.05(6)$ to $-2.56(8)$~$\mHa$ over five states whose projected energies agree within 0.12~$\mHa$. For (30e, 20o) at $K=4$, two states 0.68~$\mHa$ apart in projected energy give $-9.11(14)$ and $-2.25(12)$~$\mHa$. None of these trials brings AFQMC within chemical accuracy. This shows the limitation of compact trials optimized classically for their energy: a lower variational energy, whether from more determinants or from a lower-lying state at the same $K$, does not reliably give a smaller AFQMC error, and among the classical trials considered here chemical accuracy is reached by selected-CI expansions of about $3\times10^5$ determinants or more (right column).

\begin{figure}
    \centering
    \includegraphics[width=\linewidth]{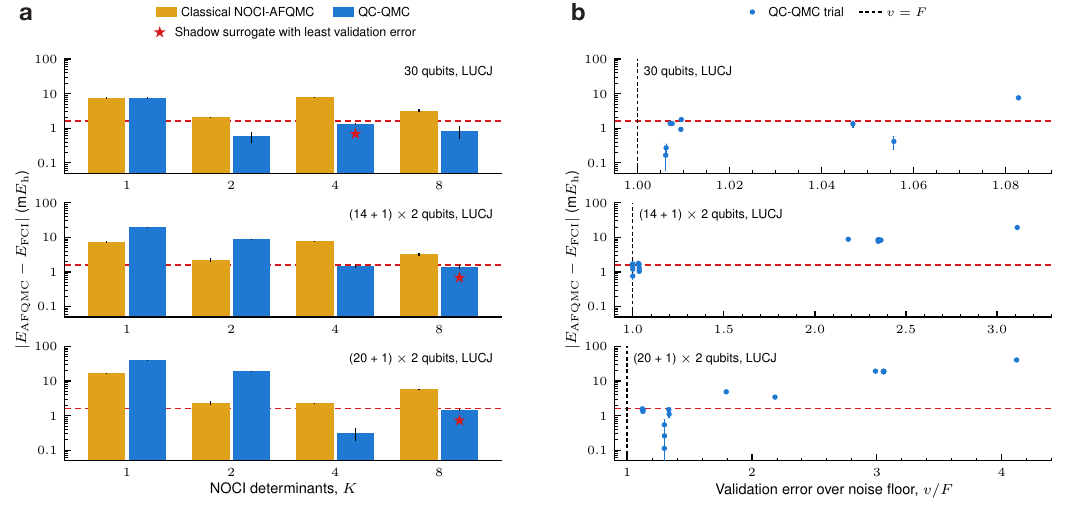}
    \caption{\textbf{QC-QMC for \FeS{} with surrogates trained on the LUCJ hardware data
    } 
    (a), Absolute AFQMC error relative to FCI, on a logarithmic axis, for the three LUCJ circuits.
    Classical AFQMC results are from classical VAP-NOCI trials.
    QC-QMC: AFQMC with the surrogate trained from that experimental data. Red dashed line: chemical accuracy (1.6~$\mHa$).
    (b), QC-QMC error of the shadow surrogates against their validation error.
    Each point is a QC-QMC surrogate, trained with different seeds or starting states.
    Horizontal axis: validation error $v$ (Eq.~\eqref{eq:v}) over the noise floor $F$ of the labels, with a separate range in each row; the dashed black line is $v=F$, below which the labels cannot show a surrogate to be closer to the circuit state.
    }\label{fig:si_afqmc_lucj_K}
\end{figure}

\Cref{fig:si_afqmc_lucj_K}a compares, for each LUCJ circuit and number of determinants $K$, AFQMC with classical VAP-NOCI trials and QC-QMC with surrogates trained on the hardware data; panel b relates the QC-QMC error of each surrogate to its validation error relative to the noise floor.

\subsection{High-spin state of \FeS{}}\label{app:fes-highspin}
\begin{figure}
    \centering
    \includegraphics[width=0.8\linewidth]{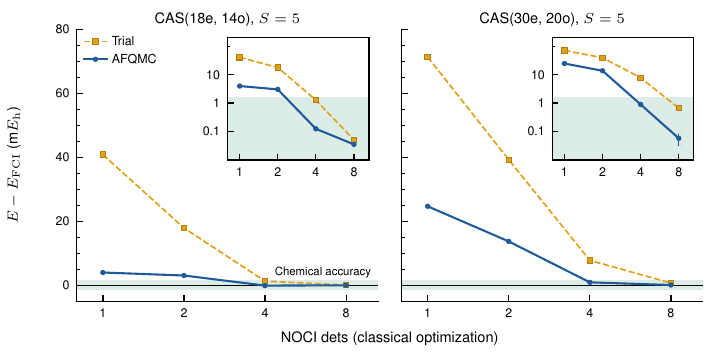}
    \caption{\textbf{QMC for \FeS{} in the high-spin ($S=5$) state with classical NOCI trials.} Trial (variational) energy (squares, dashed line) and AFQMC energy with that trial (circles, solid line) relative to FCI of the $S=5$ state, against the number of NOCI determinants $K$, for (18e, 14o) (left) and (30e, 20o) (right). Insets are on a logarithmic axis. $K=1$ is the UHF determinant; for $K\ge2$ the trials are optimized classically by minimizing the energy. The green band marks chemical accuracy ($\pm1.6$~$\mHa$; below 1.6~$\mHa$ in the insets).}
    \label{fig:noci_fes_s5}
\end{figure}

In contrast to the low-spin state where classical NOCI struggles, in the high-spin $S=5$ state (\cref{fig:noci_fes_s5}), we found that classical NOCI serves as a good QMC trial.
In the $M_S=5$ sector, both active spaces have at most ten open shells ($2\times14-18=2\times20-30=10$), the $\alpha$ shell is completely filled and a determinant with $M_S=5$ is already an $S=5$ eigenstate, so the problem reduces to the $\beta$ electrons alone (1{,}001 and 184{,}756 determinants for (18e, 14o) and (30e, 20o), versus $\sim\!4\times10^{6}$ and $\sim\!2\times10^{8}$ in the singlet sector).
The VAP-NOCI trials were grown by doubling $K$ from the UHF determinant. In contrast with the singlet, the trial energy is here a reliable guide to the AFQMC error. The variational error falls monotonically with $K$, and the AFQMC error follows it. With the UHF trial, AFQMC lies $+4.61(24)$ and $+24.68(38)$~$\mHa$ above FCI; from $K=4$ it is within chemical accuracy for both active spaces, $-0.131(12)$ and $+0.89(7)$~$\mHa$, and from $K=8$ within 0.06~$\mHa$ of FCI.
We can see this only because FCI is available for both active spaces. Without a reference, one could not tell in advance whether a classical trial is reliable for a given state, and the singlet shows that the variational energy alone can mislead (\cref{fig:noci_fes}). For the spin gaps (Fig.~\ref{fig:fes}d and Fig.~\ref{fig:fes_gap_40q}) we therefore treat the high-spin state exactly like the singlet, with a surrogate trained on measurements of the LUCJ circuit, and use no classical NOCI trial for the spin gap reported in the main text.
\Cref{tab:s5_14o} compares classical VAP-NOCI trials with NOCI trials learned from the hardware shadows of the (18e, 14o) high-spin circuit.

\begin{table*}[htbp]
\centering
\caption{\textbf{High-spin ($S=5$) \FeS{}, (18e, 14o): classical and quantum-learned NOCI trials.}
VAP: classical NOCI optimized on the Hamiltonian only. Quantum-learned: NOCI trained on hardware shadow labels, starting from the
VAP state of the same $K$, with labels from $3\times10^{4}$ random-basis circuits ($7.52\times10^{7}$ post-selected shots; noise
floor $F = 0.0116$) or $6\times10^{4}$ circuits ($1.49\times10^{8}$ shots; $F = 0.0080$). The VAP trials do not depend on the
labels; only their $v/F$ is evaluated against each label set.
$v/F$: validation error on held-out walkers divided by the noise floor of that label set.
$E_\mathrm{var}$: variational energy above FCI.
QMC: QMC energy minus FCI, one chain per distinct trial.
Quantum-learned values are means over four VAP starting seeds; their QMC uncertainty is the standard error over seeds, except at
$K=4$, where all seeds converge to one state.
Energies in mE$_\mathrm{h}$.}
\label{tab:s5_14o}
\setlength{\tabcolsep}{5pt}
\begin{tabular}{r rrrr rrr rrr}
\hline\hline
 & \multicolumn{4}{c}{VAP (classical NOCI)}
 & \multicolumn{3}{c}{Quantum-learned, $3\times10^{4}$ circuits}
 & \multicolumn{3}{c}{Quantum-learned, $6\times10^{4}$ circuits} \\
 & & & & & \multicolumn{3}{c}{($7.5\times10^{7}$ shots)} & \multicolumn{3}{c}{($1.5\times10^{8}$ shots)} \\
\cline{2-5}\cline{6-8}\cline{9-11}
$K$ & $v/F$ ($3{\times}10^{4}$) & $v/F$ ($6{\times}10^{4}$) & $E_\mathrm{var}$ & QMC
 & $v/F$ & $E_\mathrm{var}$ & QMC & $v/F$ & $E_\mathrm{var}$ & QMC \\
\hline
1 & 10.28 & 14.62 & $40.85$ & $+4.61(24)$ & 8.95 & $43.34$ & $+15.42(24)$ & --- & --- & --- \\
2 & 5.53 & 7.84 & $17.90$ & $+3.20(10)$ & 4.16 & $20.95$ & $+6.06(4)$ & 5.83 & $20.95$ & $+6.20(6)$ \\
4 & 1.90 & 2.70 & $1.28$ & $-0.131(12)$ & 1.00 & $2.55$ & $+0.226(22)$ & 1.20 & $3.24$ & $+0.331(22)$ \\
8 & 1.95 & 2.80 & $0.05$ & $-0.039(4)$ & 0.99 & $0.87$ & $+0.072(19)$ & 1.00 & $1.18$ & $+0.117(13)$ \\
\hline\hline
\end{tabular}
\end{table*}

% \clearpage
\section{Improved algorithms for matchgate shadow overlaps}
\label{app:overlap-algorithm}

Here we introduce our cubic-scaling algorithm for estimating walker overlaps from matchgate shadows, improving on a prior quartic-scaling algorithm~\cite{wan2023matchgate}. Let us first review the basic premise of classical shadows~\cite{huang2020predicting}. Given an $n$-qubit quantum state $\rho$ and a distribution of unitaries $\mathcal{D}$, a \emph{randomized measurement primitive} is the following procedure:
\begin{enumerate}
    \item Draw a random $U \sim \mathcal{D}$;
    \item Apply $\rho \mapsto U\rho U^\dagger$;
    \item Measure in the computational basis, obtaining an $n$-bit string $b$;
    \item Store an efficient classical description of the postmeasurement state $U^\dagger \ketbra{b}{b} U$.
\end{enumerate}
This defines a quantum channel
\begin{equation}
    \Mchan : \rho \mapsto \E_{U \sim \mathcal{D}} \sum_{b \in \{0, 1\}^n} \braket{b | U\rho U^\dagger | b} U^\dagger \ketbra{b}{b} U.
\end{equation}
Let $\Mchan^{-1}$ be the inverse over the range of $\Mchan$; for every operator $O \in \mathrm{ran}\,\Mchan$, by linearity and self-adjointness of $\Mchan$,
\begin{equation}
    \tr(O\rho) = \tr(\Mchan^{-1}(O) \Mchan(\rho)) = \E \, \braket{b | U \Mchan^{-1}(O) U^\dagger | b}
\end{equation}
where the expectation is over outcomes and random unitaries. Thus one has an unbiased estimate of the expectation value of $O$ with respect to the state $\rho$, provided a procedure to compute the quantity $\braket{b | U \Mchan^{-1}(O) U^\dagger | b}$.

\subsection{Problem statement and main theorem}
Let $a_j$ be the fermionic annihilation operators and define the Majorana operators
\begin{equation}
\gamma_{2j-1}=a_j+a_j^\dagger,
\qquad \gamma_{2j}=-i(a_j-a_j^\dagger).
\end{equation}
For an ordered set $S=\{s_1<\cdots<s_{2\ell}\}$, we define Hermitian Majorana monomials $\gamma_S=(-i)^\ell\gamma_{s_1}\cdots\gamma_{s_{2\ell}}$.
They form an orthogonal basis of the even operator space $\Gamma_\even = \spn\{\gamma_S : S \subseteq [2n],\ |S| \text{ even}\}$;~we write $\Pdeg_{2\ell}$ for the orthogonal projection onto this space.
A matchgate circuit (fermionic Gaussian unitary) $U_Q$ is defined by the adjoint action
\begin{equation}\label{eq:majorana_action}
U_Q^\dagger\gamma_\mu U_Q=\sum_\nu Q_{\mu\nu}\gamma_\nu,
\qquad Q\in\Orth(2n).
\end{equation}
The orthogonal group is defined as $\Orth(2n) = \{Q \in \R^{2n \times 2n} : QQ^\T = \I\}$. The signed permutation group $\B(2n) = \{\pm 1\}^{2n} \rtimes \Sym(2n)$, where $\Sym(2n)$ is the symmetric group, corresponds to matchgates that are also Clifford. The matchgate shadows protocol was originally defined for determinant-$1$ transformations~\cite{zhao2021fermionic}, and it was later shown that extending to either $\Orth(2n)$ or $\B(2n)$ yields the same channel and second-moment statistics~\cite{wan2023matchgate}. In fact, for Clifford matchgates, the distribution of signs is inconsequential for the measurement outcome statistics~\cite{ogorman2022fermionic,heyraud2025unified}.

\begin{proposition}\label{prop:shadow_channel}
The measurement channel for matchgate shadows over the Haar measure on $\Orth(2n)$ is
\begin{equation}\label{eq:shadow_channel}
\Mchan \coloneqq \E_{Q \sim \Orth(2n)}\l[ \mathcal{U}_Q^\dagger \Mchan_Z \mathcal{U}_Q \r] 
=\sum_{\ell=0}^n f_\ell\Pdeg_{2\ell},
\qquad f_\ell=\l.\binom {n}{\ell} \middle/ \binom{2n}{2\ell} \r.,
\end{equation}
where $\mathcal{U}_Q(\rho) = U_Q \rho U_Q^\dagger$ and $\Mchan_Z(\rho) = \sum_{b \in \{0,1\}^n} \braket{b | \rho | b} \ketbra{b}{b}$ denotes measurement in the computational basis. Its inverse over $\Gamma_\even$ is $\Mchan^{-1}=\sum_\ell f_\ell^{-1}\Pdeg_{2\ell}$.
This holds even when replacing $\Orth(2n)$ with any distribution on $\B(2n)$ such that the measure over the marginal of unsigned permutations is uniform.
\end{proposition}

This also holds for variances;~it is simply a consequence of the fact that $\Mchan_Z$ is invariant under conjugation by $\mathcal{U}_S$ for any $S \in \diag(\{\pm 1\}^{2n})$.

The covariance matrix of a state $\rho$ is
\begin{equation}\label{eq:covariance_convention}
(C_\rho)_{\mu\nu}=-\frac{i}{2}\tr(\rho[\gamma_\mu,\gamma_\nu]).
\end{equation}
In particular, for any $n$-bit string $b \in \{0,1\}^n$,
\begin{equation}\label{eq:basis_covariance}
C_{\ket b}=\bigoplus_{j=1}^n(-1)^{b_j}
\begin{pmatrix}0&1\\-1&0\end{pmatrix},
\qquad \text{and in particular } C_{\ket{\vac}}=J_n.
\end{equation}
Here and below $J_k$ denotes the symplectic form:~$k$ copies of this $2\times2$ block.
The covariance matrix of a Gaussian-rotated state $U_Q \rho U_Q^\dagger$ is $QC_{\rho}Q^\T$. In particular, for a matchgate shadow outcome $(Q,b)$ (a ``snapshot'') of $\rho$, the state before channel inversion is a Gaussian state $\varrho(Q, b) = U_Q^\dagger\ketbra{b}{b}U_Q$. Thus, its covariance matrix
\begin{equation}\label{eq:snapshot}
C_{\varrho(Q, b)}=Q^\T C_{\ket{b}}Q
\end{equation}
is a complete description of the snapshot.

The QC-QMC overlap with an arbitrary Slater determinant $\ket{\varphi}$ is given by the quantity
\begin{equation}\label{eq:overlap_estimator}
X_\varphi(Q,b)=2\tr\l(\ketbra{\varphi}{\vac} \Mchan^{-1}(U_Q^\dagger\ketbra{b}{b}U_Q)\r),
\end{equation}
where $(Q, b)$ is a snapshot of the state $\frac{\ket{\vac} + \ket{\Psi_\T}}{\sqrt{2}}$. By linearity, it obeys $\E[X_\varphi] = \braket{\Psi_T | \varphi}$.
The key insight of Wan \textit{et al.}~\cite{wan2023matchgate} is that \cref{eq:overlap_estimator} can be evaluated in polynomial time by classically processing the covariance matrix $C_{\varrho(Q, b)}$. 

Before recalling their result, we establish a few conventions. 
For even $\eta \leq n$, let
\begin{equation}\label{eq:W_matrix}
W=\bigoplus_{j=1}^\eta\frac1{\sqrt2}
\begin{pmatrix}1&-i\\1&i\end{pmatrix}\ \oplus\ \I_{2(n-\eta)},
\qquad
\Sbar_\eta=[2n]\setminus\{1,3,\ldots,2\eta-1\}.
\end{equation}
Let $2r = 2n - \eta$ be the size of $\Sbar_\eta$ and $\Pi_{\Sbar_\eta}$ be the $2r \times 2n$ matrix that selects the rows given by $\Sbar_\eta$. We parametrize Slater determinants $\ket{\varphi} \equiv \ket{\varphi(V)}$ via
\begin{equation}\label{eq:slater_convention}
\ket{\varphi(V)}=\widetilde{a}_1^\dagger\cdots \widetilde{a}_\eta^\dagger\ket{\vac}, \qquad \widetilde{a}_j^\dagger=\sum_{k=1}^n V_{jk}^* a_k^\dagger
\end{equation}
where the rows of $V$ are orthonormal. 
An orthogonal completion always exists for the unspecified $n-\eta$ rows, so we shall assume that $V \in \U(n)$.
We embed $\U(n) \hookrightarrow \SO(2n)$ as
\begin{equation}\label{eq:SO(2n)_embedding}
\Qtilde(V) = \begin{pmatrix}
    R_{11} & \cdots & R_{1n}\\
    \vdots & \ddots & \vdots\\
    R_{n1} & \cdots & R_{nn}
\end{pmatrix} \in \SO(2n),
\qquad R_{jk}=\begin{pmatrix}
\Re V_{jk}&-\Im V_{jk}\\
\Im V_{jk}&\Re V_{jk}
\end{pmatrix}.
\end{equation}
To any given shadow--walker pair $(C_\varrho, V)$, we define the following $2r \times 2r$ skew-symmetric matrices:
\begin{equation}\label{eq:BC_definitions}
\begin{split}
C&=L C_\varrho L^\T \qquad \text{where } L=\Pi_{\Sbar_\eta}W^*\Qtilde(V),
\\
B&=\begin{pmatrix}\zero_{\eta\times\eta}&0\\0&J_{n-\eta}\end{pmatrix}
\end{split}
\end{equation}
where $\zero$ denotes the zero matrix of a specified size. Finally, the Pfaffian of a skew-symmetric matrix $A \in \C^{2r \times 2r}$ is
\begin{equation}\label{eq:pfaffian_definition}
    \pf(A) = \frac{1}{2^r r!} \sum_{\sigma \in \Sym(2r)} (-1)^{\sigma} \prod_{j=1}^r A_{\sigma(2j-1), \sigma(2j)}
\end{equation}
where $\Sym(2r)$ is the symmetric group and $(-1)^{\sigma}$ is the parity of the permutation $\sigma$.

\begin{proposition}[{Wan \emph{et al.}~\cite[Theorem 3]{wan2023matchgate}}]\label{prop:overlap_polynomial}
Let $\ket{\varphi}$ be an $n$-mode, $\eta$-fermion Slater determinant with $\eta$ even. For the covariance matrix $C_\varrho$ of a matchgate shadow snapshot, define the polynomial
\begin{equation}\label{eq:overlap_polynomial}
q(z)=\frac{i^{\eta/2}}{2^r}\pf(B+zC)
  =\sum_{\ell=0}^r q_\ell z^\ell
\end{equation}
where $B$ and $C$ are defined as in \cref{eq:BC_definitions}.
Then the estimator from \cref{eq:overlap_estimator} obeys
\begin{equation}\label{eq:coefficient_contraction}
X_\varphi=2\sum_{\ell=0}^r f_\ell^{-1}q_\ell,
\end{equation}
and the coefficients $(q_0, q_1, \ldots, q_r)$ can be computed with $\calO(r^4)$ arithmetic operations.
\end{proposition}

Wan \textit{et al.}~\cite{wan2023matchgate} also give strong numerical evidence that the variance of $X_\varphi$ is asymptotically bounded by the variance of $X_{\ket{\vac}}$, the latter of which they rigorously bound as $\V[X_{\ket{\vac}}] = \calO(\sqrt{n} \log n)$. 
While their algorithm is formally efficient, it involves degree-$r$ polynomial interpolation in the monomial basis, which is numerically unstable and requires $\calO(r)$-bit precision to stay below machine precision~\cite{kiser2024cost}. 
Furthermore, their algorithm scales quartically in the linear dimension of matrices $B$ and $C$, where $2r \leq 2n$. On the other hand, the Pfaffian can be evaluated (with numerical stability) in $\calO(r^3)$ arithmetic operations~\cite{wimmer2012algorithm,rubow2011factorization}, hinting at the possibility of a cubic-time algorithm for the overlap estimator. 

We introduce two such algorithms here. The first is enabled by applying a complex skew-symmetric variant of the Paige--Van Loan
reduction~\cite{paige1981schur,van1984symplectic} to the matrix pencil $(B, C)$. Along with a Schur complement transformation, this reduces the Pfaffian to a determinant using a formula for certain structured block matrices~\cite{okada2019pfaffian}. We then compute the polynomial coefficients $q_\ell$ quickly using recurrence relations for determinants of Hessenberg matrices~\cite{wilkinson1965algebraic}. 
We call this method the \textit{PVL method}. However, this method relies on the invertibility of a certain block of $C$, which does not hold for all instances.\footnote{Empirically, we have observed that such instances occur with a frequency of less than $10^{-6}$ in actual QC-QMC runs.} Thus we include a second method, based on the QZ algorithm for solving generalized eigenvalue problems~\cite{moler1973algorithm,golub2013matrix}. This version applies to every $C$, but features a slightly larger prefactor in the $r^3$ scaling. Appropriately, we refer to it as the \textit{GEV method}. As such, we record the following result.

\begin{theorem}[Cubic-time overlap computation]\label{thm:cubic_time_overlaps}
    For any skew-symmetric $B, C \in \C^{2r \times 2r}$, the polynomial coefficients of $\pf(B+zC)$ can be computed in $\calO(r^3)$ arithmetic operations.
\end{theorem}

\paragraph{Note on concurrent work.} Independently of our results, Nappi \textit{et al.}~\cite{nappi2026practicability} introduced an $\calO(r^3)$ method also based on the Schur complement;~as such, it also suffers the same invertibility issue as ours. Their algorithm does not perform a Hessenberg reduction; instead, it adapts the derivative-based approach in Appendix D~\cite{wan2023matchgate}. 
To our knowledge, our GEV method is the first algorithm to solve the problem in full generality in cubic time.

\subsection{The PVL method}

The key result and algorithm describing the PVL method are as follows. We always assume that the matrices $B, C$ (and hence the polynomial $q(z)$) are defined with respect to a given shadow--walker pair, and that the particle number $\eta$ is even.

\begin{theorem}[PVL algorithm]\label{thm:block_overlap}
    Given a covariance matrix $C_\varrho$ and Slater determinant $\ket{\varphi(V)}$, suppose the leading $\eta\times\eta$ block $C_{11}$ of
    \begin{equation}
        C = \begin{pmatrix}
            C_{11} & C_{12}\\
            -C_{12}^\T & C_{22}
        \end{pmatrix}
    \end{equation}
    is invertible. Then
    Algorithm~\ref{alg:block_overlap} returns the $r+1$ complex coefficients of the overlap polynomial $q(z)$ in $\calO(r^3)$ arithmetic operations and $\calO(r^2)$ storage, where $r=n-\eta/2$. Specifically, it constructs an upper Hessenberg matrix $H\in\C^{s\times s}$,
    $s=n-\eta$, such that
    \begin{equation}\label{eq:block_headline}
        q(z)=\frac{i^{\eta/2}\pf(C_{11})}{2^r} z^{\eta/2}\det(\I_s-zH).
    \end{equation}
    The coefficients below degree $\eta/2$ are necessarily zero, and if we write $\det(\I_s-zH)=\sum_{j=0}^s d_jz^j$, then the remaining coefficients are $q_{\eta/2+j}=i^{\eta/2}2^{-r}\pf(C_{11})d_j$.
\end{theorem}

\begin{algorithm}[H]
    \caption{Overlap polynomial coefficients (PVL method)}\label{alg:block_overlap}
    \setcounter{AlgoLine}{0}
    \KwInput{A covariance matrix $C_\varrho \in \R^{2n \times 2n}$ and a Slater matrix
    $V \in \U(n)$, with even integer $0 \leq \eta \leq n$.}
    \KwPromise{$C_{11}$ is invertible.}
    \KwOutput{The coefficients $(q_0,q_1,\ldots,q_r) \in \C^{r+1}$ of $q(z)$.}
    
    \BlankLine
    
    Construct $C \gets (C_\varrho, V)$ and $B$ as in \cref{eq:BC_definitions};
    
    Permute rows and columns, $C'=RCR^\T$ as in \cref{eq:block_grouped_pencil};
    
    Linear solve for $X \gets C_{11}^{-1}C_{12}'$;
    
    Let $M\gets C_{22}'+C_{12}^{\prime\T} X$;
    
    $H\gets$ symplectic Hessenberg reduction of $M$; \hfill \Comment{Algorithm~\ref{alg:block_symplectic_reduction}}
    
    $(d_0,d_1,\ldots,d_s)\gets$ coefficients of $\det(\I_s-zH)$; \hfill \Comment{Algorithm~\ref{alg:block_coefficients}}

    Let $\alpha\gets i^{\eta/2}2^{-r}\pf(C_{11})$;
    
    Set $q_0,q_1,\ldots,q_r\gets0$;
    
    \For{$j=0,1,\ldots,s$}{
        $q_{\eta/2+j}\gets\alpha d_j$;
    }
    \Return{$(q_0,q_1,\ldots,q_r)$}
\end{algorithm}

The rest of the section proves this result.
Throughout, we set
\begin{equation}
    r=n-\eta/2,\qquad s=n-\eta,\qquad
    \kappa=\frac{i^{\eta/2}}{2^r}.
\end{equation}
Let $P_s$ be the $2s \times 2s$ permutation matrix satisfying
\begin{equation}
    P_s\begin{pmatrix}x_1\\y_1\\\vdots\\x_s\\y_s\end{pmatrix}
    =\begin{pmatrix}x_1\\\vdots\\x_s\\y_1\\\vdots\\y_s\end{pmatrix}.
\end{equation}
Its determinant is $\det(P_s)=\epsilon_s=(-1)^{s(s-1)/2}$. The $2r \times 2r$ permutation matrix $R$ is then defined as $R=\I_\eta\oplus P_s$, and for the pair $(B, C)$ we set
\begin{equation}\label{eq:block_grouped_pencil}
\begin{split}
    B' &= RBR^{\T} = \begin{pmatrix}\zero_{\eta\times\eta}&0\\0&J'_{s}\end{pmatrix}, \qquad J_s'=\begin{pmatrix}0&\I_s\\-\I_s&0\end{pmatrix},\\
    C' &= RCR^{\T}=
    \begin{pmatrix}C_{11}&C_{12}'\\-C_{12}^{\prime\T}&C_{22}'\end{pmatrix}.
\end{split}
\end{equation}
The blocks of $C'$ conform with the structure of $B'$, i.e., $C_{11} \in \C^{\eta\times\eta}$ and $C_{22}' \in \C^{2s \times 2s}$. Note that $R$ leaves the leading $\eta \times \eta$ block unchanged. The following lemma reduces the overlap polynomial to a \textit{Pfaffian variant} of the characteristic polynomial~\cite{baer2021faddeev}. A subsequent reduction that we call \textit{symplectic Hessenberg} further reduces this Pfaffian to an ordinary determinant of half the linear matrix dimension.

\begin{lemma}\label{lem:block_schur}
    Suppose $C_{11}$ is invertible, and define the skew Schur complement of $C'$:
    \begin{equation}\label{eq:block_schur_complement}
        M=C_{22}'+C_{12}^{\prime\T}C_{11}^{-1}C_{12}'.
    \end{equation}
    Then
    \begin{equation}\label{eq:block_pfaffian_schur}
        q(z)=\kappa\epsilon_s z^{\eta/2}\pf(C_{11})\pf(J_s'+zM)
    \end{equation}
    identically as a polynomial over $z\in\C$.
\end{lemma}

\begin{proof}
    Set
    \begin{equation}
        T=\begin{pmatrix}
            \I_\eta&0\\ C_{12}^{\T}C_{11}^{-1}&\I_{2s}
        \end{pmatrix}.
    \end{equation}
    Direct block-matrix multiplication reveals that
    \begin{equation}
        T (B' + z C') T^\T = T\begin{pmatrix}
            zC_{11}&zC_{12}'\\-zC_{12}^{\prime\T}&J_s'+zC_{22}'
        \end{pmatrix}T^{\T} = \begin{pmatrix}
            zC_{11} & 0\\
            0 & J_s'+zM
        \end{pmatrix}.
    \end{equation}
    The matrix $T$ has determinant $1$ since it is lower triangular with identities along the diagonal. Using the identity $\pf(QAQ^{\T})=\det(Q)\pf(A)$ and the fact that the Pfaffian of a block-diagonal matrix factorizes, we obtain
    \begin{equation}
    \begin{split}
        \pf(B+zC) &= \pf\l[ R^\T (B'+zC') R \r]\\
        &= \pf\l[ (TR)^\T \begin{pmatrix}
            zC_{11} & 0\\
            0 & J_s'+zM
        \end{pmatrix} TR \r]\\
        &=\epsilon_s z^{\eta/2}\pf(C_{11})\pf(J_s'+zM).
    \end{split}
    \end{equation}
    Multiplying by $\kappa$ proves the claim.
\end{proof}

Note that we can solve the linear equation $C_{11}X=C_{12}$ to form $M=C_{22}+C_{12}^{\T}X$, rather than explicitly constructing a matrix inverse. The sign $\epsilon_s$ in \cref{eq:block_pfaffian_schur} is essential;~in the following reduction, a second copy of the same sign will appear, thereby canceling the two.

\paragraph{Symplectic Hessenberg reduction.}
The following reduction is the main matrix subroutine to bring $\pf(J_s' + zM)$ to a convenient form. It is essentially equivalent to the Paige--Van Loan reduction for Hamiltonian and skew-Hamiltonian matrices~\cite{paige1981schur,van1984symplectic}, up to an appropriate transformation. We give a specialized treatment for our context and provide an explicit, self-contained description of the algorithm involving complex symplectic Givens rotations.

Recall that a matrix is said to be (upper) Hessenberg if it has all zero entries below the first subdiagonal.

\begin{theorem}[Symplectic Hessenberg reduction]\label{thm:block_symplectic_reduction}
    For every complex skew-symmetric $M\in\C^{2s\times2s}$ there is a product $U$ of symplectic Givens rotations (\cref{def:symplectic_givens_rotations}) such that
    \begin{equation}\label{eq:block_reduced_matrix}
        UMU^{\T}=\begin{pmatrix}0&-H^{\T}\\H&Z\end{pmatrix},
        \qquad H\in\C^{s\times s}\text{ is upper Hessenberg},\quad Z=-Z^{\T}.
    \end{equation}
    Moreover, $UJ_s'U^{\T}=J_s'$ (hence $\det(U)=1$). The reduction uses $\calO(s^3)$ arithmetic operations and $\calO(s^2)$ storage.
\end{theorem}

The procedure to compute $U$ is given in Algorithm~\ref{alg:block_symplectic_reduction}.

\begin{algorithm}[H]
    \caption{Symplectic Hessenberg (PVL) reduction}\label{alg:block_symplectic_reduction}
    \setcounter{AlgoLine}{0}
    \KwInput{A complex skew-symmetric matrix $M\in\C^{2s\times2s}$.}
    \KwOutput{An upper Hessenberg matrix $H\in\C^{s\times s}$ satisfying \cref{eq:block_reduced_matrix}.}
    
    \BlankLine
    
    $A\gets M$;
    
    \For{$j=1,2,\ldots,s-1$}{
        \For{$i=s,s-1,\ldots,j+1$}{
            Choose $G$ sending $\begin{pmatrix}A_{ij}\\A_{s+i,j}\end{pmatrix}$ to $\begin{pmatrix}0\\v\end{pmatrix}$; \hfill \Comment{$|v| = \sqrt{|A_{ij}|^2 + |A_{s+i,j}|^2}$}
            
            $A\gets G_{\inter}^{(i)} A G_{\inter}^{(i),\T}$;
        }
        \For{$i=s,s-1,\ldots,j+2$}{
            Choose $G$ so that $G^*$ sends $\begin{pmatrix}A_{s+i-1,j}\\A_{s+i,j}\end{pmatrix}$ to $\begin{pmatrix}v\\0\end{pmatrix}$; \hfill \Comment{$|v| = \sqrt{|A_{s+i-1,j}|^2 + |A_{s+i,j}|^2}$}
            
            $A\gets G_{\intra}^{(i-1,i)} A G_{\intra}^{(i-1,i),\T}$;
        }
    }
    \Return{$H\gets A_{s+1:2s,\,1:s}$}
\end{algorithm}

We use symplectic unitary transformations $U$, i.e., satisfying $UJ_s'U^{\T}=J_s'$ and $UU^\dagger = \I$. Note the distinct uses of transpose and conjugate transpose:~unitary congruence $A \to UAU^\T$ is essential to preserve skew symmetry, rather than conjugation. 
Our elementary rotations $G$ are defined as follows.

\begin{definition}[Symplectic Givens rotations]\label{def:symplectic_givens_rotations}
    Consider complex Givens rotations of the form
    \begin{equation}\label{eq:block_complex_givens}
        G=\begin{pmatrix}c&u\\-u^*&c\end{pmatrix},\qquad
        c\in\R,\quad c^2+|u|^2=1.
    \end{equation}
    Given such a $G$, we define two ways to embed it into the $2s$ coordinates:
    \begin{itemize}
        \item An \emph{inter-block rotation} $G_{\inter}^{(i)}$ applies $G$ on coordinates $(i,s+i)$.
        \item An \emph{intra-block rotation} $G_{\intra}^{(i,j)}$ applies $G$ on coordinates $(i,j)$ and $G^*$ on coordinates $(s+i,s+j)$.
    \end{itemize}
\end{definition}

The transformations $G_{\inter}^{(i)}$ and $G_{\intra}^{(i,j)}$ are unitary and symplectic with respect to $J_s'$. 
Given $a,b\in\C$, the Givens rotation is chosen to map $(a,b)^{\T}$ either to $(0,v)^{\T}$ or to $(v,0)^{\T}$, where $|v|^2=|a|^2+|b|^2$. For example, writing $w=e^{i(\arg a-\arg b)}$ and $d=\sqrt{|a|^2+|b|^2}$, the first choice has $c=|b|/d$, $u=-w|a|/d$, while the second has $c=|a|/d$, $u=w|b|/d$. If both inputs vanish, we take $G=\I_2$ with an arbitrary phase choice.

\begin{proof}[Proof (of \cref{thm:block_symplectic_reduction})]
    We prove the zero pattern by induction on $j$. Before processing column $j$, every earlier column $k<j$ has a zero upper half, and its lower half vanishes at rows $s+i$ with $i>k+1$. This is vacuous at $j=1$.

    In the first loop, we have $i>j$. At a previous column $k<j$, both entries mixed by $G_{\inter}^{(i)}$ vanish:~the upper entry is in a previously zeroed column of the upper-left block, and the lower entry has $i>k+1$. Thus the left multiplication preserves these entries. Right multiplication only changes columns $i$ and $s+i$, so it leaves columns $1,\ldots,j$ unchanged. Consequently, we may zero out all entries $A_{ij}$ for $i>j$ without disturbing the completed columns or earlier zeros in the current column.

    In the second loop, the affected upper indices $(i-1,i)$ are both greater than $j$. Thus, they contain zeros in column $j$ after the first loop. At previous columns $k<j$, their lower partners also vanish because $i-1\geq j+1>k+1$. The rotations preserve all of these zeros. Right multiplication again leaves columns $1,\ldots,j$ unchanged. Processing the lower entries from bottom to top leaves only rows $s+1,\ldots,s+j+1$ possibly nonzero in the lower half of column $j$.

    By skew symmetry, all the corresponding rows are also zeroed out, and the diagonal of $M$ is automatically zero. As such, after column $s-1$, the remaining upper-left entry is diagonal, hence zero automatically. The lower-left block has precisely the upper Hessenberg pattern, while the lower-right block remains generally dense and skew-symmetric. Every rotation preserves $J_s'$ and has determinant one, so the accumulated transformation has the claimed properties.
    
    To bound the complexity, observe that there are $s(s-1)/2$ inter-block rotations and $(s-1)(s-2)/2$ intra-block rotations, and each congruence transformation updates a constant number of rows and columns in $\calO(s)$ time.
\end{proof}

\begin{corollary}\label{cor:block_pfaffian_determinant}
    Let $H$ be the output of Algorithm~\ref{alg:block_symplectic_reduction}. Then
    \begin{equation}\label{eq:block_residual_determinant}
        \pf(J_s'+zM)=\epsilon_s\det(\I_s-zH),
    \end{equation}
    and hence
    \begin{equation}\label{eq:block_final_polynomial}
        q(z)=\kappa\pf(C_{11}) z^{\eta/2}\det(\I_s-zH).
    \end{equation}
\end{corollary}

\begin{proof}
    Since $\det(U)=1$, congruence by $U$ gives
    \begin{equation}
        \pf(J_s'+zM)=\pf\mathopen{}\begin{pmatrix}
            0&\I_s-zH^{\T}\\zH-\I_s&zZ
        \end{pmatrix}.
    \end{equation}
    For any square $A$ and skew-symmetric $Y$ of the same size, we use the following identity for Pfaffians of block matrices (Corollary 2.4)~\cite{okada2019pfaffian}:
    \begin{equation}
        \pf\mathopen{}\begin{pmatrix}0&A\\-A^{\T}&Y\end{pmatrix}
        =(-1)^{s(s-1)/2}\det(A).
    \end{equation}
    Applying this with $A=\I_s-zH^{\T}$ proves \cref{eq:block_residual_determinant}. Substitution into \cref{eq:block_pfaffian_schur} cancels the sign $\epsilon_s$, proving \cref{eq:block_final_polynomial}.
\end{proof}

\paragraph{Recurrence relation for Hessenberg determinants.}
Given the upper Hessenberg block $H$, the next subroutine returns all $s+1$ coefficients of $\det(\I_s-zH)$ in cubic time. We use a classic recurrence relation for Hessenberg matrices; see, for instance, Chapter 6, Eq.~(57.1) of Wilkinson~\cite{wilkinson1965algebraic}. Computing the determinant of an $R$-valued Hessenberg matrix in this way costs $\calO(s^2)$ $R$-operations. In our case, $R = \C[z]$ of degree at most $s$, which takes $\calO(s)$ arithmetic operations per entry to manipulate.

\begin{proposition}[{Wilkinson~\cite[Chapter 6, Section 57]{wilkinson1965algebraic}}]\label{prop:block_hessenberg_recurrence}
    Let $H\in\C^{s\times s}$ be upper Hessenberg, and let
    \begin{equation}
        p_0(t)=1,\qquad p_k(t)=\det(t\I_k-H_{1:k,1:k}) \quad \text{for } 1 \leq k \leq s.
    \end{equation}
    For $k=1,\ldots,s$, we have the recurrence relation
    \begin{equation}\label{eq:block_hessenberg_recurrence}
        p_k(t)=(t-H_{kk})p_{k-1}(t)
        -\sum_{i=1}^{k-1}H_{ik}
        \l(\prod_{j=i}^{k-1}H_{j+1,j}\r)p_{i-1}(t).
    \end{equation}
\end{proposition}

\begin{corollary}\label{cor:hessenberg_recurrence_polynomial}
    If we write $p_s(t)=\sum_{j=0}^s a_j t^j$, then
    \begin{equation}\label{eq:block_coefficient_reversal}
        \det(\I_s-zH)=\sum_{\ell=0}^s a_{s-\ell}z^\ell.
    \end{equation}
    The coefficients $a_j$ can be computed using $\calO(s^3)$ arithmetic operations without divisions, and $\calO(s^2)$ storage.
\end{corollary}

\begin{proof}
    For $z\neq0$, we have $\det(\I_s-zH)=z^sp_s(z^{-1})$. Reversing the coefficients proves \cref{eq:block_coefficient_reversal}, including at $z=0$ by equality of polynomials. To bound the complexity, observe that to compute the recurrence, we first tabulate the products of consecutive subdiagonal entries in $\calO(s^2)$ time. Each step combines at most $k$ coefficient arrays of length at most $k+1$, so its cost is $\calO(k^2)$. Summing over $k$ steps gives $\calO(s^3)$ time. The arrays for $p_0,p_1,\ldots,p_s \in \C[t]$ and the subdiagonal products occupy $\calO(s^2)$ total storage. Finally, by \cref{eq:block_coefficient_reversal} we do not need any divisions.
\end{proof}

This subroutine is summarized in Algorithm~\ref{alg:block_coefficients}.

\begin{algorithm}[H]
    \caption{Coefficients of $\det(\I_s-zH)$}\label{alg:block_coefficients}
    \setcounter{AlgoLine}{0}
    \KwInput{An upper Hessenberg matrix $H\in\C^{s\times s}$.}
    \KwOutput{Complex coefficients $d_0,d_1,\ldots,d_s$ such that $\det(\I_s-zH)=\sum_{\ell=0}^s d_\ell z^\ell$.}
    
    \BlankLine
    
    $p_0(t)\gets1$; \hfill \Comment{Polynomials stored as length-$(s+1)$ arrays}
    
    Precompute $b_{ik}\gets\prod_{j=i}^{k-1}h_{j+1,j}$ for $1\leq i<k\leq s$;
    
    \For{$k=1,2,\ldots,s$}{
        $p_k(t)\gets(t-h_{kk})p_{k-1}(t)$;
        
        \For{$i=1,2,\ldots,k-1$}{
            $p_k(t)\gets p_k(t)-h_{ik}b_{ik}p_{i-1}(t)$;
        }
    }
    Write $p_s(t)=a_0+a_1t+\cdots+a_st^s$;
    
    \Return{$(d_0,\ldots,d_s)\gets(a_s,a_{s-1},\ldots,a_0)$}
\end{algorithm}

Putting these pieces together, we now prove the main result of the PVL method.

\begin{proof}[Proof (of \cref{thm:block_overlap})]
    Correctness is \cref{eq:block_final_polynomial} followed by \cref{eq:block_coefficient_reversal} and placement of the coefficients starting at degree $\eta/2$. Let us verify the complexity. Evaluating the Pfaffian of $C_{11}$ costs $\calO(\eta^3)$ arithmetic operations, and solving $C_{11} X = C_{12}'$ where $C_{12}' \in \C^{\eta \times 2s}$ costs $\calO(\eta^2s)$. Forming $C_{12}^{\prime\T}X$ costs $\calO(\eta s^2)$, and by \cref{thm:block_symplectic_reduction,cor:hessenberg_recurrence_polynomial}, the symplectic Hessenberg reduction and the recurrence computation each cost $\calO(s^3)$;~placing the resulting coefficients in the output vector costs $\calO(s)$. Finally, all matrices and coefficient arrays have at most quadratic size.
\end{proof}

\begin{remark}[Generality of the PVL method]
    The approach presented above specializes to finding the polynomial $\pf(B + zC)$ where $B$ has the particular structure $B = \zero_{\eta\times\eta} \oplus J_{n-\eta}$. This applies more broadly:~suppose instead that $B$ is an arbitrary skew-symmetric matrix of rank $2r-\eta$. Then there always exists an invertible matrix $U$ such that $U^\T B U = \zero_{\eta\times\eta} \oplus J_{n-\eta}$~\cite{horn2012matrix}. The PVL algorithm can then be applied if the leading $\eta \times \eta$ block of $U^\T C U$ is invertible.
\end{remark}

\subsection{The GEV method}
The PVL method crucially requires that $C_{11}$ be invertible. Here we describe an $\calO(r^3)$-time algorithm which applies to \textit{any} $(B, C)$, based on a reduction to a generalized eigenvalue problem (which in turn is solved by the QZ algorithm of Moler and Stewart)~\cite{moler1973algorithm}.

\begin{theorem}[GEV algorithm]\label{thm:gev_coefficients}
    Given a covariance matrix $C_\varrho\in\R^{2n\times2n}$ and a Slater determinant $\ket{\varphi(V)}$, Algorithm~\ref{alg:gev_coefficients} returns the $r+1$ complex coefficients of the overlap polynomial $q(z)$ in $\calO(r^3)$ arithmetic operations and $\calO(r^2)$ storage, where $r = n-\eta/2$. Specifically, it constructs upper triangular matrices $T,S \in \C^{2r\times 2r}$ and computes a scalar $\alpha \in \C$ such that
    \begin{equation}\label{eq:gev_polynomial_thm}
        q(z) = \alpha \prod_{\mu=1}^r (t_\mu - zs_\mu),
    \end{equation}
    where $\{(t_\mu, s_\mu) : \mu \in [r]\}$ are the pairs of diagonal entries $(T_{jj}, S_{jj})$ at half multiplicity.
\end{theorem}

\begin{algorithm}[H]
    \caption{Overlap polynomial coefficients (GEV method)}\label{alg:gev_coefficients}
    \setcounter{AlgoLine}{0}
    \KwInput{A covariance matrix $C_\varrho$ and a Slater matrix
    $V\in\U(n)$, with even integer $0 \leq \eta \leq n$.}
    \KwOutput{The coefficients $(q_0, q_1, \ldots, q_r) \in \C^{r+1}$ of $q(z)$.}
    
    \BlankLine
    
    Construct $C \gets (C_\varrho, V)$ and $B$ as in \cref{eq:BC_definitions};
    
    $(S,T) \gets $ generalized Schur decomposition; \hfill \Comment{\cref{prop:qz_algorithm}}
    
    \BlankLine
    
    \If{some diagonal pair satisfies $T_{jj}=S_{jj}=0$}{
        \Return{$(0,\ldots,0)$}
    }
    
    \BlankLine
    
    Form the finite roots $T_{jj}/S_{jj}$ for $S_{jj}\neq0$;
    
    Take half of each root's multiplicity, obtaining
    $\lambda_1,\ldots,\lambda_d$;
    
    Compute $g(z)\gets\prod_{j=1}^d(z-\lambda_j)$;
    
    Choose any $z_0\in\C\setminus\{\lambda_1,\ldots,\lambda_d\}$; \hfill \Comment{See \cref{rem:z0_heuristic} for a practical heuristic}
    
    $\alpha\gets i^{\eta/2}2^{-r}\pf(B+z_0C)/g(z_0)$;
    
    \Return{the coefficients of $\alpha g(z)$, padded with zeros
    as necessary}
\end{algorithm}

The GEV method is based on two key observations:~first, that $\pf(A)^2 = \det(A)$ allows us to reduce a Pfaffian to a determinant up to a sign;~and second, that $B + zC$ is a \textit{linear matrix pencil} whose characteristic polynomial $\det(B + zC)$ can be determined from the generalized eigenvalue equation $Bv = -\lambda Cv$.

\paragraph{The QZ algorithm.}
First we review some basic theory of matrix pencils. Let $A, B$ be square matrices. A matrix pencil is the ray $A - \lambda B$, where $\lambda \in \C$. We also write a pencil as the pair $(A,B)$. The characteristic polynomial of $(A,B)$ is
\begin{equation}
p(\lambda)=\det(A - \lambda B).
\end{equation}
The pencil is called \emph{regular} if $p \in \C[\lambda]$ is not the zero polynomial, and
\emph{singular} otherwise.

The generalized eigenvalues of a matrix pencil are given by the set
\begin{equation}
    \lambda(A, B) = \{z \in \C : p(z) = 0\}.
\end{equation}
The pencil $(A,B)$ is singular if and only if $\lambda(A,B)=\C$. Otherwise, $|\lambda(A,B)|\leq\rank(B)$. The generalized eigenvalues can be determined from the generalized Schur decomposition of the pencil~\cite{moler1973algorithm,golub2013matrix}.

\begin{proposition}[{Generalized Schur decomposition~\cite[Theorem 7.7.1]{golub2013matrix}}]
    For any matrix pencil $(A, B)$, there exist unitary matrices $Q, Z$ such that
    \begin{equation}\label{eq:generalized_schur}
        A = QTZ^\dagger, \qquad B = QSZ^\dagger
    \end{equation}
    for upper triangular $T,S$. If there exists some $j$ such that $T_{jj}=S_{jj}=0$, then $\lambda(A,B)=\C$. Otherwise,
    \begin{equation}
        \lambda(A,B) = \{T_{jj}/S_{jj} : S_{jj} \neq 0\}.
    \end{equation}
\end{proposition}

There are also ``infinite eigenvalues'' of $(A,B)$, corresponding to when $S_{jj}=0$ but $T_{jj}\neq 0$. These merely contribute to the characteristic polynomial as an overall scale factor, as can be seen from
\begin{equation}\label{eq:polynomial_QZ_expanded}
    p(\lambda) = \det(QZ^\dagger) \prod_{j} (T_{jj} - \lambda S_{jj}).
\end{equation}

The standard procedure to bring a pencil into generalized Schur form is the QZ algorithm.

\begin{proposition}[QZ algorithm~\cite{moler1973algorithm,golub2013matrix}]\label{prop:qz_algorithm}
    The generalized Schur decomposition of an $N \times N$ matrix pencil can be determined in $\calO(N^3)$ arithmetic operations.
\end{proposition}

\paragraph{Reconstructing coefficients from roots.}
Now let $B, C$ be the skew-symmetric overlap matrices from \cref{eq:BC_definitions}. The overall strategy of Algorithm~\ref{alg:gev_coefficients} is to compute the generalized eigenvalues $\lambda(B, -C)$. This is the set of roots of the polynomial $\det(B + zC)$, which is the square of the desired polynomial $\pf(B + zC)$. A polynomial is equivalently specified by either its coefficients or roots, up to a nonzero scalar factor. Thus, with some clean-up (whose cost is subleading compared to the cost of the QZ algorithm), we can recover $q_\ell$ from the roots.

\begin{lemma}\label{lem:pencil_roots}
Suppose the pencil $(B, -C)$ is regular, and that $q(z)$ has degree $d\leq r$.
Its characteristic polynomial has $2d$ finite roots, counted with
multiplicity, each with even multiplicity. Take half the multiplicity
of each finite root to form a list $\lambda_1,\ldots,\lambda_{d}$.
For any $z_0$ with $p(z_0)\neq0$,
\begin{equation}\label{eq:pencil_reconstruction}
q(z)=\frac{i^{\eta/2}}{2^r}\pf(B+z_0C)\prod_{j=1}^{d}\frac{z-\lambda_j}{z_0-\lambda_j}.
\end{equation}
The remaining $2r-2d$ roots are at infinity.
\end{lemma}

\begin{proof}
The identity $\pf(A)^2=\det(A)$ implies that every root of $\det(B+zC)$ has even multiplicity. Hence $\lambda(B, -C) = \{T_{jj}/S_{jj} : S_{jj} \neq 0\}$ inherits this multiplicity, where $T, S$ are the upper triangular matrices of the generalized Schur decomposition.
As in \cref{eq:polynomial_QZ_expanded}, we can write
\begin{equation}
    \det(B+zC) = \alpha \prod_{j=1}^{2d} (T_{jj} - zS_{jj}) = \alpha \prod_{i=1}^{2d} S_{ii} \prod_{j=1}^{2d} \l( \frac{T_{jj}}{S_{jj}} - z \r) = \alpha' \prod_{j=1}^d (z - \lambda_j)^2,
\end{equation}
where we have absorbed all scalar factors into $\alpha'$. Hence $\pf(B+zC) = \alpha'' \prod_{j=1}^d (z-\lambda_j)$ for another scalar $\alpha''$. To determine that scalar, we simply evaluate the polynomial at some non-root $z_0$.
\end{proof}

Note that because $(B,-C)$ is a regular pencil, there are only finitely many roots and so almost all $z_0 \in \C$ are not roots. Thus one could randomly sample in a region, for example the unit disc, and obtain a valid reference point almost surely.

\begin{remark}[Heuristic for finding non-roots]\label{rem:z0_heuristic}
    In practice, for better numerical stability, we find $z_0$ by a small numerical optimization heuristic. Parametrize $z_0 = R\,e^{i\theta}$ for some small fixed $R\sim1$. We run bounded Brent minimization on the cost function $\mathcal{L}(\theta) = -\sum_{j=1}^d \log|R\,e^{i\theta} - \lambda_j|$ over $\theta \in [-\pi, \pi)$. This returns a reasonably large value of $|p(R\,e^{i\theta})|$, which boosts numerical stability since we divide by $p(R\,e^{i\theta})$. Recall that we are guaranteed that only finitely many zeros may exist on this circle, if any.
\end{remark}

We can now prove the main result of the GEV method.

\begin{proof}[Proof (of \cref{thm:gev_coefficients})]
If $(B, -C)$ is singular, then Algorithm~\ref{alg:gev_coefficients} correctly returns all zeros for the polynomial.
Otherwise, the pencil is regular and so \cref{lem:pencil_roots} proves correctness of reconstructing the polynomial from its roots.
As noted in \cref{rem:z0_heuristic} and the comment directly above, there are only finitely many forbidden values for $z_0$. 
The single Pfaffian evaluation fixes the overall scalar, and multiplication by $i^{\eta/2}2^{-r}$ gives $q$.

The numerical linear algebraic primitives---QZ algorithm and Pfaffian evaluation---have $\calO(r^3)$ cost. The remaining steps have at most quadratic cost:
\begin{enumerate}
    \item Pairing roots can be accomplished in the worst case by enumerating and checking all $\frac{r(r-1)}{2}$ pairs.
    
    \item Ordinary multiplication of the factors $(z-\lambda_j)$ costs $\calO(r^2)$ to reconstruct the polynomial coefficients, and this can be improved to $\calO(r\log^2 r)$ using FFT multiplication for polynomials~\cite{gathen2013modern}.
\end{enumerate}
Finally, we only need $\calO(r^2)$ space to store the constant number of matrices throughout the computation.
\end{proof}

\begin{remark}[Normal forms for skew-symmetric matrix pencils]
    While the QZ algorithm applies to any pencil, it does not use the skew-symmetric structure of our problem. This requires operating on a matrix roughly twice the size needed and identifying the double-multiplicity roots. Thompson~\cite{thompson1991pencils} proved the existence of a skew-symmetric Kronecker normal form, which is a pencil analog of the Jordan normal form. In principle, this affords the desired reductions;~in practice, this is not an algorithmic result, and besides, the Kronecker normal form is numerically unstable (just as the Jordan normal form also is). The theory of stratification~\cite{dmytryshyn2017structure} gives a systematic way to probe such instabilities, but does not comport with the high throughput required of QC-QMC. Our symplectic Hessenberg reduction is therefore a possible numerically stable skew-symmetric pencil normal form, although it is not fully general, as it requires one of the matrices to be invertible.
\end{remark}

\subsection{Fast practical runtime implementation}
\label{app:overlap-implementation}

Here we describe some bespoke modifications to our implementation to maximize performance in practice. 
Although $n = 40$ qubits is large compared to prior QC-QMC experiments, each individual linear-algebraic step on matrices of size $2n-\eta$ is fairly modest by modern high-performance computing standards. 
We find that a mixture of the skew-Schur complement representation from the PVL method and the polynomial interpolation approach from the original work~\cite{wan2023matchgate} strikes a good balance of practical speed and precision. 
Note that, by virtue of using interpolation, the asymptotic cost here scales as $\calO(r^4)$; nonetheless, the smaller constant prefactors remain advantageous at the scales we probe.
See \cref{fig:gpu_timimg} for the wall-time gains.

The key improvement that we make here is to use Fourier interpolation to mitigate numerical ill-conditioning.
Recall from the proof of \cref{lem:block_schur} that, assuming $C_{11}$ is invertible, we can write
\begin{equation}
    \pf(B + zC) = z^{\eta/2} \pf(C_{11}) \pf(J_s + zM), \qquad M = C_{22} + C_{12}^\T C_{11}^{-1} C_{12},
\end{equation}
where, for simplicity, we do not permute the rows and columns according to $P_s$. Let $p(z) = \pf(J_s + zM)$. 
From this form, we know that $p(z)$ has degree exactly $s = r - \eta/2$, and we can perform polynomial interpolation by evaluating $p(z)$ at $s+1$ points. 
To ameliorate the poor conditioning of polynomial interpolation in the monomial basis, we can work in the Fourier basis. 
The idea is that we do not need to compute the polynomial coefficients necessarily in the monomial basis; any basis suffices as long as we transform the coefficients $f_\ell^{-1}$ in the same way.

Let $\omega = e^{2\pi i/(s+1)}$ be a root of unity and choose nodes $z_j = R\,\omega^j$ for some radius $R > 0$ (we choose $R=1$ in our implementation). 
If $p(z) = \sum_{k=0}^s a_k z^k$ in the monomial basis, then by Fourier orthogonality we have
\begin{equation}\label{eq:a_k_coeff_fourier}
    a_k = \frac{1}{R^k (s+1)} \sum_{j=0}^{s} p(z_j) \omega^{-jk}.
\end{equation}
Thus, for any fixed coefficients $b_k$, we have
\begin{equation}
    \sum_{k=0}^s b_k a_k = \sum_{j=0}^s w_j p(z_j), \qquad \text{where } w_j = \frac{1}{s+1} \sum_{k=0}^s b_k R^{-k} \omega^{-jk}.
\end{equation}
Indeed, our goal is to compute the contraction (recall \cref{eq:coefficient_contraction,thm:block_overlap})
\begin{equation}
    X = 2\kappa \pf(C_{11}) \sum_{k=0}^s f_{k+\eta/2}^{-1} a_k,
\end{equation}
so taking $b_k = f_{k+\eta/2}^{-1}$ we can precompute all the new weights $w_j$, applicable to every shadow and every walker. Technically, we precompute the modified weights  $\widetilde{w}_j$ defined below: substituting the expression for $a_k$ from \cref{eq:a_k_coeff_fourier}, we get
\begin{equation}
    X = \pf(C_{11}) \sum_{j=0}^s \widetilde{w}_j \pf(J_s + z_j M), \qquad \text{where } \widetilde{w}_j = \frac{2\kappa}{s+1} \sum_{k=0}^s f_{k+\eta/2}^{-1} R^{-k} e^{-2\pi ijk/(s+1)}.
\end{equation}
We then evaluate the Pfaffian at the $s+1$ roots of unity $z_j$, each costing $\calO(s^3)$ operations. Unlike naive interpolation (say, with Chebyshev nodes), this Fourier transform has an $\calO(1)$ condition number.

Note that this approach requires an invertible $C_{11}$, just as in the PVL method. In the rare cases where the block is singular, we also fall back to evaluating the full pencil directly.

\subsection{Analytical derivatives}\label{app:shadow-derivatives}
The force bias and local energy in Eq.~\eqref{eq:app-local-energy} require first and second derivatives of the overlap with respect to the walker orbitals. We evaluate these derivatives using an alternative factorization of the polynomial $q(z)$ from Eq.~\eqref{eq:overlap_polynomial}, which allows us to separate the snapshot dependence from the walkers. This approach also does not require an invertible $C_{11}$. As in the previous subsection, we use a fast numerical implementation of interpolation rather than a formally cubic-scaling primitive.

\paragraph{Alternative factorization of the polynomial.}
To match the convention of \cref{app:afqmc}, let $\Phi\in\C^{n\times\eta}$ be the walker coefficients, corresponding to the first $\eta$ rows of $V^\dagger$ as defined in Eq.~\eqref{eq:slater_convention}.
For each snapshot covariance matrix $C_\varrho$, define the following skew-symmetric matrix:
\begin{equation}
    F(z)=-i\l[\I_n\otimes\begin{pmatrix}
        1 & -i
    \end{pmatrix}\r]
    (J_n+zC_\varrho)^{-1}
    \l[\I_n\otimes\begin{pmatrix}
        1 & -i
    \end{pmatrix}\r{]^\T}\in\C^{n\times n}.
    \label{eq:shadow-pairing-matrix}
\end{equation}
Note that because $J_n$ is full rank, $(J_n+zC_\varrho)^{-1}$ exists everywhere except for finitely many values of $z$, so throughout we may assume that $z$ is outside of those singularities. Equivalently, one may treat $F(z)$ as a fixed matrix whose entries are taken from the field of rational functions.
The rectangular matrix $\I_n\otimes(\begin{matrix}1 & -i\end{matrix})$ simply applies the same row $(1,-i)$ to every canonical Majorana pair $(2j-1, 2j)$ in the covariance matrix space. These are the creation-operator rows from $W$ in Eq.~\eqref{eq:W_matrix}, without the $1/\sqrt{2}$ normalization. In particular, $F(z)$ depends on the snapshot $C_\varrho$ and polynomial indeterminate $z$, while all walker dependence will enter separately through $\Phi$.

To derive the factorization, we will use the following identity for any invertible skew-symmetric matrix $A$ (Theorem 2.6 of ref.~\citenum{ishikawa2006applications}):
\begin{equation}
    \pf\l(A[\overline{X}]\r)
    =(-1)^{\sum_{j\in X} j} \, \pf(A)
       \pf\l(A^{-1}[X]\r), \qquad X \subseteq [2n],
    \label{eq:shadow-complementary-minor-identity}
\end{equation}
where $A[X]$ is the principal submatrix of $A$ given by the rows and columns indexed by $X$.
In our case, $X = S_\eta$ is the complement of $\Sbar_\eta$ and $\sum_{j\in X} j=1+3+\cdots+(2\eta-1)=\eta^2$ is always even.
Now introduce the pencil
\begin{equation}
    A(z) = W^*\Qtilde(V)(J_n+zC_\varrho)\Qtilde(V)^\T W^\dagger,
    \label{eq:shadow-full-matrix}
\end{equation}
which recovers $A(z)[\Sbar_\eta] = B+zC$. Using \cref{eq:shadow-complementary-minor-identity}, we have
\begin{equation}
    \pf(B+zC)=\pf(A(z)) \pf\l(A(z)^{-1}[S_\eta]\r).
    \label{eq:shadow-two-factors}
\end{equation}
We will evaluate these two factors separately.

For the first factor, we use the classic identity $\pf(PMP^\T)=\det(P)\pf(M)$. In Eq.~\eqref{eq:shadow-full-matrix}, this is $P = W^*\Qtilde(V)$ and $M = J_n+zC_\varrho$. Each of the $\eta$ nontrivial $2\times2$ blocks of
$W^*$ has determinant $-i$, and $\det\Qtilde(V)=1$. Thus
\begin{equation}
    \pf(A(z))
    =\det(W^*\Qtilde(V))\pf(J_n+zC_\varrho)
    =(-i)^\eta\pf(J_n+zC_\varrho).
    \label{eq:shadow-full-pfaffian}
\end{equation}
For the second factor, first invert the full matrix in Eq.~\eqref{eq:shadow-full-matrix}. Since $W$ is unitary and $\Qtilde(V)$ is real orthogonal, the inverse is
\begin{equation}
    A(z)^{-1}
    =W\Qtilde(V)(J_n+zC_\varrho)^{-1}\Qtilde(V)^\T W^\T.
    \label{eq:shadow-full-inverse}
\end{equation}
Let $\Pi_{S_\eta}$ be the analogue of $\Pi_{\Sbar_\eta}$. The rows of
$W\Qtilde(V)$ selected by $\Pi_{S_\eta}$ can be computed one block at a time: for row
$j$ and column $k$, the block multiplication is
\begin{equation}
    \frac{1}{\sqrt{2}}\begin{pmatrix}
        1 & -i
    \end{pmatrix}
    \begin{pmatrix}
        \Re V_{jk}&-\Im V_{jk}\\
        \Im V_{jk}&\Re V_{jk}
    \end{pmatrix}
    =\frac{V_{jk}^*}{\sqrt{2}}\begin{pmatrix}
        1 & -i
    \end{pmatrix}
    =\frac{\Phi_{kj}}{\sqrt{2}}\begin{pmatrix}
        1 & -i
    \end{pmatrix}.
\end{equation}
Putting these blocks together gives
\begin{equation}
    \Pi_{S_\eta} W\Qtilde(V)
    =\frac{1}{\sqrt{2}}\Phi^\T\l[\I_n\otimes\begin{pmatrix}
        1 & -i
    \end{pmatrix}\r].
    \label{eq:shadow-deleted-rows}
\end{equation}
Extracting rows and columns $S_\eta$ in Eq.~\eqref{eq:shadow-full-inverse} therefore yields
\begin{align}
    A(z)^{-1}[S_\eta]
    &=(\Pi_{S_\eta} W\Qtilde(V))(J_n+zC_\varrho)^{-1}
       (\Pi_{S_\eta} W\Qtilde(V))^\T\notag\\
    &=\frac{1}{2}\Phi^\T\l[\I_n\otimes\begin{pmatrix}
        1 & -i
    \end{pmatrix}\r]
       (J_n+zC_\varrho)^{-1}\l[\I_n\otimes\begin{pmatrix}
        1 & -i
    \end{pmatrix}\r{]^\T}\Phi\notag\\
    &=\frac i2\Phi^\T F(z)\Phi,
    \label{eq:shadow-deleted-inverse}
\end{align}
where $F(z)$ was defined above in Eq.~\eqref{eq:shadow-pairing-matrix}.

Substituting Eqs.~\eqref{eq:shadow-full-pfaffian} and
\eqref{eq:shadow-deleted-inverse} into
Eq.~\eqref{eq:shadow-two-factors} yields
\begin{equation}
    \pf(B+zC)=(-i)^\eta\pf(J_n+zC_\varrho)
                   \pf\!\left(\frac{i}{2}\Phi^\T F(z)\Phi\right).
    \label{eq:shadow-complementary-pfaffian}
\end{equation}
The matrix $\Phi^\T F(z)\Phi$ has size $\eta\times\eta$, so the second factor simplifies to
\begin{equation}
    \pf\l(\frac{i}{2}\Phi^\T F(z)\Phi\r)
    =\l(\frac{i}{2}\r{)^{\eta/2}}
       \pf\l(\Phi^\T F(z)\Phi\r).
\end{equation}
Finally, recall that $q(z)=i^{\eta/2}2^{-r}\pf(B+zC)$. The scalar prefactor reduces to
\begin{equation}
    \frac{i^{\eta/2}}{2^r}(-i)^\eta
       \left(\frac{i}{2}\right)^{\eta/2}
    =\frac{1}{2^{r+\eta/2}}=2^{-n},
\end{equation}
hence
\begin{equation}
    q(z)=2^{-n}\pf(J_n+zC_\varrho)
          \pf\!\big(\Phi^\T F(z)\Phi\big).
    \label{eq:shadow-occupied-polynomial}
\end{equation}
This is exactly the polynomial in Eq.~\eqref{eq:overlap_polynomial}.
Note that the inverse in Eq.~\eqref{eq:shadow-pairing-matrix} requires $\det(J_n+zC_\varrho)\neq0$. But because both $J_n$ and $C_\varrho$ are orthogonal matrices, the product $J_n^\T C_\varrho$ is also orthogonal and therefore its eigenvalues lie on the unit circle. Hence the zeros of
\begin{equation}
    \det(J_n+zC_\varrho) = \det(\I_{2n} + z J_n^\T C_\varrho)
\end{equation}
may only lie on the unit circle as well. 
As long as we work away from $|z|=1$, we guarantee the necessary
invertibility condition. This factor is independent of $\Phi$ and so can be fixed for each snapshot; we show how to exploit this structure below.

We now apply the Fourier contraction from before.
Recall $r=n-\eta/2$ and $s=n-\eta$. Since $q_\ell=0$ for all $\ell<\eta/2$, the polynomial $q(z)/z^{\eta/2}$ has degree at most $s$.
At the $s+1$ nodes
$z_j=Re^{i\theta}\omega^j$, with $\omega=e^{2\pi i/(s+1)}$, Fourier orthogonality gives
\begin{equation}
    q_\ell=\frac{1}{s+1}\sum_{j=0}^s q(z_j)z_j^{-\ell},
    \qquad \eta/2\leq\ell\leq r.
    \label{eq:shadow-fourier-coefficients}
\end{equation}
We use $R=0.8$ here to guarantee invertibility. For each $j$th node let us define
\begin{equation}
    F_j = F(z_j), \qquad S_j = \Phi^\T F_j \Phi.
\end{equation}
Substituting Eq.~\eqref{eq:shadow-occupied-polynomial} into $X_\varphi=2\sum_\ell f_\ell^{-1}q_\ell$ yields
\begin{equation}
    X_\varphi = \sum_{j=0}^s c_j\pf(S_j), \qquad \text{where } c_j = \frac{\pf(J_n+z_jC_\varrho)}{2^{n-1} (s+1)}
        \sum_{\ell=\eta/2}^r f_\ell^{-1}z_j^{-\ell}.
    \label{eq:shadow-occupied-estimator}
\end{equation}
The coefficients $c_j$ and matrices $F_j$ can be reused across different walkers.
After forming $S_j$, the Pfaffian and the inverse required for the
derivatives have dimension $\eta$, rather than $2n-\eta$, reducing the costs noticeably.

\paragraph{Derivative formulas.}
For a fixed snapshot $C_\varrho$ and node $j$, consider the walker
evolved by some one-body transformation generated by $H\in\C^{n\times n}$:
$S_j(t)=(e^{tH}\Phi)^\T F_j(e^{tH}\Phi)$. The derivatives to compute
the force bias and local energy are with respect to this perturbation
in $t$. We can compute the derivatives of $S_j(t)$ using the product
rule. Henceforth, all derivatives are evaluated at $t=0$ and we omit
the explicit dependence on $t$ for notational brevity:
\begin{align}
    \partial_t S_j
    &=(H\Phi)^\T F_j\Phi+\Phi^\T F_jH\Phi,\\
    \partial_t^2 S_j
    &=(H^2\Phi)^\T F_j\Phi
      +2(H\Phi)^\T F_jH\Phi+\Phi^\T F_jH^2\Phi.
\end{align}
Assume first that $S_j$ is invertible. Jacobi's formula for the derivative of the Pfaffian is
\begin{align}
    \partial_t\pf(S_j)
    &=\frac{1}{2}\pf(S_j)\tr(S_j^{-1}\partial_t S_j)\notag\\
    &=\frac{1}{2}\pf(S_j)\l[
        \tr(S_j^{-1}(H\Phi)^\T F_j\Phi)
        +\tr(S_j^{-1}\Phi^\T F_jH\Phi)\r]\notag\\
    &=\pf(S_j)\tr(S_j^{-1}\Phi^\T F_jH\Phi).
    \label{eq:shadow-first-derivative}
\end{align}
The two traces in the second line are equal: transposing the first
and using $F_j^\T=-F_j$ and $(S_j^{-1})^\T=-S_j^{-1}$ gives the
second, after cycling the factors inside the trace.

For the second derivative, differentiating $S_jS_j^{-1}=\I$ gives
\begin{equation}
    \partial_t S_j^{-1}
    =-S_j^{-1}(\partial_t S_j)S_j^{-1}.
\end{equation}
Applying the product rule to Eq.~\eqref{eq:shadow-first-derivative}
then gives
\begin{equation}
\begin{split}
    \partial_t^2\pf(S_j) &= \partial_t\pf(S_j)
            \tr(S_j^{-1}\Phi^\T F_jH\Phi)\\
            &+ \pf(S_j)\tr\l[
           -S_j^{-1}(\partial_t S_j)S_j^{-1}\Phi^\T F_jH\Phi + S_j^{-1}(H\Phi)^\T F_jH\Phi
           +S_j^{-1}\Phi^\T F_jH^2\Phi\r].
\end{split}
\end{equation}
Substituting the expressions for $\partial_t S_j$ and
$\partial_t\pf(S_j)$ and expanding yields
\begin{equation}
\begin{split}
    \partial_t^2\pf(S_j)
    =\pf(S_j)&\l\{[\tr(S_j^{-1}\Phi^\T F_jH\Phi)]^2\r.\\
        &-\tr\l[(S_j^{-1}\Phi^\T F_jH\Phi)^2\r]\\
        &-\tr\l[S_j^{-1}\Phi^\T F_jH\Phi
                     S_j^{-1}(H\Phi)^\T F_j\Phi\r]\\
        &+\tr\l[S_j^{-1}(H\Phi)^\T F_jH\Phi\r]\\
        & +\l. \tr\l[S_j^{-1}\Phi^\T F_jH^2\Phi\r]\r\}.
    \label{eq:shadow-second-derivative}
\end{split}
\end{equation}

\paragraph{Assembling the estimates with reuse.}
These formulas allow the Hamiltonian contractions to be shared across snapshots. For the first derivative, observe that
\begin{equation}
    \partial_t X_\varphi=\sum_{j=0}^s c_j\pf(S_j)
        \tr(S_j^{-1}\Phi^\T F_jH\Phi) =\tr\l[H\Phi\sum_{j=0}^s
        c_j\pf(S_j)S_j^{-1}\Phi^\T F_j\r].
\end{equation}
Thus we can sum over the nodes and snapshots before contracting with $H$. Let an overline denote the average over the measured snapshots, and define
\begin{equation}
    G=\overline{\sum_{j=0}^s
        c_j\pf(S_j)S_j^{-1}\Phi^\T F_j}\in\C^{\eta\times n}.
\end{equation}
The averaged first derivative is then
\begin{equation}
    \overline{\partial_t X_\varphi}=\tr(H\Phi G).
    \label{eq:shadow-averaged-first-derivative}
\end{equation}
For each walker, we form $\Phi G$ once and contract it with the $N_\gamma$ Cholesky vectors $L_\gamma$, which costs $\calO(N_\gamma n^2)$ operations. With $v_\gamma=iL_\gamma$, the force bias and local energy in Eq.~\eqref{eq:app-local-energy} become
\begin{align}
    \bar x_\gamma
    &=-i\sqrt{\Delta\tau}\,
      \frac{\tr(L_\gamma\Phi G)}{\overline{X_\varphi}},\\
    E_{\mathrm L}
    &=E_{\mathrm{core}}
      +\frac{\tr(v_0\Phi G)}{\overline{X_\varphi}}
      +\frac{1}{2\overline{X_\varphi}}
         \sum_\gamma
         \left.\overline{\partial_t^2 X_\varphi}\right|_{H=L_\gamma}.
\end{align}
The second-derivative term has a plus sign because $v_\gamma=iL_\gamma$. We average both the numerator and denominator over snapshots separately before taking their ratio. It remains to show how to evaluate $\overline{\partial_t^2 X_\varphi}$ efficiently.

The first four terms in the second derivative,
Eq.~\eqref{eq:shadow-second-derivative}, are quadratic in $H\Phi$. Setting $H=L_\gamma$, we can therefore collect the sum over $\gamma$ by defining the array
\begin{equation}
    \Omega_{pa,qb}
    =\sum_\gamma(L_\gamma\Phi)_{pa}(L_\gamma\Phi)_{qb},
    \qquad p,q\in[n],\quad a,b\in[\eta].
    \label{eq:shadow-walker-tensor}
\end{equation}
For example, the square of the trace in the first term gives
\begin{equation}
    \sum_\gamma
    [\tr(S_j^{-1}\Phi^\T F_jL_\gamma\Phi)]^2
    =
    \sum_{p,q=1}^n
    \sum_{a,b=1}^\eta 
    \Omega_{pa,qb}
    (S_j^{-1}\Phi^\T F_j)_{ap}
    (S_j^{-1}\Phi^\T F_j)_{bq}.
\end{equation}
Expanding the other three traces in the same way, and using
$F_j^\T=-F_j$ and $(S_j^{-1})^\T=-S_j^{-1}$, gives the full sum
\begin{equation}
\begin{aligned}
    \sum_\gamma
       \left.\partial_t^2\pf(S_j)\right|_{H=L_\gamma} &=\pf(S_j)\Biggl\{
       \sum_{p,q=1}^n\sum_{a,b=1}^\eta\Omega_{pa,qb}\Bigl[
          (S_j^{-1}\Phi^\T F_j)_{ap}
          (S_j^{-1}\Phi^\T F_j)_{bq}\\
    &\qquad\qquad
         -(S_j^{-1}\Phi^\T F_j)_{bp}
          (S_j^{-1}\Phi^\T F_j)_{aq}\\
    &\qquad\qquad
         +(F_j\Phi S_j^{-1}\Phi^\T F_j-F_j)_{pq}
           (S_j^{-1})_{ab}\Bigr]\\
    &\qquad\qquad
         +\tr\!\left[S_j^{-1}\Phi^\T F_j
             \left(\sum_\gamma L_\gamma^2\right)\Phi\right]\Biggr\}.
    \label{eq:shadow-summed-second-derivative}
\end{aligned}
\end{equation}
The last trace follows by summing the $H^2\Phi$ term separately.
Multiplying this expression by $c_j$, summing over nodes $j = 0, 1, \ldots, s$, and averaging over snapshots gives the second-derivative term in $E_{\mathrm L}$.

The array $\Omega$ has $n^2 \eta^2$ entries and depends only on the walker and the Cholesky vectors. It costs $\calO(N_\gamma n^2\eta^2)$ operations to form and can be reused for every snapshot for a fixed walker $\Phi$. The matrix $\sum_\gamma L_\gamma^2$ depends only on the Hamiltonian, so we can precompute it once and apply it to each walker $\Phi$ before processing the snapshots. Thus the contractions in Eq.~\eqref{eq:shadow-summed-second-derivative} cost $\calO(n^2\eta^2)$ per walker--snapshot pair and node, independently of $N_\gamma$.

\paragraph{Handling singularities.}
Finally, we must handle cases where $S_j$ is singular or its inverse is numerically unstable. The Pfaffian is a polynomial in the matrix entries, so its derivatives exist even when $S_j$ is singular. Likewise, the entries of $\pf(S_j)S_j^{-1}$ are (signed) Pfaffians of smaller principal submatrices, and therefore analytically extend to polynomials even where the inverse itself is undefined.

We evaluate these polynomials by shifting the matrix to nonsingular values. Choose some fixed invertible skew-symmetric matrix $\Delta$ and replace
\begin{equation}
    S_j(t)\mapsto S_j(t)+\epsilon\Delta.
\end{equation}
The shift is independent of $t$, so $\partial_t S_j$ and $\partial_t^2 S_j$ remain unchanged. Observe that the Pfaffian, its first two $t$-derivatives, and $\pf(S_j+\epsilon\Delta)(S_j+\epsilon\Delta)^{-1}\Phi^\T F_j$ are all polynomials in $\epsilon$ of degree at most $\eta/2$.
Thus we can choose $\eta/2+1$ shifts on a circle centered at zero,
\begin{equation}
    \epsilon_k=\epsilon_0 e^{2\pi i k/(\eta/2+1)},
    \qquad k=0,\ldots,\eta/2,
\end{equation}
with $\epsilon_0\neq0$ chosen so that the shifted matrices are invertible. For each power appearing in these polynomials, we have the convenient identity
\begin{equation}
    \frac{1}{\eta/2+1}\sum_{k=0}^{\eta/2}\epsilon_k^d
    =\begin{cases}
        1,&d=0,\\
        0,&1\leq d\leq\eta/2.
    \end{cases}
\end{equation}
Averaging the shifted values therefore allows us to extract the constant term, which is the desired value at $\epsilon=0$. The same averaging is applied to the Pfaffian and both derivatives.

As an implementation note, we use this procedure when $S_j$ is exactly singular, or obeys the heuristic conditioning estimate
\begin{equation}
    \max_{ab}|(S_j)_{ab}|\times\max_{ab}|(S_j^{-1})_{ab}| > 10^{10}.
\end{equation}
Using AMD GPUs, we evaluate the Pfaffian, matrix inverses, and ``small'' matrix products (i.e., smaller than the size of $\Omega$) in a fused kernel. Meanwhile, the contractions with $\Omega$ use batched matrix products.
Supplementary Fig.~\ref{fig:gpu_timimg} compares the implementation described here with automatic differentiation of the same estimator (one reverse pass for the force bias and one second-order pass per Cholesky vector for the local energy), showing a speedup of up to $80\times$ and agreement to within $5\times10^{-15}$.

\section{Improved matchgate circuit implementations}
\label{app:circuits}

\subsection{Optimized compilation of Clifford matchgate shadows}
\label{app:matchgate_compilation}

Let $\B(2n)$ be the group of signed permutation matrices. We present a new procedure to compile the Clifford matchgate circuit implementing the transformation $Q\in\B(2n)$, on a one-dimensional geometry of qubits. This has improved gate count and circuit depth compared to prior state-of-the-art~\cite{zhao2024group} when restricted to signed permutations, which was implemented in previous QC-QMC experiments~\cite{zhao2026trapped}.
We display a spin-blocked example (i.e., running the same algorithm for two independently drawn permutations; see \cref{app:spin-block}) in \cref{fig:matchgate_circuit_40q}.

\begin{theorem}[Clifford matchgate compilation for shadows]\label{thm:compiler_permutation}
    Let $n\geq2$. Given a permutation $\pi_0\in\Sym(2n)$, Algorithm~\ref{alg:compile_shadow_local} runs in $\calO(n^2)$ classical time and outputs a circuit
    \begin{equation}\label{eq:compiler_headline}
        U=V_{p_{m}q_{m}}\cdots V_{p_1q_1},\qquad \text{where }
        V_{pq}=\exp\l(-\frac\pi4\gamma_p\gamma_q\r),
    \end{equation}
    together with a new permutation $\pi$ and signs $s\in\{\pm1\}^{2n}$ such that $U^\dagger\gamma_jU=s_j\gamma_{\pi(j)}$. Here $\pi$ defines the same measurement basis as $\pi_0$, and it is obtained by ordering the pairs of $\pi_0$ as in Algorithm~\ref{alg:compiler_pair_order}. Each pair $(p_k,q_k)$ belongs to either one qubit or two neighboring qubits on a one-dimensional line. The total $V_{pq}$ count obeys
    \begin{equation}\label{eq:compiler_local_count}
        m\leq\inv(\pi)\leq n(n-1),
    \end{equation}
    with two-qubit $\CZ$ depth at most $2n$, and total circuit depth at most $4n+1$. 
    Moreover, for a uniformly random input $\pi_0$, the expected number of $\CZ$ gates, denoted $N_{\CZ}$, satisfies
    \begin{equation}\label{eq:compiler_local_expected}
        \E[N_{\CZ}]\leq\E[m]\leq\frac{n(n-1)}2.
    \end{equation}
    Each $V_{pq}$ involving two qubits requires one $\CZ$ and at most four
    single-qubit Clifford gates.
\end{theorem}

The signs $s$ are recorded only for postprocessing (e.g., to compute $Q^\T C_\varrho Q$ in the shadow-inversion step). They are a consequence of the fact that we only input an \textit{unsigned} permutation $\pi_0$, which compiles to \textit{some} circuit that implements a \textit{signed} permutation $(\pi, s)$ whose measurement basis is equivalent to that of $\pi_0$, up to reordering the bits.

\begin{algorithm}[H]
    \caption{Clifford matchgate measurement compilation:~$\CompileMatchgatePermutation(\pi_0)$}\label{alg:compile_shadow_local}
    \setcounter{AlgoLine}{0}
    \KwInput{A permutation $\pi_0 \in \Sym(2n)$ specified as an array ($n\geq2$).}
    \KwOutput{A Clifford matchgate circuit $U$, a permutation $\pi$, and signs $s$ describing the Majorana transformation $D_sP_\pi$ of $U$.}
    
    \BlankLine
    
    $\pi\gets\ReducePairs(\pi_0)$; \hfill \Comment{Algorithm~\ref{alg:compiler_pair_order}}
    
    $\scrL\gets\SortPermutation(\pi)$; \hfill \Comment{Algorithm~\ref{alg:compile_permutation}}
    
    Reverse the order of the pairs in $\scrL$;
    
    $(U,s)\gets\CompileSwaps(n,\scrL)$; \hfill \Comment{Algorithm~\ref{alg:compiler_swap_circuit}}
    
    \Return{$(U,\pi,s)$}
\end{algorithm}

Compared to prior designs that adopt the continuous $\Orth(2n)$ compilation algorithm, our gate savings arise because we are promised a discrete Majorana permutation. The main ingredients for our improvement are:
\begin{itemize}
    \item \emph{Discrete block synthesis.}
    A general $4\times4$ orthogonal block uses six Givens rotations~\cite{zhao2024group}. For a permutation, the primitive $\SortFour$ produces at most three Majorana swaps, at most two of which involve two-qubit gates (\cref{lem:compiler_two_CZ}).

    \item \emph{Direct Clifford quadratic gates.}
    We allow all six Majorana pairs within two neighboring qubits, including the $YX$, $XY$, and $YY$ generators, in addition to the commonly used $XX$ and $Z$ matchgate generators. Each cross-qubit exponential is compiled directly with one $\CZ$ (\cref{eq:compiler_one_CZ}), rather than two $\CZ$ gates needed for a generic matchgate.

    \item \emph{Ordering the measured pairs.}
    We employ the perfect matching pair reduction ($\ReducePairs$) of Heyraud \emph{et al.}~\cite{heyraud2025unified}:~order each pair $(\pi_0(2j-1), \pi_0(2j))$ and then order the pairs by their first entries. This reduces the inversion number $\inv(\pi)$ before compilation while preserving the shadow measurement basis (\cref{prop:compiler_pair_order}).

    \item \emph{Classical sign tracking.}
    We track the signs generated by the sequence of swaps and apply them in classical postprocessing rather than implementing them as quantum gates. It was previously established that these signs do not affect the shadow measurement channel~\cite{zhao2021fermionic,ogorman2022fermionic,heyraud2025unified}, so for matchgate shadows we only need to generate a uniformly random unsigned permutation $\pi_0 \in \Sym(2n)$.
\end{itemize}

\begin{remark}[Complete Clifford matchgate circuit compilation]
    Our protocol is streamlined for matchgate shadows and the invariance of the measurements;~if instead one wants to compile the matchgate circuit itself corresponding to $D_tP_{\pi_0} \in \B(2n)$, then one modifies Algorithm~\ref{alg:compile_shadow_local} by skipping $\ReducePairs$ and applying a single layer of Pauli gates at the end that implements the signs $D_{ts}$ according to Supplemental Note 4, Eq.~(127)~\cite{zhao2024group}.
\end{remark}

We now explain the correctness and compiled circuit size of our construction.

\paragraph{Conventions and elementary Majorana swaps.}
We work in the following Jordan--Wigner convention
\begin{equation}
    \gamma_{2j-1}=Z_1\cdots Z_{j-1}X_j,
    \qquad
    \gamma_{2j}=Z_1\cdots Z_{j-1}Y_j,
    \qquad j=1,\ldots,n.
\end{equation}
Since $\B(2n) = \{\pm 1\}^{2n} \rtimes \Sym(2n)$, every element $Q \in \B(2n)$ can be decomposed into permutation-and-sign form:~there exist $\pi \in \Sym(2n)$ and $s \in \{\pm 1\}^{2n}$ such that
\begin{equation}\label{eq:compiler_signed_permutation}
    Q=D_sP_\pi,
    \qquad \text{where } (P_\pi)_{jk}=\delta_{k,\pi(j)},
    \quad \text{and} \quad D_s = \diag(s).
\end{equation}
In practice, we represent $\pi = (\pi(1), \ldots, \pi(2n))$ as an array that records the nonzero column in each row.

For $p<q$, define the Majorana swap gate
\begin{equation}\label{eq:compiler_quarter_turn}
    V_{pq}=\exp\l(-\frac{\pi}{4}\gamma_p\gamma_q\r)
\end{equation}
which has the $\B(2n)$-action $\begin{pmatrix}0&-1\\1&0\end{pmatrix}$ between Majoranas $\gamma_p$ and $\gamma_q$, and acts trivially on the rest:
\begin{equation}
    V_{pq}^\dagger\gamma_pV_{pq}=-\gamma_q, \qquad V_{pq}^\dagger\gamma_qV_{pq}=\gamma_p.
\end{equation}
Note that this signed permutation has determinant $1$. We record two useful lemmas for our circuit construction.

\begin{lemma}[Nearest-neighbor Majorana swap gates]\label{lem:compiler_local_swaps}
    A swap of two Majoranas in the set $\{2j-1,2j,2j+1,2j+2\}$ can be implemented, up to a known sign, by a Clifford circuit on qubits $j,j+1$. We can implement each primitive swap with at most one $\CZ$ gate and four single-qubit Clifford gates. In the case of either $(2j-1,2j)$ or $(2j+1,2j+2)$, the unitary is simply a phase gate $S$.
\end{lemma}

\begin{proof}
    Write $V_{pq}=\exp(-i\pi P_{pq}/4)$ where $P_{pq}=-i\gamma_p\gamma_q$. The six cases are
    \begin{equation}\label{eq:compiler_local_paulis}
    \begin{array}{c|cccccc}
        (p-2j+1,q-2j+1)&(0,1)&(2,3)&(1,2)&(0,2)&(1,3)&(0,3)\\\hline
        P_{pq}&Z_j&Z_{j+1}&X_jX_{j+1}&-Y_jX_{j+1}&X_jY_{j+1}&-Y_jY_{j+1}.
    \end{array}
    \end{equation}
    The first two give a phase gate $S$. Each remaining Pauli can be mapped to $\pm Z_jZ_{j+1}$ by single-qubit Clifford gates. Since $\exp(-i\pi Z_jZ_{j+1}/4)$ equals $(S_jS_{j+1})\CZ_{j,j+1}$ up to global phase, we only need one $\CZ$.
\end{proof}

\begin{lemma}[Updating signs]\label{lem:compiler_sign_update}
    Suppose a Clifford matchgate circuit $U$ currently implements $D_sP$. The updated circuit $V_{pq}U$ implements $D_{s'}T_{pq}P$, where
    \begin{equation}
        s'_j = \begin{cases}
            -s_q & j = p,\\
            \phantom{-}s_p & j = q,\\
            \phantom{-}s_j & \text{else}.
        \end{cases}
    \end{equation}
    The matrix $T_{pq} \in \B(2n)$ is the unsigned transposition that maps $p \leftrightarrow q$.
\end{lemma}

\begin{proof}
    Let $D^{(p)} = \diag(1,\ldots,1,-1,1,\ldots,1)$ with the $-1$ in the $p$th entry. The primitive $V_{pq}$ implements $R_{pq}=D^{(p)}T_{pq}$. Hence $V_{pq}U$ implements
    \begin{equation}
        R_{pq}(D_sP)
        =D^{(p)}(T_{pq}D_sT_{pq})(T_{pq}P).
    \end{equation}
    Conjugation by $T_{pq}$ exchanges the two entries of $s$, after which $D^{(p)}$ flips the entry at $p$.
\end{proof}

\paragraph{Ordering the measured pairs.}
Before sorting, the shadow generator applies the pair reduction of Algorithm 1~\cite{heyraud2025unified}, which exploits the freedom to relabel the outcomes of computational basis measurements. 
For a permutation $\pi_0$, the measured Majorana pairs are $(\pi_0(2j-1),\pi_0(2j))$, for $j=1,\ldots,n$. 
We record the pair-ordering reduction in Algorithm~\ref{alg:compiler_pair_order}, which first orders each pair internally, then orders the pairs by their first entries. 
The resulting permutation $\pi$ gives an equivalent measurement basis to $\pi_0$, except with a minimized inversion number (which affects the gate count when compiled to a one-dimensional line of qubits).

\begin{algorithm}[H]
    \caption{Ordering canonical pairs:~$\ReducePairs(\pi_0)$}\label{alg:compiler_pair_order}
    \setcounter{AlgoLine}{0}
    \KwInput{A permutation array $\pi_0$ of length $2n$.}
    \KwOutput{A permutation $\pi$ with the same unordered Majorana pairs.}
    
    \BlankLine
    
    $\scrM\gets$ empty list;
    
    \For{$j=1,\ldots,n$}{
        $(p,q)\gets(\pi_0(2j-1),\pi_0(2j))$;
        
        Append $(\min\{p,q\},\max\{p,q\})$ to $\scrM$;
    }
    
    Sort $\scrM$ by the first entry of each pair;
    
    $\pi\gets$ the array obtained by concatenating the pairs in $\scrM$;
    
    \Return{$\pi$}
\end{algorithm}

It was previously shown that this pair reduction yields an equivalent measurement basis while improving the permutation's complexity.

\begin{proposition}[Pair ordering~\cite{heyraud2025unified}]\label{prop:compiler_pair_order}
    Algorithm~\ref{alg:compiler_pair_order} preserves the measurement basis up to outcome labels and satisfies $\inv(\pi)\leq\inv(\pi_0)$. Hence, for uniform $\pi_0$, the matchgate shadow channel and variances are preserved.
\end{proposition}

\paragraph{Sorting four Majoranas at a time.}
Our construction uses the same grouping of four Majoranas per two neighboring qubits, as in previous work~\cite{zhao2024group}. The additional promise that $Q$ is a signed permutation lets us replace their continuous block elimination by instead sorting four integer labels on each neighboring qubit pair. $\SortPermutation$ (Algorithm~\ref{alg:compile_permutation}) records the swaps that sort $\pi$ to the identity. Note that the final protocol $\CompileMatchgatePermutation$ (Algorithm~\ref{alg:compile_shadow_local})
reverses this list before compiling it, which is necessary to implement the desired permutation $\pi$ (rather than $\pi^{-1}$).

\begin{algorithm}[H]
    \caption{Blocked four-Majorana sorting network:~$\SortPermutation(\pi)$}\label{alg:compile_permutation}
    \setcounter{AlgoLine}{0}
    \KwInput{A permutation array $\pi=(\pi(1),\ldots,\pi(2n))$, with $n\geq2$.}
    \KwOutput{An ordered list $\scrL$ of position swaps that sorts $\pi$ to $(1,\ldots,2n)$.}
    
    \BlankLine
    
    $a\gets\pi$;
    
    $\scrL\gets$ empty list;
    
    \Repeat{$\changed=\false$}{
        $\changed\gets\false$;
        
        \For{$d=0,1$}{
            \For{$j=2-d,4-d,\ldots$ with $j<n$}{
                $b\gets2j-1$;
                
                $(y,\scrR)\gets\SortFour(a_b,a_{b+1},a_{b+2},a_{b+3})$;
                
                $(a_b,a_{b+1},a_{b+2},a_{b+3})\gets y$;
                
                \For{each $(p,q)$ in $\scrR$, in the returned order}{
                    Append $(b+p-1,b+q-1)$ to $\scrL$;
                }
                
                \If{$\scrR$ is nonempty}{$\changed\gets\true$}
            }
        }
    }
    
    \Return{$\scrL$}
\end{algorithm}

For an array $a=(a_1,\ldots,a_{2n})$, the qubit pair $(j,j+1)$ contains the four positions $2j-1,2j,2j+1,2j+2$. We alternate the disjoint
pairs
\begin{equation}\label{eq:compiler_brickwork}
    (2,3),(4,5),\ldots;
    \qquad (1,2),(3,4),\ldots
\end{equation}
which is a block version of odd--even transposition sorting~\cite{habermann1972parallel}.
This framework was introduced previously for compiling continuous matchgate circuits~\cite{zhao2024group}; our version simplifies it by assuming the input is a matchgate permutation. Rather than block eliminations used there (a four-dimensional generalization of Givens rotations eliminations), each block of size-four permutations is locally sorted to yield a minimally sized sequence of nearest-neighboring Majorana swaps.
This local operation $\SortFour$ is described in Algorithm~\ref{alg:compiler_local_sort}.

\begin{algorithm}[H]
    \caption{Sorting one four-Majorana block:~$\SortFour(x)$}\label{alg:compiler_local_sort}
    \setcounter{AlgoLine}{0}
    \KwInput{Four distinct integer labels $x=(x_1,x_2,x_3,x_4)$.}
    \KwOutput{The sorted tuple $x$, and an ordered list $\scrR$ of swaps on its four positions.}
    
    \BlankLine
    
    $y\gets$ the entries of $x$ in increasing order;
    
    $\scrR\gets$ empty list;
    
    \For{$p=1,2,3,4$}{
        \While{$x_p\neq y_p$}{
            Let $q$ be the unique position with $y_q=x_p$;
            
            Exchange $x_p$ and $x_q$;
            
            Append $(p,q)$ to $\scrR$;
        }
    }
    
    \Return{$(x,\scrR)$}
\end{algorithm}

\begin{lemma}[Cost of a local four-label sort]\label{lem:compiler_two_CZ}
    Algorithm~\ref{alg:compiler_local_sort} sorts its input with
    at most three swaps, at most two of which couple two qubits. Its swap count is at most the inversion number of its input, and its reversed swap list can be implemented with
    at most two $\CZ$ gates.
\end{lemma}

\begin{proof}
    The correctness of sorting is standard:~at the start of the iteration for $p$, all positions below $p$ already contain their target labels. If $x_p\neq y_p$, the destination $q$ of $x_p$ therefore satisfies $q>p$. Exchanging positions $p,q$ puts the label $x_p$ in its final position $q$. This leaves every previously fixed position unchanged, and so repeating this step eventually puts $y_p$ at $p$ and hence sorts the block.

    To count the swaps, we replace the four labels by their ranks $1,2,3,4$. At a fixed $p$, the while loop follows the cycle containing $p$ and removes one position from that cycle at each swap. A cycle of length $\ell$ therefore gives $\ell-1$ swaps, all involving its smallest position $p$. This is the standard minimal transposition factorization;~if the local permutation has $c$ cycles, including fixed points, then the total is $4-c\leq3$ (Sec.~26.13)~\cite{nistpermutations}. Sorting by adjacent inversions is another transposition factorization, with exactly $\inv(x)$ swaps, so we have $4-c\leq\inv(x)$.

    Only a four-cycle requires three swaps:~it uses the smallest position as its common endpoint and exchanges it once with each of the other three positions. Exactly one of those positions is on the same qubit, and the other two are on the neighboring qubit. Meanwhile, every other cycle type requires at most two swaps in total. The $\CZ$ bound therefore follows from
    \cref{lem:compiler_local_swaps}.
\end{proof}

\paragraph{Compiling the swap list.}
After $\SortPermutation$ (Algorithm~\ref{alg:compile_permutation}), we reverse the list $\scrL$ to bring it to the correct order for the quantum circuit. Then Algorithm~\ref{alg:compiler_swap_circuit} below compiles it into $\CZ$ and single-qubit Clifford gates. It assumes input pairs satisfy $p<q$ by convention and lie within one two-qubit local block. For a two-qubit pair, write the entry of \cref{eq:compiler_local_paulis} as $P_{pq}=\epsilon A_jB_{j+1}$, where $\epsilon\in\{\pm1\}$ and $A,B\in\{X,Y\}$. Let $H$ denote the Hadamard gate, and write
\begin{equation}\label{eq:compiler_basis_changes}
    C_X=H,\qquad C_Y=SH,
    \qquad C_A Z C_A^\dagger=A.
\end{equation}
The following explicit form combines the local basis changes with the phase gates:
\begin{equation}\label{eq:compiler_one_CZ}
    e^{-i\epsilon\pi A\otimes B/4}
    =e^{-i\epsilon\pi/4}
      (C_AS^\epsilon\otimes C_BS^\epsilon)
      \CZ
      (C_A^\dagger\otimes C_B^\dagger).
\end{equation}
Thus, each such swap has two single-qubit Clifford gates before
the $\CZ$ and two after it. We suppress all global phases throughout, although they can in principle be tracked if desired (e.g., for different applications that involve the controlled circuit).

Algorithm~\ref{alg:compiler_swap_circuit} produces a sign vector $s$ from the swap list, describing the circuit's actual action $Q=D_sP_\pi$. Thus, even if one begins with an unsigned permutation as input (as we do in Algorithm~\ref{alg:compile_shadow_local}), we obtain a sign vector to track.
In particular,
\begin{equation}\label{eq:compiler_measured_observable}
    U^\dagger Z_jU
    =s_{2j-1}s_{2j}
      \l(-i\gamma_{\pi(2j-1)}\gamma_{\pi(2j)}\r).
\end{equation}

\begin{algorithm}[H]
    \caption{$\CZ$ gate set compilation and sign tracking:~$\CompileSwaps(n,\scrL)$}\label{alg:compiler_swap_circuit}
    \setcounter{AlgoLine}{0}
    \KwInput{The number of qubits $n$ and a list $\scrL$ of pairs $(p,q)$, in circuit order, with $p<q$.}
    \KwPromise{Every pair $(p,q)$ spans a neighboring qubit pair in a one-dimensional line.}
    \KwOutput{A circuit $U$ for these Majorana swaps and its accumulated sign vector $s$.}
    
    \BlankLine
    
    $U\gets$ empty circuit;
    
    $s\gets(1,\ldots,1)$;
    
    \For{each $(p,q)$ in $\scrL$, in order}{
        $j\gets\lceil p/2\rceil$;
        
        \eIf{$q=2j$}{
            Append $S_j$ to $U$;
        }{
            Read $\epsilon,A,B$ for $(p,q)$ from \cref{eq:compiler_local_paulis};
            
            Append $C_A^\dagger$ on qubit $j$ and $C_B^\dagger$ on qubit $j+1$;
            
            Append $\CZ_{j,j+1}$;
            
            Append $C_AS^\epsilon$ on qubit $j$ and $C_BS^\epsilon$ on qubit $j+1$;
        }
        Update $(s_p,s_q)\gets(-s_q,s_p)$;
    }
    
    Merge consecutive single-qubit Clifford gates on each wire;
    
    \Return{$(U,s)$}
\end{algorithm}

\paragraph{Correctness and resource bounds.}
We now prove the correctness of Algorithm~\ref{alg:compile_shadow_local} and establish the final circuit size. 

\begin{proof}[Proof (of \cref{thm:compiler_permutation})]
    Write the sorting list as $\scrL=((a_1,b_1),\ldots,(a_m,b_m))$ and put $T_k=T_{a_kb_k}$. Each swap exchanges two rows of $P_\pi$, and the sorted array represents the identity. Thus
    \begin{equation}\label{eq:compiler_reverse_order}
        T_m\cdots T_1P_\pi=\I,\qquad
        P_\pi=T_1\cdots T_m.
    \end{equation}
    Reading $\scrL$ backward sets each $(p_k,q_k)=(a_{m+1-k},b_{m+1-k})$ in \cref{eq:compiler_headline} and produces the desired permutation.
    \Cref{lem:compiler_local_swaps,lem:compiler_sign_update}
    and Algorithm~\ref{alg:compiler_swap_circuit} give a circuit
    $U$ implementing $D_sP_\pi$, with exactly the signs returned
    by the algorithm. \Cref{prop:compiler_pair_order} shows that
    its measurement basis agrees with that specified by $\pi_0$.
    
    To bound the number of swaps, recall the standard inversion
    count:
    \[
        \inv(a)=|\{(p,q):p<q,\ a_p>a_q\}|.
    \]
    Completely sorting a contiguous four-label block decreases the global inversion count by exactly the inversion count of that block before sorting (\cref{lem:compiler_two_CZ}). This removes its internal inversions, while its multiset of labels, and hence all inversions with positions \textit{outside} the block, are unchanged. By the same lemma, the number of swaps this block incurs is at most this decrease. There are no inversions within an ordered pair, and since $\ReducePairs$ already ordered pairs by their first entries, each unordered pair of matched pairs contributes at most two inversions. Thus we have
    \begin{equation}
        m\leq\inv(\pi)\leq2\binom n2=n(n-1).
    \end{equation}

    To bound the depth, first observe that Algorithm~\ref{alg:compile_permutation} sorts any permutation within at most $n$ sweeps of the four-Majorana blocks. Indeed, this is nothing but the well-known odd--even sort~\cite{habermann1972parallel}, regarding each block of four labels as a pair of pairs. This also implies that the classical work is $\calO(n^2)$, dominated by sorting sweeps. By \cref{lem:compiler_two_CZ}, each block requires two $\CZ$ gates, implying a depth of at most $2n$.
    All other gates within a block are single-qubit Cliffords, which we can merge to yield the structure
    \begin{equation}
        M_{\mathrm{block}} = (C_1 \otimes C_2) \CZ (C_3 \otimes C_4) \CZ (C_5 \otimes C_6).
    \end{equation}
    The last single-qubit layer of one sweep merges with the first of the next, implying a single-qubit depth bound of at most $2n+1$.
    
    For a uniform input $\pi_0$, the reduced matching is uniform~\cite{ogorman2022fermionic,heyraud2025unified}. Choose two of its matched pairs uniformly and condition on their four endpoints
    $a<b<c<d$. The three possible pairings
    \begin{equation}
        \{\{a,b\},\{c,d\}\},\qquad
        \{\{a,c\},\{b,d\}\},\qquad
        \{\{a,d\},\{b,c\}\}
    \end{equation}
    are equally likely. After $\ReducePairs$, they contribute respectively $0$, $1$, and $2$ inversions. There are no inversions within an ordered pair, so summing over the $\binom n2$ unordered pairs of matched pairs gives
    \begin{equation}
        \E[\inv(\pi)]
        =\binom n2\frac{0+1+2}{3}
        =\frac{n(n-1)}2,
    \end{equation}
    recovering the expectation used in ref.~\citenum{heyraud2025unified}. Each two-qubit Majorana swap requires one $\CZ$, while each single-qubit swap requires none. Hence $N_{\CZ}\leq m\leq\inv(\pi)$, proving \cref{eq:compiler_local_expected}.

    It is well known that the odd--even sort yields an $\calO(n)$-depth swap network~\cite{habermann1972parallel};~our blocked version expands each nearest neighbor into $\calO(1)$-sized blocks, so that argument also applies to our scheme.
    We only use $\calO(n^2)$ classical work, dominated by the cost of sorting length-$\calO(n)$ lists. This proves all claims.
\end{proof}

\paragraph{Comparison to optimal Haar-random constructions.}
The work of Braccia \textit{et al.}~\cite{braccia2025optimal} provides a gate- and depth-optimal template for random matchgate circuits;~this is advantageous over prior designs by fixing the circuit template (only needing to adjust rotation angles) and requiring $\calO(n^2)$ classical effort to generate a random instance, rather than previous $\calO(n^3)$-time algorithms. We opted for our new construction instead, which we observed yields smaller circuit sizes while still requiring only quadratic classical cost. We attribute this observation to a few properties:
\begin{itemize}
    \item While they prove optimality for continuous matchgates, the authors explicitly acknowledge that the depth of the Clifford matchgate template is not optimal~\cite{braccia2025optimal}. Specifically, they use a triangular structure which has total and two-qubit depth at most $6n-7$ and $4n-6$, respectively (see their Lemma 3). Our design uses a more parallelized structure, achieving total and two-qubit depth bounds of $4n+1$ and $2n$, respectively.
    
    \item Optimality of their fixed template is in expectation over the uniform measure;~each specific instance may have a shorter-depth realization, which is what our construction targets. This is significant when implementing on noisy quantum hardware.
    
    \item They prove that the expected number of two-qubit gates in their design is $\frac{1}{2} n (n-\frac{1}{2})$ and essentially optimal, which is nearly identical to our \cref{eq:compiler_local_expected}.
\end{itemize}

\subsection{Use of spin-blocked matchgate shadows}
\label{app:spin-block}
We use \emph{spin-blocked} matchgate shadows as introduced by Zhao and Miyake~\cite{zhao2024group}. This nearly halves the circuit depth and gate count, since they scale linearly and quadratically with the effective system size, respectively. This applies to the walker overlaps that we calculate in our chemistry experiments, since the associated Slater determinants factorize into $\alpha$ and $\beta$ spin sectors. 
In \cref{fig:matchgate_circuit_40q} we show a compiled example for 40 qubits:~the $\alpha$ register (qubits 0--19) and the $\beta$ register (qubits 20--39) are acted on by independent random circuits, with no two-qubit gate between them.
Due to the smaller effective system size and shallower depth, these circuits admit faster classical postprocessing ($n \to n/2$), as well as experimentally lower overlap error measured from the noisy hardware (\cref{fig:SI_hchain_conv}).

We do not introduce new theoretical results in this section, as we already carried out the spin-blocked analysis in previous work~\cite{zhao2024group}. 
Rather, here we review and clarify the theory and validity of spin-blocked shadows for the overlap estimation problem.

\paragraph{Formalism.}
Let $n_\alpha,n_\beta$ and $\eta_\alpha,\eta_\beta$ be the number of orbitals and electrons per spin, respectively. Order all $\alpha$ modes before all $\beta$ modes. A spin-blocked matchgate transformation is block diagonal,
\begin{equation}
    Q = \begin{pmatrix}
        Q_\alpha & 0\\
        0 & Q_\beta
    \end{pmatrix} \in \Orth(2n_\alpha) \oplus \Orth(2n_\beta),
\end{equation}
and a Slater
matrix has the form
\begin{equation}\label{eq:spin_blocked_slater}
    A=\begin{pmatrix}A_\alpha&0\\0&A_\beta\end{pmatrix},\qquad
    A_\sigma\in\C^{\eta_\sigma\times n_\sigma},
\end{equation}
with orthonormal rows. With occupied $\alpha$ rows placed before occupied $\beta$ rows, its creation operators give
\begin{equation}\label{eq:spin_blocked_factorization}
    \ket{\varphi}=
    \left(\prod_{j=1}^{\eta_\alpha}d_{\alpha,j}^\dagger\right)
    \left(\prod_{k=1}^{\eta_\beta}d_{\beta,k}^\dagger\right)
    \ket{\vac}
    =\ket{\varphi_\alpha}\otimes\ket{\varphi_\beta}, \qquad d_{\sigma,j}^\dagger = \sum_{k=1}^{n_\sigma} (A_{\sigma})_{jk}^* a_k^\dagger.
\end{equation}
The vacuum also factorizes, so
\begin{equation}\label{eq:spin_transition_factorization}
    \ketbra{\varphi}{\vac}=\ketbra{\varphi_\alpha}{\vac_\alpha} \otimes \ketbra{\varphi_\beta}{\vac_\beta}.
\end{equation}
For even $\eta_\alpha,\eta_\beta$, both factors lie in the even operator subspace $\Gamma_{\even,\sigma}$ and hence are in the image of each spin sector's shadow measurement channel
\begin{equation}
    \Mchan_{\alpha,\beta} = \Mchan_{\alpha} \otimes \Mchan_{\beta}.
\end{equation}
Let $\varrho_\alpha(Q_\alpha, b_\alpha) \otimes \varrho_\beta(Q_\beta, b_\beta)$ be an i.i.d.\ spin-blocked snapshot of $\frac{\ket{\vac}+\ket{\Psi_\T}}{\sqrt{2}}$. The inverse on the joint outcome yields
\begin{equation}\label{eq:spin_opening_estimator}
    2\tr\l[\ketbra{\varphi}{\vac}
       (\Mchan_\alpha^{-1}\otimes\Mchan_\beta^{-1})
       (S_\alpha\otimes S_\beta)\r]
    =\frac{X_\alpha X_\beta}{2}, \qquad X_\sigma=2\tr\l[\ketbra{\varphi_\sigma}{\vac_\sigma} \Mchan_\sigma^{-1}(\varrho_\sigma(Q_\sigma, b_\sigma)) \r].
\end{equation}
Thus $\E\l[X_\alpha X_\beta/2\r]=\braket{\Psi_\T | \varphi}$. Importantly, this does not require any symmetry of the quantum state, although each $\eta_\sigma$ must be even (as is the case with standard matchgate shadow overlaps). We can handle odd electron numbers by appending a dummy occupied orbital to the trial state, e.g., $\ket{\Psi_\T} \to \ket{1}_\alpha\ket{1}_\beta\ket{\Psi_\T}$. The Slater coefficient matrices must also be extended via
\begin{equation}
    A_\sigma'=\begin{pmatrix}1 & 0\\
        0 & A_\sigma\end{pmatrix}
\end{equation}
Finally, we can apply the same polynomial subroutines from \cref{app:overlap-algorithm} separately to the two spin blocks without any modification. The classical computational cost therefore improves from $\calO((n_\alpha+n_\beta)^3)$ to $\calO(n_\alpha^3 + n_\beta^3)$.
\SupplementaryReferences

\end{document}